\documentclass[12pt,a4paper]{article}

\usepackage[utf8]{inputenc}
\usepackage{a4wide}                              
\usepackage{float}                               
\usepackage{amsmath,amssymb}
\numberwithin{equation}{section}
\usepackage{graphicx}                            
\usepackage{setspace}                            
\usepackage{tikz}
\usetikzlibrary{math}                            
\usepackage{subcaption}                          
\usepackage[ruled,vlined,linesnumbered]{algorithm2e}
\usepackage{tabularx,booktabs}                   
\usepackage{threeparttable}                      
\usepackage{cite}
\usepackage{hyperref}                            
\usepackage{cleveref}                            

\crefname{algocf}{algorithm}{algorithms}
\Crefname{algocf}{Algorithm}{Algorithms}
\crefname{appendix}{appendix}{appendices}
\Crefname{appendix}{Appendix}{Appendices}

\newcommand{\code}[1]{\texttt{#1}}
\begin{document}
	
	\pagestyle{plain}
	\pagestyle{empty}
	
	\vspace{0.5cm}
	
	\begin{center}
		
		{\LARGE \bf{Sampling Triangulations and Calabi-Yau Threefolds with Autoregressive GNNs}
    \\[10mm]}
	\end{center}

	\begin{center}
		\scalebox{0.95}[0.95]{{\fontsize{14}{30}\selectfont Nate MacFadden$^{a}$}}
	\end{center}

	\begin{center}
		\vspace{0.25 cm}
		\textsl{$^{a}$Department of Physics, Cornell University, Ithaca, NY 14853 USA}\\
		
		\vspace{1cm}
		\normalsize{\bf Abstract} \\[8mm]

	\end{center}
        We introduce `dualGNN', an autoregressive message-passing GNN for sampling fine, regular triangulations of lattice polytopes. dualGNN operates on a generalization of the dual graph of a triangulation, with edges labeled by `signed circuits' --- combinatorial invariants from the theory of oriented matroids. We show that these circuits are necessary and sufficient to determine a triangulation's regularity from the graph, provided certain magnitude information is retained. 
        The model is independent of the polytope's point count and invariant under its orientation-preserving symmetries ($\mathrm{SL}(d,\mathbb{Z}) \ltimes \mathbb{Z}^d$), and our masking procedure further guarantees that every rollout produces a fine triangulation (in 2D). 
        On unseen polygons with $N_\mathrm{pts} \leq 40$, dualGNN is the only sampler we tested that is consistent with uniform sampling across all our diagnostics (KL divergence from uniformity, collision counts, and sample autocorrelation). 
        The model is small ($\sim92$k parameters) and trains in $\sim7.5$ hours on a single consumer GPU. We apply dualGNN to string theory, sampling Calabi-Yau threefolds uniformly at $h^{1,1}=86$; we also sample CYs at $h^{1,1}=128$, observing no deviations from uniformity, but our diagnostics are weaker here. Code, training scripts, and pretrained models are available at \url{https://github.com/natemacfadden/dualGNN} (\code{pip install dualgnn}), and dualGNN is integrated into CYTools.
	\begin{center}
		\begin{minipage}[h]{15.0cm}

		\end{minipage}
	\end{center}
	\newpage
	\setcounter{page}{1}
	\pagestyle{plain}
	\renewcommand{\thefootnote}{\arabic{footnote}}
	\setcounter{footnote}{0}
	%
	%
	\tableofcontents
	\newpage

\section{Introduction}
\label{sec:intro}

Sampling fine, regular triangulations (FRTs) of lattice polytopes is a long-studied (over two decades~\cite{meyer2002enumeration,Kaibel_Ziegler_2003}) discrete, combinatorial problem with rich structure. It involves
\begin{enumerate}
    \item large scale (the largest polygon relevant to our applications has $\geq3.9\times10^{167}$ FRTs~\cite{macfadden2026boundingkreuzerskarkelandscape}),
    \item local constraints (triangulation fineness and validity),
    \item global constraints (triangulation regularity), and
    \item nontrivial symmetries (lattice polytopes are $\mathrm{GL}(d,\mathbb{Z}) \ltimes \mathbb{Z}^d$ invariant).
\end{enumerate}
There are a wide variety of approaches to this problem\footnote{Subsequent to our original arXiv posting, another approach \cite{wang2026trisearchlearningoptimizetriangulations} was also posted to arXiv. We do not discuss it here other than noting that it operates on $4$D triangulations directly, so the redundancy concerns we later raise for CYTransformer in this section apply, and could likewise be addressed by the 2-face reduction of \cite{macfadden2026efficientalgorithmgeneratinghomotopy}. I.e., restrict to flips modifying the $2$-face triangulations. This can be done using existing code in CYTools.} including, but not limited to those discussed in \cite{meyer2002enumeration,Kaibel_Ziegler_2003,demirtas2020boundingkreuzerskarkelandscape,yip2025transformingcalabiyauconstructionsgenerating} as well as \code{grow2d} from CYTools. Despite this long study and rich literature, all existing approaches tend to struggle with scale, generality, or bias; there is no efficient\footnote{We return to \cite{Kaibel_Ziegler_2003} later.} algorithm that provides (even approximately) uniform FRT samples for general polygons. There is need for such capabilities: these triangulations define Calabi-Yau threefolds, so such samples of triangulations provide much needed statistical probes into the landscape of string theory. This was the motivation of \cite{demirtas2020boundingkreuzerskarkelandscape,yip2025transformingcalabiyauconstructionsgenerating}.

To address this sampling problem, we introduce dualGNN, an autoregressive message-passing GNN~\cite{gilmer2017neuralmessagepassingquantum} 
built on the theory of oriented matroids, a general combinatorial structure underlying disparate areas of discrete mathematics including that of triangulations of a polytope. Exploiting this structure, dualGNN is the only sampler we tested that is consistent with uniformity across all our diagnostics, generalizes (zero-shot!) to unseen polygons, and scales to large polygons. Our studied application of dualGNN is for sampling Calabi-Yau manifolds up to $h^{1,1}=128$, but this problem can be viewed more generally as learning to sample certain subsets of signed circuits of the matroid, as we briefly discuss later.

\subsection{Triangulations}

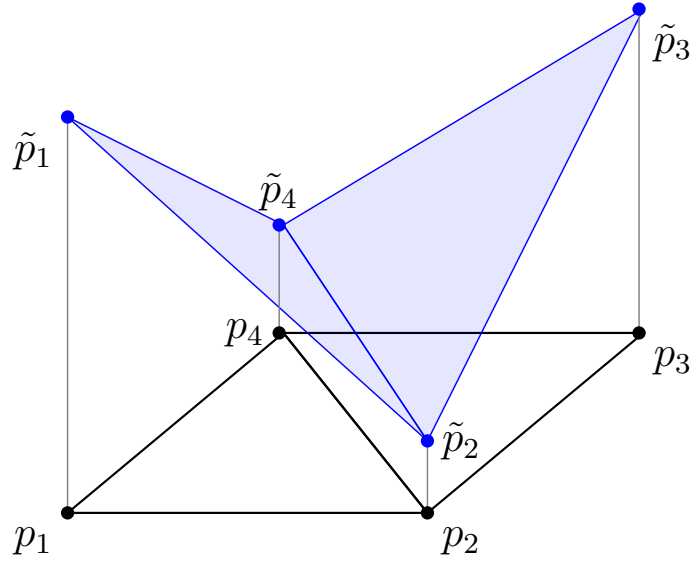
\begin{figure}[ht]
    \centering
    \resizebox{0.6\textwidth}{!}{
    \begin{tikzpicture}[scale=3.5]
        \draw[black, semithick] (0, 0) -- (1, 0) -- (0+0.6, 0.5) -- cycle;
        \draw[black, semithick] (1+0.6, 0.5) -- (1, 0) -- (0+0.6, 0.5) -- cycle;
    
        \fill[blue,opacity=0.1] (0, 0+1.1) -- (1, 0+0.2) -- (0+0.6, 0.5+0.3) -- cycle;
        \fill[blue,opacity=0.1] (1+0.6, 0.5+0.9) -- (1, 0+0.2) -- (0+0.6, 0.5+0.3) -- cycle;
        \draw[blue] (0, 0+1.1) -- (1, 0+0.2) -- (0+0.6, 0.5+0.3) -- cycle;
        \draw[blue] (1+0.6, 0.5+0.9) -- (1, 0+0.2) -- (0+0.6, 0.5+0.3) -- cycle;
    
        \foreach \x in {0,1} {
            \foreach \y in {0,0.5} {
                \pgfmathsetmacro\height{
                    ifthenelse(\x==0 && \y==0, 1.1,
                    ifthenelse(\x==1 && \y==0, 0.2,
                    ifthenelse(\x==1 && \y==0.5, 0.9, 
                    ifthenelse(\x==0 && \y==0.5, 0.3, ))))
                }
                \draw[thin, gray] (\x+1.176*\y,\y) -- (\x+1.176*\y,\y+\height);
            }
        }
    
        \node[fill=black, circle, inner sep=1.3pt] at (0,0) {};
        \node[fill=black, circle, inner sep=1.3pt] at (1,0) {};
        \node[fill=black, circle, inner sep=1.3pt] at (1+1.176*0.5,0.5) {};
        \node[fill=black, circle, inner sep=1.3pt] at (0+1.176*0.5,0.5) {};

        \node[below left] at (0,0) {$p_1$};
        \node[below right] at (1,0) {$p_2$};
        \node[below right] at (1+1.176*0.5,0.5) {$p_3$};
        \node[left] at (0+1.176*0.5,0.5) {$p_4$};
    
        \node[fill=blue, circle, inner sep=1.3pt] at (0,0+1.1) {};
        \node[fill=blue, circle, inner sep=1.3pt] at (1,0+0.2) {};
        \node[fill=blue, circle, inner sep=1.3pt] at (1+1.176*0.5,0.5+0.9) {};
        \node[fill=blue, circle, inner sep=1.3pt] at (0+1.176*0.5,0.5+0.3) {};

        \node[below left] at (0,0+1.1) {$\tilde{p}_1$};
        \node[right] at (1,0+0.2) {$\tilde{p}_2$};
        \node[below right] at (1+1.176*0.5,0.5+0.9) {$\tilde{p}_3$};
        \node[above] at (0+1.176*0.5,0.5+0.3) {$\tilde{p}_4$};

    \end{tikzpicture}
    }
    \caption{Diagram of the `lifting' procedure defining regular triangulations. The points $p_1$, $p_2$, $p_3$, and $p_4$ are embedded into $\mathbb{R}^3$ and then lifted by heights $h_1=1.1$, $h_2=0.2$, $h_3=0.9$, and $h_4=0.3$. The convex hull of the lifted point configuration is a $3$-simplex whose lower faces are plotted in blue. Projecting out the lifted coordinate generates the regular triangulation plotted in black. Figure modified from ref.~\cite{macfadden2026efficientalgorithmgeneratinghomotopy}.}
    \label{fig:lifting}
\end{figure}

\begin{figure}
    \centering
    \tikzset{
        latticepoint/.style={draw, circle, inner sep=1.2pt, fill},
        trilinestyle/.style={line width=0.25mm}
    }
    \newcommand{\tri}[6]{%
        \draw[trilinestyle]
            (#1*\scale,#2*\scale) -- (#3*\scale,#4*\scale) --
            (#5*\scale,#6*\scale) -- cycle;
    }
    \newcommand{\latticepoints}[2]{%
        \foreach \x in {#1,...,#2}{
            \foreach \y in {0,...,4}{
                \node[latticepoint] at (\x*\scale,\y*\scale) {};
            }
        }
    }
    \newcommand{\tritrianglesL}{
        \tri{0}{0}{1}{0}{0}{1}  \tri{1}{0}{1}{1}{0}{1}
        \tri{1}{1}{1}{0}{2}{0}  \tri{2}{1}{2}{0}{1}{1}
        \tri{0}{1}{0}{2}{1}{1}  \tri{0}{2}{1}{1}{2}{1}
        \tri{0}{2}{2}{1}{1}{2}  \tri{1}{2}{2}{2}{2}{1}
        \tri{0}{2}{0}{3}{1}{3}  \tri{0}{2}{1}{2}{1}{3}
        \tri{0}{3}{0}{4}{1}{4}  \tri{0}{3}{1}{3}{1}{4}
        \tri{1}{2}{2}{2}{2}{3}  \tri{1}{2}{2}{3}{2}{4}
        \tri{1}{2}{2}{4}{1}{3}  \tri{1}{3}{1}{4}{2}{4}
    }
    \newcommand{\tritrianglesR}{
        \tri{2}{0}{3}{0}{3}{1}  \tri{2}{0}{3}{1}{3}{2}
        \tri{2}{0}{2}{1}{3}{2}  \tri{2}{1}{2}{2}{3}{2}
        \tri{3}{0}{4}{0}{4}{1}  \tri{3}{0}{4}{1}{3}{1}
        \tri{3}{1}{4}{1}{4}{2}  \tri{3}{1}{3}{2}{4}{2}
        \tri{2}{2}{2}{3}{3}{2}  \tri{2}{3}{3}{2}{4}{2}
        \tri{2}{3}{3}{3}{4}{2}  \tri{3}{3}{4}{3}{4}{2}
        \tri{2}{3}{2}{4}{3}{3}  \tri{2}{4}{3}{4}{3}{3}
        \tri{3}{3}{3}{4}{4}{3}  \tri{3}{4}{4}{3}{4}{4}
    }
    \begin{minipage}{0.15\textwidth}
        \begin{tikzpicture}
            \tikzmath{ \scale = 1.0; }
            \latticepoints{0}{2}
            \tritrianglesL
        \end{tikzpicture}
    \end{minipage}
    \begin{minipage}{0.02\textwidth}\end{minipage}
    \begin{minipage}{0.15\textwidth}
        \begin{tikzpicture}
            \tikzmath{ \scale = 1.0; }
            \latticepoints{2}{4}
            \tritrianglesR
        \end{tikzpicture}
    \end{minipage}
    \begin{minipage}{0.05\textwidth}$\longmapsto$\end{minipage}
    \begin{minipage}{0.26\textwidth}
        \begin{tikzpicture}
            \tikzmath{ \scale = 1.0; }
            \latticepoints{0}{4}
            \tritrianglesL
            \tritrianglesR
        \end{tikzpicture}
    \end{minipage}

    \caption{Two fine regular triangulations of $[0,2]\times[0,4]$ being `patched' to a single triangulation of $[0,4]^2$. The two triangulations on the left are both regular while the triangulation on the right is irregular. This example was originally found by Francisco Santos and appears in ref.~\cite{Kaibel_Ziegler_2003}. Figure modified from \cite{macfadden2026boundingkreuzerskarkelandscape}.}
    \label{fig:patching}
\end{figure}
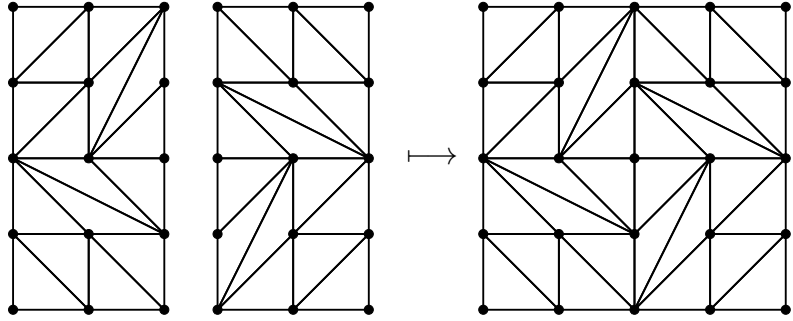

A triangulation $\mathcal{T}$ of a lattice polytope $\Delta$ is a decomposition of $\Delta$ into simplices (obeying certain compatibility constraints~\cite{DeLoera2010}) with vertices taken from $\Delta \cap \mathbb{Z}^d$. See \cref{fig:lifting,fig:patching}. We say that a triangulation is `fine' (denoted FT) if, for each $p\in\Delta\cap\mathbb{Z}^d$, there exists a simplex $\sigma\in\mathcal{T}$ such that $p$ is a vertex of $\sigma$. In $2$D, `fine' is synonymous with `unimodular'. Fineness can be viewed as a local constraint, verified by checking that each simplex $\sigma \in \mathcal{T}$ contains exactly $d+1$ lattice points in its support (in $2$D this is an area computation).

Regularity is the more interesting constraint. A triangulation $\mathcal{T}$ is `regular' (denoted FRT if also fine) if and only if it can be defined by the lifting procedure in \cref{alg:lifting}. For $2$D, this is to embed every lattice point $p_i = (x_i, y_i)$ of the polygon into $\mathbb{R}^3$ as $(x_i, y_i, h_i)$ for some choice of $h_i \in \mathbb{R}$, construct the convex hull of these lifted points, and then project the lower envelope of this hull to $\mathbb{R}^2$ as the triangulation (see \cref{fig:lifting}). The height vector $h = (h_1, \dots, h_{N_\mathrm{pts}}) \in \mathbb{R}^{N_\mathrm{pts}}$ is not unique: $\mathcal{T}$ is equivalently generated by any vector in the interior of the `secondary cone' $\{h \in \mathbb{R}^{N_\mathrm{pts}} : Hh > 0\}$, where $H$ is a matrix determined by the simplices of $\mathcal{T}$. The existence of a height vector depends on the global structure of the triangulation; \cref{fig:patching} gives an example of two FRTs which become irregular after `patching' them together.

\begin{algorithm}[ht]
\DontPrintSemicolon
\SetKwInOut{Input}{Input}
\SetKwInOut{Output}{Output}
\SetKwFunction{LowerHull}{LowerFacets}
\SetKwFunction{Project}{Project}

\Input{Lattice polygon $\Delta = \mathrm{conv}\{x_i \in \mathbf{A}\}$}
\Input{Height vector $h\in\mathbb{R}^{|\mathbf{A}|}$}
\Output{Triangulation $\mathcal{T}$ of $\Delta$}
\BlankLine
$\tilde{\mathbf{A}} \gets \{(x_i, h_i) : x_i \in \mathbf{A}\}$ \tcp*{lift points}
$\tilde{\Delta} \gets \mathrm{conv}(\tilde{\mathbf{A}})$\;
$F \gets \LowerHull(\tilde{\Delta})$ \tcp*{facets with inward normal $n_{d+1} > 0$}
$\mathcal{T} \gets \emptyset$\;
\ForEach{facet $\tilde{\sigma} \in F$}{
    $\sigma \gets \Project(\tilde{\sigma})$ \tcp*{drop last coordinate}
    $\mathcal{T} \gets \mathcal{T} \cup \{\sigma\}$\;
}
\Return $\mathcal{T}$\;
\caption{Lifting Procedure}
\label{alg:lifting}
\end{algorithm}

FRTs of lattice polygons arise naturally in string theory, where they provide an efficient route to enumerating Calabi-Yau threefolds. A central goal in string theory is to find Calabi-Yau threefolds (CYs) which give rise to certain desired physics (de Sitter geometry~\cite{mcallister2024candidatesittervacua}, standard model embeddings\cite{Cveti__2001,Aldazabal_2000,Cveti__2019,Blumenhagen_2009}, \dots). These searches cannot be done exhaustively: the largest-known collection contains up to $10^{296}$~\cite{macfadden2026boundingkreuzerskarkelandscape} CYs\footnote{More pointedly, the current best algorithm for enumerating these CYs~\cite{macfadden2026efficientalgorithmgeneratinghomotopy} would require iterating over at least $10^{276}$ items to generate this entire collection~\cite{macfadden2026boundingkreuzerskarkelandscape}.}, each constructed from a certain fine, regular, `star' triangulation (denoted FRST) of a $4$D polytope~\cite{Batyrev:1993oya}. There are $\leq 10^{928}$ FRSTs~\cite{demirtas2020boundingkreuzerskarkelandscape}, making the map $\mathrm{FRSTs}\to\mathrm{CYs}$ many-to-one. The redundancy has a simple combinatorial description: if two FRSTs $\mathcal{T}_1$ and $\mathcal{T}_2$ of a lattice polytope $\Delta$ define the same triangulations on the $2$-faces of $\Delta$, then they generate homotopy-equivalent CYs. In our prior work~\cite{macfadden2026efficientalgorithmgeneratinghomotopy} we developed an algorithm that turns this redundancy into a constructive tool: given a set of triangulated $2$-faces, one can directly construct a compatible FRST if one exists, or obtain a certificate that no such FRST exists. This reduces CY enumeration to the generation of FRTs of polygons ($2$D), sidestepping the exponentially-redundant FRST space and motivating the focus on polygons in this work.

\subsection{The dualGNN Representation}

\begin{figure}
\centering
\begin{subfigure}{0.45\textwidth}
\centering
\begin{tikzpicture}[scale=0.9]
    \draw[help lines, gray!40] (-1,0) grid (2,3);
    
    \draw[thick] (0,1) -- (1,1) -- (0,2) -- cycle;
    \draw[thick] (0,1) -- (1,1) -- (1,2) -- cycle;
    
    \draw[very thick] (0,1) -- (1,1);
    
    \fill (0,1) circle (2pt);
    \fill (1,1) circle (2pt);
    \fill (0,2) circle (2pt);
    \fill (1,2) circle (2pt);
\end{tikzpicture}
\end{subfigure}
\hfill
\begin{subfigure}{0.45\textwidth}
\centering
\begin{tikzpicture}[scale=0.9]
    \draw[help lines, gray!40] (-1,-1) grid (2,2);
    
    \draw[thick] (0,0) -- (1,0) -- (0,1) -- cycle;
    \draw[thick] (0,0) -- (1,0) -- (1,-1) -- cycle;
    
    \draw[very thick] (0,0) -- (1,0);
    
    \fill (0,0) circle (2pt);
    \fill (1,0) circle (2pt);
    \fill (0,1) circle (2pt);
    \fill (1,-1) circle (2pt);
    
    \node[above left] at (0,0) {$e_k$};
    \node[above right] at (1,0) {$e_\ell$};
    \node[above] at (0,1) {$v_j$};
    \node[below] at (1,-1) {$v_i$};
\end{tikzpicture}
\end{subfigure}
\caption{Left: two simplices share a facet but in an invalid way. The simplices have a solid ($2$D) intersection. Right: two simplices share a facet in a proper way. Their intersection is only on the facet $\mathrm{conv}(\{e_k, e_\ell\})$. Edges are drawn between pairs of simplices with such proper intersections only.}
\label{fig:gnn_edges}
\end{figure}
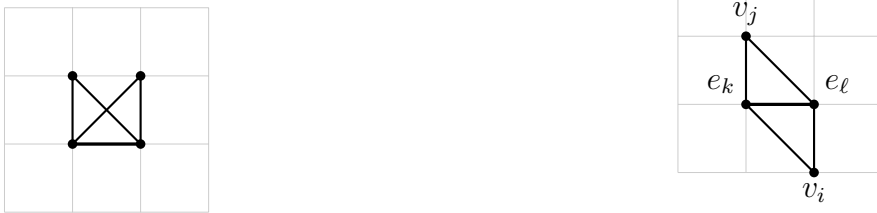

To sample FRTs of polytopes (particularly, polygons), dualGNN operates on a graph whose nodes are candidate simplices and whose edges connect pairs of simplices that share a facet with no overlapping interior. See right side of \cref{fig:gnn_edges}. As discussed in \cref{sec:dualgnn}, each edge carries a fixed feature vector $\lambda$, called a `signed circuit', which encodes a linear dependency among the lattice points in the adjacent simplices $\sigma\cup\sigma'$. Writing $\mathbf{A}_{\sigma\cup\sigma'}$ for the matrix whose columns are these points, $\lambda$ satisfies
\begin{equation}
    \mathbf{A}_{\sigma\cup\sigma'}\, \lambda = 0 \quad \text{and} \quad \sum_i \lambda_i = 0.
\end{equation}
We will loosely call $\lambda$ itself a `circuit' despite it having strictly more information\footnote{A signed circuit would more honestly correspond to the decomposition of $\lambda$ into positive and negative indices, without magnitudes. We carry the relative magnitude of components since those are crucial, e.g., to determine the regularity of a triangulation --- see \cite{DeLoera2010} section 7.1.1.} than an actual signed circuit from oriented matroid\cite{Ziegler1999-bp} theory. Of note:
\begin{enumerate}
    \item Circuits encode the oriented matroid of the polytope $\Delta$, which determines the combinatorial structure of all triangulations of $\Delta$.
    \item For a regular triangulation, these vectors $\lambda$ are exactly the hyperplane normals $H$ defining its secondary cone, so edges directly encode the regularity constraints.
    \item As we demonstrate in \cref{sec:dualgnn}, circuits are invariant under the symmetries of the polytope (translation and unimodular maps). dualGNN is therefore invariant under the orientation-preserving subgroup\footnote{The orientation-reversing case is addressed in \cref{sec:dualgnn}.} $\mathrm{SL}(2,\mathbb{Z}) \ltimes \mathbb{Z}^2$ of these symmetries.
\end{enumerate}
Our circuit-based network dualGNN is autoregressive in the style of a Pointer Network~\cite{vinyals2017pointernetworks}, predicting probability distributions over its inputs instead of a fixed vocabulary. More explicitly, at each step, $K$ rounds of message-passing on the graph produce a probability distribution over candidate simplices, a simplex $\sigma$ is sampled, and then other simplices that overlap $\sigma$ are masked. The initial construction of the graph enforces fineness while our masking enforces validity. This implies, for a $2$D polytope, every rollout will produce a valid, fine triangulation. We utilize both supervised learning and RL. We need RL since the cross-entropy objective in our supervised learning teaches conditional probabilities according to our \textit{estimated} targets (when we are bootstrapping a pool of triangulations), so bias in the training pool propagates to the sampler. We correct this by fine-tuning with REINFORCE~\cite{Williams1992} under an entropy-maximizing reward, directly optimizing the rollout's distribution toward uniformity (\cref{subsubsec:multi}).

Empirically, dualGNN matches a uniform sampler more closely than any baseline we tested across our diagnostics. Uniformity is surprisingly subtle to measure here, primarily because some polygons we study have up to $10^{21}$ FRTs while our computational budget limits us to $\lesssim 10^7$ samples. When the FRT count is $\lesssim 10^6$, we use diagnostics like KL divergence to the uniform (flat) distribution; for larger FRT counts, KL divergence becomes less informative due to low (but often nonzero) collision counts. In such cases, just comparing the empirical collision count to theoretical predictions (the birthday problem in probability theory) proves more useful. Additionally, for polygons of all sizes, we measure the autocorrelation between consecutive samples, analogous to standard tests for pseudorandom number generators~\cite{Knuth1997-eo}. Our single $92$k-parameter model, trained in $\sim7.5$ hours on an NVIDIA RTX 5060 Ti, generalizes to held-out polygons with $N_\mathrm{pts}$ as large as $40$, the maximum we tested. This model even runs on common laptops like an M1 MacBook Pro.

\subsection{Prior Work}

The problem of sampling FRTs has an extensive literature. A standard approach is a flip-walk on the space of triangulations, which converges to the uniform distribution but whose mixing is poorly understood~\cite{Kaibel_Ziegler_2003,314161.314262}. The authors of \cite{Kaibel_Ziegler_2003} also provide the first (to our knowledge) autoregressive sampler of fine triangulations, which is \textit{provably uniform}, but it scales exponentially on the `width' of their polygons (for $m\times n$ lattice rectangles, their algorithm requires at most $10^m(n + 1)^{m+1}$ arithmetic operations). This could be improved by the methods of \cite{orevkov2022counting} but the exponential scaling would persist. Other non-learned methods are discussed in \cite{demirtas2020boundingkreuzerskarkelandscape}, including \code{random\_triangulations\_fast} and \code{random\_triangulations\_fair}, both implemented in CYTools\cite{Demirtas:2022hqf}. Additionally, with the publishing of \cite{macfadden2026efficientalgorithmgeneratinghomotopy}, we released an autoregressive simplex-based method \code{grow2d} with CYTools (commit \code{9f5ef24}). We also introduce an algorithm \code{pushing} in this work, inspired by TOPCOM~\cite{Rambau2002}.

There are also a collection of relevant learned methods. Broadly, there is work on sampling from combinatorial spaces, including discrete normalizing flows~\cite{tran2019discreteflowsinvertiblegenerative,hoogeboom2019integerdiscreteflowslossless} and GFlowNets~\cite{bengio2021flownetworkbasedgenerative} among others. GFlowNets are especially relevant: with reward $R(\mathcal{T})=1$ for regular $\mathcal{T}$ (and $0$ otherwise), a GFlowNet converges to the uniform distribution over FRTs; we discuss the relation to our training objective in \cref{subsubsec:multi}. More tailored to FRT sampling, there is CYTransformer~\cite{yip2025transformingcalabiyauconstructionsgenerating}, which is an encoder-decoder that autoregressively generates simplices conditioned on encoded polytope vertices. Finally, a distinct but complementary line of learned methods optimizes functions over CYs via reinforcement learning and genetic algorithms\cite{berglund2024generatingtriangulationsfibrationsreinforcement,macfadden2025dnacalabiyauhypersurfaces}. These works are complementary in that they consume FRTs of polygons, rather than generating them.

In particular, we will make heavy use of the classical baselines \code{random\_triangulations\allowbreak\_fast} (which we abbreviate to \code{fast})\footnote{We note that \code{random\_triangulations\_fast} was not advertised by~\cite{demirtas2020boundingkreuzerskarkelandscape} as a uniform sampler. We include it as a biased reference.}, \code{pushing}, and \code{grow2d}. We also compare against random walks in flip space (\code{flip\_walk}; algorithm \#1 of \cite{demirtas2020boundingkreuzerskarkelandscape}). We notably exclude \code{random\allowbreak\_triangulations\_fair} despite it being designed for uniformity because this algorithm adds height-modification steps to \code{flip\_walk} which make the sampler prohibitively slow for our work.

The most-similar previous work is CYTransformer~\cite{yip2025transformingcalabiyauconstructionsgenerating}. Like dualGNN, it seeks to autoregressively generate fine, regular triangulations of lattice polytopes for applications in CY generation. \cite{yip2025transformingcalabiyauconstructionsgenerating} utilizes an encoder-decoder transformer to achieve strong performance in generating regular triangulations (something we partially recreate, albeit for decoder-only transformers, in \cref{sec:transformer}). dualGNN shares this learned-autoregressive framing but differs in architecture and intermediate representation, as described below.
\begin{enumerate}
    \item \cite{yip2025transformingcalabiyauconstructionsgenerating} trained their CYTransformer encoder-decoder model for each $N_\mathrm{vert}$ of interest (analogous to $N_\mathrm{pts}$ in our work), with token vocabulary of size $\binom{N_\mathrm{vert}-1}{4}$; we train a \textit{single} dualGNN model to operate on general $N_\mathrm{pts}$ since there is no fixed vocabulary,
    \item CYTransformer utilizes a coordinate-based representation of the polytope, which is not symmetry invariant. They address point relabeling via data augmentation but did not address other symmetries; dualGNN benefits~\cite{bronstein2021geometricdeeplearninggrids} from having symmetries built into the architecture, and
    \item CYTransformer samples CYs via triangulations of $4$D polytopes; we instead use the $2$-face decomposition of \cite{macfadden2026efficientalgorithmgeneratinghomotopy} to more directly sample the (potentially) homotopy-inequivalent ones.
\end{enumerate}
The first two differences allow dualGNN, unlike CYTransformer, to generalize zero-shot across polytopes of different sizes, shapes, and $N_\mathrm{pts}$. The third difference enables us to sample CYs at significantly higher $h^{1,1}$: by generating triangulated 2-face data directly, we avoid the exponential FRST redundancy ($N_\mathrm{FRST}\lesssim 10^{1.91 h^{1,1}-5.31}$ compared to $N_\mathrm{CY}\lesssim 10^{0.9 h^{1,1}-15.45}$~\cite{demirtas2020boundingkreuzerskarkelandscape}) that a 4D sampler must contend with. This enables dualGNN to sample up to $h^{1,1}=128$ compared to their $h^{1,1}\leq 10$. dualGNN is also $\sim1300\times$ smaller and significantly cheaper to train ($7.5$ hours on an RTX 5060 Ti compared to \cite{yip2025transformingcalabiyauconstructionsgenerating}'s few days per $(h^{1,1}, N_\mathrm{vert})$ on 8 NVIDIA V100 GPUs). We stress that CYTransformer could likely be made smaller, quicker to train, and operable at higher $h^{1,1}$ if it adopted the 2-face decomposition of
\cite{macfadden2026efficientalgorithmgeneratinghomotopy} (as we do here), though these do not address its difficulty in generalizing across polytopes. A direct head-to-head would require this retraining, which is outside our budget given CYTransformer's currently unavailable weights and source; we therefore compare only via their figure-$4$ diagnostic in \cref{subsec:comparison_to_cytransformer}.

\subsection{Contributions, Paper Structure}

In summary, our contributions are:
\begin{enumerate}
    \item A symmetry-respecting encoding for triangulation problems. We label the edges of a generalized dual graph with magnitude-bearing signed circuits from the polytope's oriented matroid, yielding an architecture invariant under $\mathrm{SL}(d,\mathbb{Z}) \ltimes \mathbb{Z}^d$ and independent of $N_\mathrm{pts}$ (\cref{sec:dualgnn}). Sign-only circuits are provably insufficient for regularity, while our magnitude-bearing `circuit' features coincide with the secondary-cone normals and thus the encoding exposes regularity (validated empirically in \cref{subsec:classifier}).
    \item Constraint satisfaction by construction. Graph construction enforces fineness and autoregressive masking enforces validity, so in 2D every rollout is a valid fine triangulation (\cref{subsec:sampler}).
    \item Size generalization and zero-shot transfer. A single $\sim$92k-parameter pointer-network-style sampler, trained in hours on one consumer GPU, transfers zero-shot across polygons of different sizes and shapes --- even when trained on just a single polygon (\cref{subsubsec:single,subsubsec:zeroshot}).
    \item A bias-inheritance fix. Teacher-forced training on bootstrapped pools inherits their bias; entropy-reward REINFORCE fine-tuning corrects it, yielding the only sampler we tested that is consistent with uniformity across all diagnostics (\cref{subsubsec:multi}).
    \item Application. Combined with the 2-face reduction of \cite{macfadden2026efficientalgorithmgeneratinghomotopy}, we sample Calabi-Yau threefolds uniformly at $h^{1,1}=86$ and observe no deviations from uniformity at $h^{1,1}=128$ (where our diagnostics are weaker), an order of magnitude larger than prior learned samplers (\cref{sec:cy}).
\end{enumerate}

The rest of the paper is as follows. \Cref{sec:dualgnn} introduces dualGNN in detail. \Cref{subsec:classifier} evaluates it as a regularity classifier; \cref{subsubsec:single} as an FRT sampler trained on a single polygon; \cref{subsubsec:zeroshot} demonstrates zero-shot transfer between polygons; and \cref{subsubsec:multi} as a general-purpose sampler trained across polygon configurations. \Cref{sec:cy} then applies the general-purpose sampler to Calabi-Yau enumeration up to $h^{1,1} = 128$. \Cref{sec:limits} discusses scope and scalability; \cref{sec:conclusion} concludes.

\section{dualGNN}
\label{sec:dualgnn}

\begin{figure}
    \centering
    \begin{tikzpicture}[scale=0.9]
        \draw[help lines, gray!40] (-2,-2) grid (2,2);
        \draw[thick] (0,0) -- (1,0) -- (0,-1) -- cycle;
        \draw[thick] (0,0) -- (1,0) -- (-1,1) -- cycle;
        \fill (0,0) circle (2pt);
        \fill (1,0) circle (2pt);
        \fill (0,-1) circle (2pt);
        \fill (-1,1) circle (2pt);
        \node[below right] at (0,-1) {$A$};
        \node[below left] at (0,0) {$B$};
        \node[above left] at (-1,1) {$C$};
        \node[right] at (1,0) {$D$};
    \end{tikzpicture}
    \hfill
    \begin{tikzpicture}[scale=0.9]
        \draw[help lines, gray!40] (-2,-2) grid (2,2);
        \draw[thick] (0,0) -- (1,0) -- (0,-1) -- cycle;
        \draw[thick] (0,0) -- (1,0) -- (0,1) -- cycle;
        \fill (0,0) circle (2pt);
        \fill (1,0) circle (2pt);
        \fill (0,-1) circle (2pt);
        \fill (0,1) circle (2pt);
        \node[below] at (0,-1) {$A$};
        \node[left] at (0,0) {$B$};
        \node[above] at (0,1) {$C$};
        \node[right] at (1,0) {$D$};
    \end{tikzpicture}
    \hfill
    \begin{tikzpicture}[scale=0.9]
        \draw[help lines, gray!40] (-2,-2) grid (2,2);
        \draw[thick] (0,0) -- (1,0) -- (0,-1) -- cycle;
        \draw[thick] (0,0) -- (1,0) -- (1,1) -- cycle;
        \fill (0,0) circle (2pt);
        \fill (1,0) circle (2pt);
        \fill (0,-1) circle (2pt);
        \fill (1,1) circle (2pt);
        \node[below right] at (0,-1) {$A$};
        \node[left] at (0,0) {$B$};
        \node[above left] at (1,1) {$C$};
        \node[right] at (1,0) {$D$};
    \end{tikzpicture}
    \caption{Pairs of adjacent simplices $ABD$ and $BCD$. Each pair corresponds to an edge in a dual graph, but these pairs correspond to circuits playing different roles in the oriented matroid so they must be distinguished.}
    \label{fig:different_edges}
\end{figure}
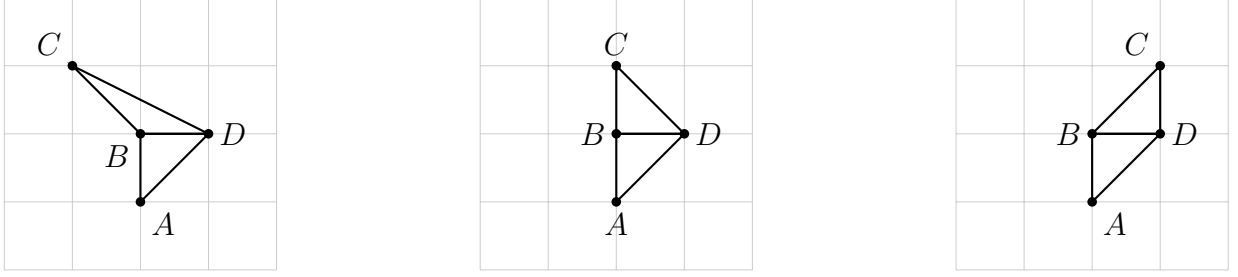

We first introduce a simplified variant of dualGNN; the more complete model will be discussed in \cref{subsec:sampler}. Consider encoding a triangulation $\mathcal{T}$ of $\Delta$ by its dual graph $G_\mathcal{T}$. That is, draw a node for each simplex of $\mathcal{T}$ and edges between any two nodes whose corresponding simplices share a facet (call such simplices `adjacent'). Such a graph encodes the triangulation as a simplicial complex, but it obscures some geometric data. For example, each pair of simplices in \cref{fig:different_edges} would correspond to an edge in $G_\mathcal{T}$, but they play different roles in a triangulation as we now describe.

The combinatorial structure of triangulations is often described in the language of oriented matroids\cite{DeLoera2010}. In this language, one characterizes the transformations a triangulation can take (`flips'), the validity of a triangulation, the total number of triangulations of $\Delta$, etc. in terms of certain objects called\footnote{Other equivalent objects like cocircuits and chirotopes also equivalently characterize the triangulation\cite{DeLoera2010}. Circuits are most-directly useful for this work, so we use them.} `signed circuits' (often just called `circuits'). If we organize the lattice points of $\Delta$ as the columns of a matrix $\mathbf{A}$, the circuits correspond to certain minimal (in terms of the count of nonzero elements) vectors
$\lambda$ satisfying
\begin{equation}
    \begin{bmatrix} \mathbf{A} \\ \mathbf{1} \end{bmatrix} \lambda = 0
    \label{eq:circuit}.
\end{equation}
Circuits are sparse, with $\leq d+2$ nonzero elements (for a $d$-dimensional polytope). Additionally, only a small subset of circuits are relevant to a triangulation $\mathcal{T}$: every flip of $\mathcal{T}$ is characterized by a circuit $\lambda$ such that $\mathrm{nonzero}(\lambda) \subseteq \sigma\cup\sigma'$ for $\sigma,\sigma'$ adjacent in $\mathcal{T}$.

Return to the pairs in \cref{fig:different_edges}. Each pair corresponds to an edge in $G_\mathcal{T}$ but such edges describe different types of flips/circuits. The leftmost two pairs describe flips in which a point is deleted (albeit in different ways) while the rightmost pair describes a diagonal flip. These are fundamentally different transformations which must be distinguished for a faithful representation of the oriented matroid, but $G_\mathcal{T}$ is blind to their differences. 
The primary idea of dualGNN is to directly inject this circuit information into $G_\mathcal{T}$ as a fixed feature of each edge, appending this feature to any message passed through it. The circuit data respects the $\mathrm{GL}(d,\mathbb{Z}) \ltimes \mathbb{Z}^d$ symmetries of the polytope:
\begin{equation}
    \left(\mathbf{A} + \begin{bmatrix}
        r & \cdots & r
    \end{bmatrix}\right)\lambda = 0 \iff \mathbf{A}\lambda = 0 \iff (\mathbf{U}\mathbf{A})\lambda = 0,
\end{equation}
for unimodular $\mathbf{U}$. In the first equality, we used that $\sum_i\lambda_i = 0$. The same circuit data, and hence the same edge-labeled graph $G_\mathcal{T}$, is therefore produced regardless of how the polytope is translated or unimodularly transformed. We stress: our above discussion on circuits applies to any dimension. The directional encoding introduced below for $d=2$ reduces this to $\mathrm{SL}(2, \mathbb{Z}) \ltimes \mathbb{Z}^2$ by sacrificing orientation-reversal invariance.

One technical subtlety: $\lambda$ above is not technically a circuit, but is related to one. A circuit is more precisely a decomposition of $\lambda$ into the indices with positive and negative coefficients (defining a `minimally dependent subconfiguration of $\Delta$'), with magnitude information discarded. We retain $\lambda$ itself, including the magnitudes, because the relative magnitudes are necessary for determining regularity (see the example in \S7.1.1 of \cite{DeLoera2010}). That example also validates the necessity of our encoding: it exhibits two triangulations with the same dual graph $G_\mathcal{T}$ and the same magnitude-stripped circuits, but different regularity (one regular, one irregular). Without $\lambda$-features, no function of $G_\mathcal{T}$ alone could distinguish them --- even augmenting $G_\mathcal{T}$ with true circuits (sign patterns of $\lambda$) is insufficient; only the full magnitude-bearing $\lambda$ exposes a regularity signal.

The only intricacy is in how one actually assigns a circuit $\lambda$ to an edge. The core data of the circuit is the map from vertex $i$ to its coefficient $\lambda_i$. We focus on $2$D problems for which we propose the following encoding:\footnote{This is obviously not the only possible way to encode the circuit to an edge.} to the edge from node $a$ to node $b$, assign a $4$D vector $\mathbf{C}_{ab} = (\lambda_{v_i}, \lambda_{v_j}, \lambda_{e_k}, \lambda_{e_\ell})$ in which (see \cref{fig:gnn_edges})
\begin{enumerate}
    \item $v_i$ is the vertex unique to node-$a$ (the sender),
    \item $v_j$ is the vertex unique to node-$b$ (the receiver),
    \item $e_k$ is the vertex `to the left', and
    \item $e_\ell$ is the vertex `to the right'.
\end{enumerate}
By `left' and `right', we mean: compute the signed areas
\begin{align}
    \mathrm{area}_k = \det\begin{bmatrix}(v_i-e_k)_1 & (v_i-e_k)_2\\(v_j-e_k)_1 & (v_j-e_k)_2\end{bmatrix} & \quad &
    \mathrm{area}_\ell = \det\begin{bmatrix}(v_i-e_\ell)_1 & (v_i-e_\ell)_2\\(v_j-e_\ell)_1 & (v_j-e_\ell)_2\end{bmatrix}
\end{align}
and enforce $\mathrm{area}_k > \mathrm{area}_\ell$. Observe: these edges are directed $\mathbf{C}_{ab}\neq \mathbf{C}_{ba}$. This ordering is chosen because it represents the circuit in a locally-meaningful way and it is invariant under any point relabeling, but it costs invariance under orientation-reversing transformations ($\mathrm{det}(\mathbf{U})=-1$). That is, the dualGNN we are now constructing is only invariant under $\mathrm{SL}(2,\mathbb{Z}) \ltimes \mathbb{Z}^2$.

Since they encode the oriented matroid (the combinatorics of a triangulation), circuits are attractive for describing a triangulation. Such an architecture is especially interesting for applications to regular triangulations. For a regular triangulation $\mathcal{T}$, the hyperplane normals of the secondary cone (i.e., rows of $H$) are exactly these dependencies $\lambda$ (with $0$-coefficients for other points). That is, the circuits \textit{are} the constraints on height space, defining the secondary cone. Such an encoding, $G_\mathcal{T}$ with edges labeled by signed circuits, then also encodes the secondary cone of $\mathcal{T}$. Since a triangulation is regular if and only if its secondary cone is full-dimensional, this encoding directly exposes regularity to the model.

\subsection{Regularity Classifier}
\label{subsec:classifier}

Before applying dualGNN as a sampler, we first verify that message-passing can read out the regularity signal in principle. We do this by configuring dualGNN as a binary classifier on complete triangulations: given $\mathcal{T}$, predict whether it is regular. Note that this task is trivial when $H$ is known since an LP can check the feasibility of $Hh>0$ directly. In this way, this task is a diagnostic, not an end application: it simply verifies that the network can convert circuit features into a regularity signal. This will not prove that the dualGNN autoregressive sampler can do so mid-rollout. We return to this gap in \cref{subsec:sampler,sec:limits}.

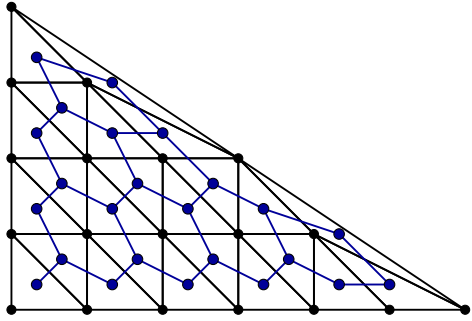
\begin{figure}
    \centering
    \tikzset{
        latticepoint/.style={draw, circle, inner sep=1.2pt, fill},
        trilinestyle/.style={line width=0.25mm},
        dualnode/.style={draw, circle, inner sep=1.4pt, fill=blue!60!black},
        dualedge/.style={line width=0.25mm, blue!60!black}
    }
    \newcommand{\tri}[6]{%
        \draw[trilinestyle]
            (#1*\scale,#2*\scale) -- (#3*\scale,#4*\scale) --
            (#5*\scale,#6*\scale) -- cycle;
    }
    \newcommand{\polytopelatticepoints}{%
        \foreach \x/\y in {%
            0/0, 1/0, 2/0, 3/0, 4/0, 5/0, 6/0,
            0/1, 1/1, 2/1, 3/1, 4/1,
            0/2, 1/2, 2/2, 3/2,
            0/3, 1/3,
            0/4%
        }{
            \node[latticepoint] at (\x*\scale,\y*\scale) {};
        }
    }
    \newcommand{\polytopetriangles}{%
        \tri{1}{1}{2}{1}{1}{2}    
        \tri{1}{1}{2}{1}{2}{0}    
        \tri{1}{1}{1}{2}{0}{2}    
        \tri{1}{1}{1}{0}{2}{0}    
        \tri{1}{1}{1}{0}{0}{1}    
        \tri{1}{1}{0}{1}{0}{2}    
        \tri{2}{1}{3}{1}{2}{2}    
        \tri{2}{1}{3}{1}{3}{0}    
        \tri{2}{1}{1}{2}{2}{2}    
        \tri{2}{1}{2}{0}{3}{0}    
        \tri{3}{1}{4}{1}{4}{0}    
        \tri{3}{1}{4}{1}{3}{2}    
        \tri{3}{1}{2}{2}{3}{2}    
        \tri{3}{1}{3}{0}{4}{0}    
        \tri{4}{1}{6}{0}{5}{0}    
        \tri{4}{1}{6}{0}{3}{2}    
        \tri{4}{1}{4}{0}{5}{0}    
        \tri{1}{2}{2}{2}{1}{3}    
        \tri{1}{2}{1}{3}{0}{3}    
        \tri{1}{2}{0}{2}{0}{3}    
        \tri{2}{2}{1}{3}{3}{2}    
        \tri{1}{3}{0}{4}{3}{2}    
        \tri{1}{3}{0}{4}{0}{3}    
        \tri{0}{1}{1}{0}{0}{0}    
    }
    \newcommand{\dualgraph}{%
        \coordinate (s0)  at (4/3*\scale,  4/3*\scale);
        \coordinate (s1)  at (5/3*\scale,  2/3*\scale);
        \coordinate (s2)  at (2/3*\scale,  5/3*\scale);
        \coordinate (s3)  at (4/3*\scale,  1/3*\scale);
        \coordinate (s4)  at (2/3*\scale,  2/3*\scale);
        \coordinate (s5)  at (1/3*\scale,  4/3*\scale);
        \coordinate (s6)  at (7/3*\scale,  4/3*\scale);
        \coordinate (s7)  at (8/3*\scale,  2/3*\scale);
        \coordinate (s8)  at (5/3*\scale,  5/3*\scale);
        \coordinate (s9)  at (7/3*\scale,  1/3*\scale);
        \coordinate (s10) at (11/3*\scale, 2/3*\scale);
        \coordinate (s11) at (10/3*\scale, 4/3*\scale);
        \coordinate (s12) at (8/3*\scale,  5/3*\scale);
        \coordinate (s13) at (10/3*\scale, 1/3*\scale);
        \coordinate (s14) at (5*\scale,    1/3*\scale);
        \coordinate (s15) at (13/3*\scale, 1*\scale);
        \coordinate (s16) at (13/3*\scale, 1/3*\scale);
        \coordinate (s17) at (4/3*\scale,  7/3*\scale);
        \coordinate (s18) at (2/3*\scale,  8/3*\scale);
        \coordinate (s19) at (1/3*\scale,  7/3*\scale);
        \coordinate (s20) at (2*\scale,    7/3*\scale);
        \coordinate (s21) at (4/3*\scale,  3*\scale);
        \coordinate (s22) at (1/3*\scale,  10/3*\scale);
        \coordinate (s23) at (1/3*\scale,  1/3*\scale);
        \foreach \a/\b in {%
            0/1, 0/2, 0/8, 1/3, 1/9, 2/5, 2/19, 3/4, 4/5, 4/23,
            6/7, 6/8, 6/12, 7/9, 7/13, 8/17, 10/11, 10/13, 10/16,
            11/12, 11/15, 12/20, 14/15, 14/16, 17/18, 17/20,
            18/19, 18/22, 20/21, 21/22%
        }{
            \draw[dualedge] (s\a) -- (s\b);
        }
        \foreach \i in {0,...,23}{
            \node[dualnode] at (s\i) {};
        }
    }
    \begin{tikzpicture}
        \tikzmath{ \scale = 1.0; }
        \polytopelatticepoints
        \polytopetriangles
        \dualgraph
    \end{tikzpicture}
    \caption{A fine triangulation $\mathcal{T}$ of the polytope $\Delta=\mathrm{conv}(\{(0,0), (0,4),(6,0)\})$ (in black) as well as its dual graph $G_\mathcal{T}$ (in blue). This polygon has $408,826$ triangulations of which all but $3,120$ are regular.}
    \label{fig:4x6tri}
\end{figure}

We configure the dualGNN with $32$-dimensional feature vectors $f_a$ on each node $n_a$ (initialized to $0$) and $K=16$ message-passing rounds. Message-passing operates by:
\begin{enumerate}
    \item normalizing the $D=32$-dimensional feature vector $f_a$,
    \item forming the message $[f_a, \mathbf{C}_{ab}]$ to node $n_b$,
    \item simultaneously sending all such messages (one from each node $n_a$ to each neighbor $n_b$), having node $n_b$ aggregate all incoming messages with sum, min, and max (aggregate length $3(D+4)$),
    \item concatenating the receiving node's own (normalized) feature vector $f_b$ (total length $3(D+4)+D$), and then
    \item running the combined vector through an MLP (linear layer mapping to dimension-$D$; GELU; linear layer mapping to $D$-dimension),
\end{enumerate}
after which the result is added to the node's current $f_b$. After $K$ rounds of message-passing, an output signal is achieved by first taking the softmin\footnote{The softmin is motivated by an LP-feasibility view of regularity, originally explored in a dual variant of this architecture. One can construct a graph whose nodes are lattice points and edges are the edges of a triangulation (inspired by the message-passing simplicial network design\cite{bodnar2021weisfeilerlehmantopologicalmessage}). This graph naturally, even without learning, propagates constraints on vertex heights by message passing between nodes mediated through circuits. For such an architecture, the regularity signal is whether the upper bound on the heights is larger than the lower bound for all nodes; one wants to take the min over $\mathrm{upper}-\mathrm{lower}$. Despite the graph used by dualGNN being dual to this variant, it still inspires the softmin choice.} over nodes and then projecting the resultant vector to a scalar which we call the `regularity logit'. We train this model via BCE on classified data ($y=1$ for regular $\mathcal{T}$; $y=0$ for irregular) with loss
\begin{equation}
    \mathrm{loss} = -[y \log(\mathrm{sigm}(\text{regularity logit})) + (1-y)\log(1-\mathrm{sigm}(\text{regularity logit}))].
\end{equation}
For all supervised training in this paper, we use the AdamW optimizer~\cite{loshchilov2019decoupledweightdecayregularization} ($\beta_1=0.9$, $\beta_2=0.95$, weight decay $0.01$) with learning rate $1\times 10^{-3}$, batch size $16$ or $32$, and gradient clipping by magnitude $1$. Hyperparameters were chosen by light manual search; we did not perform a systematic sweep. We did explore larger $D$ and $K$ values (here and for later sampling purposes), but they consistently yielded worse results despite the increased parameter count.

\begin{figure}[ht]
    \centering
    \includegraphics[width=0.8\textwidth]{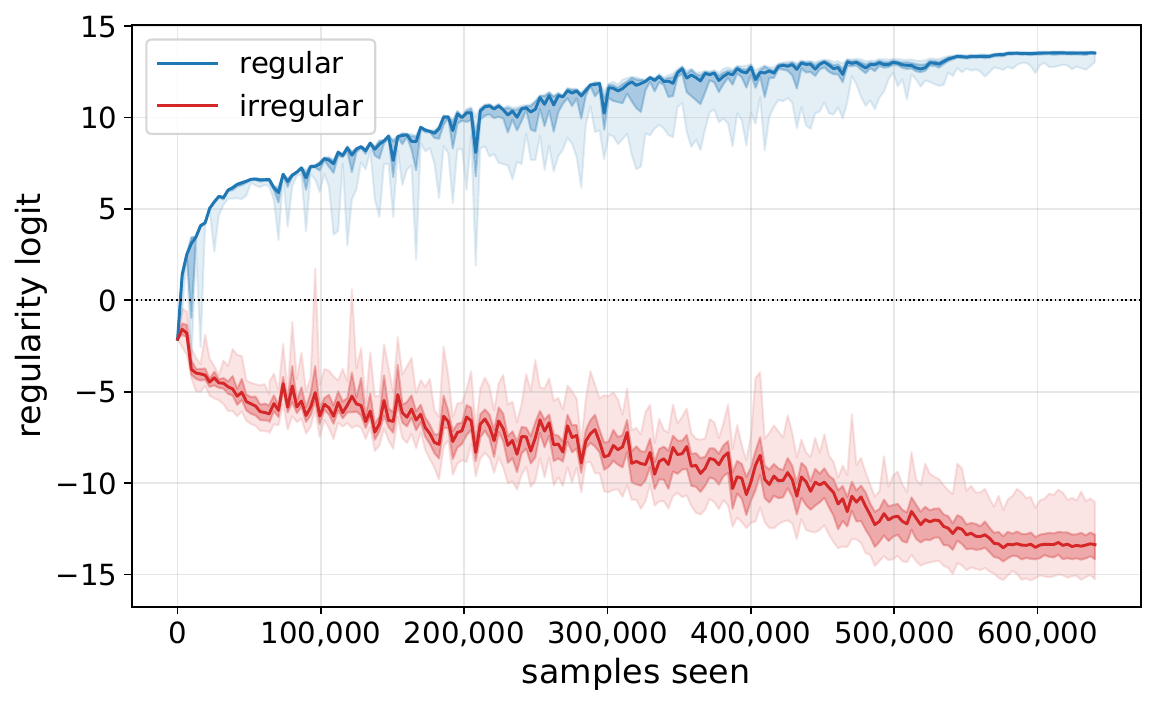}
    \caption{Performance of the regularity classifier trained and validated on the polygon in \cref{fig:4x6tri}. This evaluation is on validation data of said polygon (unseen during training; $15\%$ of the total regular/irregular triangulation pool), with each data point representing the classification of $256$ random samples from each validation pool. Note that small polygons have few irregular triangulations, that in \cref{fig:4x6tri} is no exception, so the total irregular pool here is somewhat small ($3,120$). The classifier first achieved $\geq 99\%$ accuracy on both regular and irregular triangulations by $\sim35,000$ samples seen.}
    \label{fig:regularity_learning}
\end{figure}

For this simple test, we study the polygon in \cref{fig:4x6tri} due to its relatively high fraction of fine irregular triangulations ($3,120/408,826$) despite the small count $408,826$ of fine triangulations. Of these triangulations, we split the regular and irregular ones each into training/validation pools by assigning $15\%$ of each regularity class to validation. During training, triangulations are sampled from the training pool with equal probability of regular vs. irregular. We assess the performance of dualGNN by testing it intermittently on the validation pool during training (see \cref{fig:regularity_learning}). dualGNN quickly learns regularity, achieving $>99\%$ accuracy in both regular and irregular classification by $35,000$ samples seen.

\subsection{As an Autoregressive Triangulation Sampler}
\label{subsec:sampler}

Our focus is on using dualGNN as an autoregressive sampler of FRTs. For this task, we must generalize dualGNN. First, the graph: instead of the graph $G_\mathcal{T}$ dual to some triangulation $\mathcal{T}$, we use the minimal graph\footnote{As an aside, we note that $G$ is not always connected: consider $[0,1]^2$.} $G$ which contains all $G_\mathcal{T}$ for all fine $\mathcal{T}$. This is computed by collecting, out of the $\binom{N_\mathrm{pts}}{3}$ possible simplices (in $2$D), only those containing exactly $3$ lattice points. For a matroid-flavored definition, $G$ can be defined as a certain subset (with signs) of all circuits in the matroid. A circuit $\lambda$ defines a unimodular simplex\footnote{Some care needs to be taken with homogenization. Additionally for so-called $(2,1)$-deletion circuits.} $\{j\neq i: \lambda_j\neq0\}$ for each $i$ such that $\lambda_i>0$. This simplex actually has area $\lambda_i \left| \det A_{\{j \neq i \;\mathrm{s.t.}\; \lambda_j \neq 0\}} \right|$, showing that $\lambda_i=1$ for circuits relevant to unimodular triangulations. This restriction, viewed either way, guarantees fineness of any triangulation built from these nodes.

We then draw an edge $\mathbf{C}_{ab}$ between any two nodes $n_a$ and $n_b$ whose simplices $\sigma_a, \sigma_b$ properly share a facet $f = \sigma_a\cap\sigma_b$ (right side of \cref{fig:gnn_edges}). The resulting graph is modest: despite the polygon $[0,4]^2$ in \cref{fig:patching} having $736,983,568$ fine triangulations\cite{Kaibel_Ziegler_2003}, $G$ consists of only $320$ nodes (out of the $\binom{25}{3}=2300$ possible ones). Even the most extreme polygon occurring in our applications in string theory, $\mathrm{conv}(\{(0,0),(0,7),(84,0)\})$, with up to $2.0\times10^{180}$ FRTs~\cite{macfadden2026boundingkreuzerskarkelandscape}, defines a graph with only $69,416$ nodes.

The autoregressive sampler processes $G$ in stages. At stage $n$, a set of $n$ nodes will have been selected from $G$; these nodes correspond to $n$ simplices forming a partial triangulation $\mathcal{T}_n$ being extended toward a complete triangulation. The model then predicts, for each outstanding node corresponding to a simplex $\sigma$, the fraction of complete triangulations (optionally, restricted by regularity) extending $\mathcal{T}_n$ that also include $\sigma$. This is done by projecting each node's feature vector to a scalar and then taking the softmax over such scalars (overriding the logits of already placed or masked simplices to $-\infty$). A simplex $\sigma$ is then sampled according to these weights and all other simplices $\sigma'$ with $\dim(\sigma\cap\sigma')=d$ are masked out since they cannot occur in any valid extensions. In $2$D, this guarantees that the network always generates an FT: any uncovered region admits a triangulation using only its existing lattice points, so a legal next simplex always exists. There is no guarantee in higher dimensions (see, e.g., the Sch\"onhardt polyhedron~\cite{Schnhardt1928}). Regardless, this is significantly better than the corresponding situation for a transformer, which has \textit{no} guarantees  about validity/fineness. For example, even when trained on many polygons, the transformers in \cref{sec:transformer} struggled to generalize across polygons in our experiments.

For the network to be able to do this selection, we must modify the message-passing. Specifically, we inject, between the aggregation and the MLP, the vector $s_b = (\mathrm{placed},\mathrm{legal})$ into any message sent to node $n_b$. This makes the total message length that the MLP sees $3(D+4)+D+2$. Here, `$\mathrm{placed}$' and `$\mathrm{legal}$' are binary variables indicating whether node $n_b$ has already been placed and whether it can be chosen in future rounds, respectively. We also, in contrast to \cref{subsec:classifier}, use this $s$ to initialize the node features as $f_a = \mathrm{MLP}(s_a)$.

Conceptually, dualGNN's inference loop is a learned generalization of \code{grow2d} (see \cref{sec:classical}): both pick simplices one at a time and mask out incompatible candidates. \code{grow2d} makes uniform random choices among legal continuations that share a facet; dualGNN learns the conditional probabilities of each continuation from the circuit features. dualGNN differs by not limiting attention to new simplices sharing a facet, but that is surface-level.

Since the model operates autoregressively on partial triangulations, we train it on random prefixes (subsets of complete triangulations). Specifically, each training step selects a positive integer $k$ (uniformly chosen) and then selects a subset of $k$ simplices from a triangulation $\mathcal{T}$ drawn from the training pool. 
For each prefix, the network performs a forward-pass, predicting simplex probabilities, and then it is given loss equal to the cumulative cross-entropy compared to the ground-truth probabilities. If the ground truth is not known (common for large polygons), one estimates conditional probabilities from a bootstrap pool of triangulations sampled intermittently during training (similar to the strategy in \cite{yip2025transformingcalabiyauconstructionsgenerating}). Bias in this estimate can then propagate to the trained model; we address this in \cref{subsubsec:multi} by fine-tuning with REINFORCE.

During inference, one samples a simplex $\sigma$ according to the predicted weights, masks out $\sigma'$ for which $\dim(\sigma\cap\sigma')=2$, and then repeats. This guarantees\footnote{Again, this fact is unique to $2$D.} that the network always generates an FT (but not necessarily regular). See \cref{sec:inference} for a visualization. We note that while the circuit encoding is sufficient to expose regularity (\cref{subsec:classifier}), the autoregressive sampler does not always succeed in generating a regular triangulation: some fraction of rollouts produce irregular triangulations. For our later inference on the polygons in \cref{fig:npts40-ood-polygons-grid}, this irregular sample rate is $\sim0.6\%$ on average, but this is highly dependent on polygon. This could suggest that our training signal restricting to regularity was insufficient. Independently, since similar training approaches worked for the transformers studied in \cref{sec:transformer} (see \cref{fig:rope_irreg}), the gap may also reflect an architectural limitation. Moving to a Graphormer-style architecture~\cite{graphormer}, which gives a more global simplex signal, or even a GFlowNet could resolve this.

\subsubsection{Single Polygon}

\label{subsubsec:single}

We begin by training and testing dualGNN on the polygon from \cref{fig:4x6tri} since we can fully enumerate and classify all of its FTs ($405,706$ regular triangulations and $3,120$ irregular ones). 
As points of comparison, we use four classical samplers: \code{fast}~\cite{demirtas2020boundingkreuzerskarkelandscape}, a Markov-chain sampler \code{flip\_walk} (algorithm \#1 of \cite{demirtas2020boundingkreuzerskarkelandscape}; this is closely related to the algorithm described in section 4.1 of \cite{Kaibel_Ziegler_2003}), and two new methods \code{pushing} (inspired by methods in TOPCOM~\cite{Rambau2002}) and \code{grow2d}. These are discussed in detail in \cref{sec:classical}.

\begin{figure}
    \centering
    \includegraphics[width=1.0\textwidth]{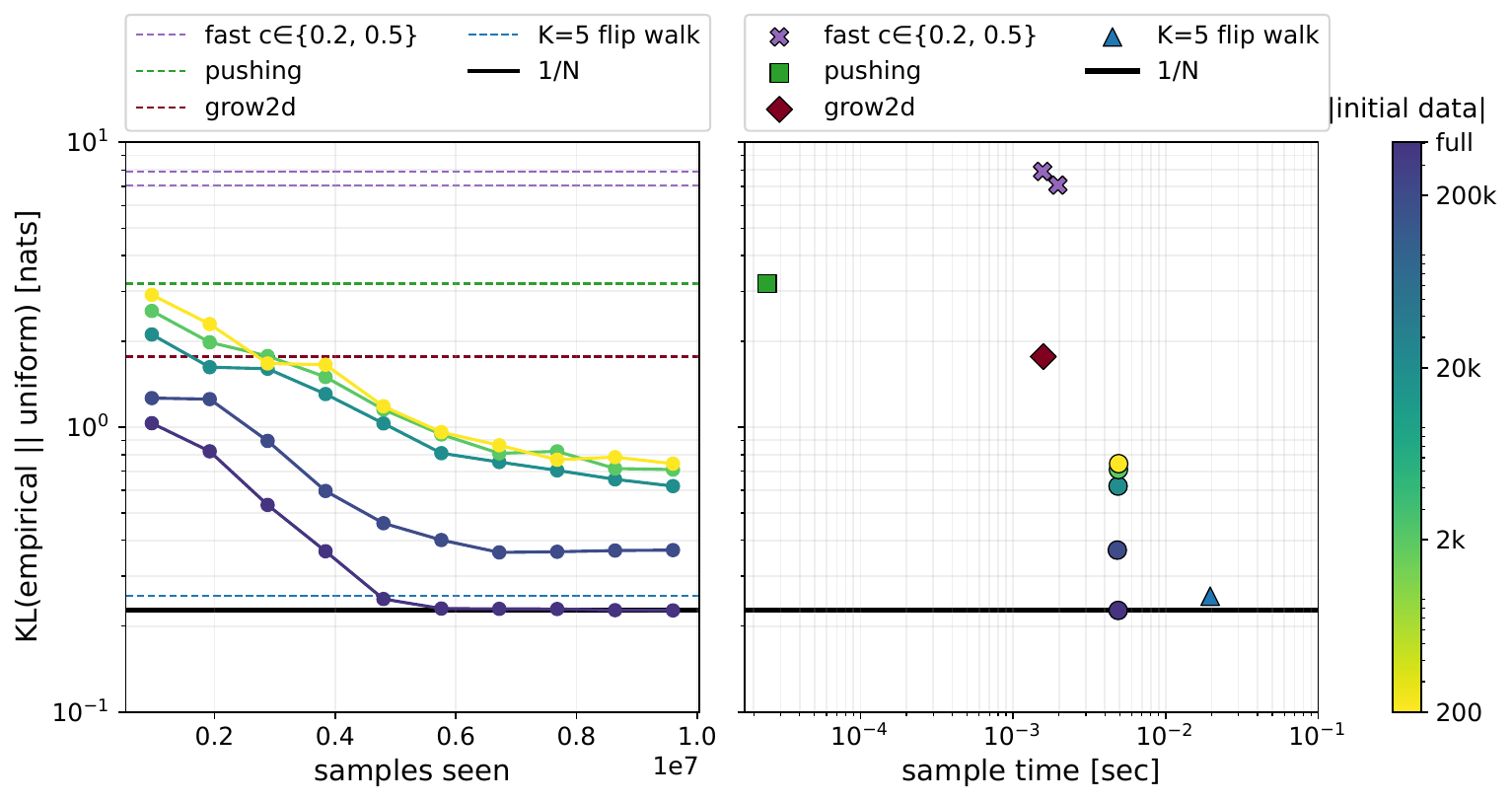}
    \caption{Performance of the autoregressive dualGNN on the polygon in \cref{fig:4x6tri}. Left: the uniformity of $10^6$ samples (total FRT count is $405,706$) drawn from checkpoints mid training for varying levels of initial training data (from a full distribution down to $200$ triangulations). dualGNN with access to all triangulations as training data generates more uniform (lower KL) samples than all other methods by $\sim5\times 10^6$ samples seen. The dualGNN variants with reduced training data beat all methods other than \code{flip\_walk}. Right: the KL divergence of all sampling methods versus the average time to draw a sample. \code{pushing} is, by far, the quickest method since it does not need to check regularity of its outputs, but it is very biased and unable to generate some regular triangulations. dualGNN is approximately four times as quick as \code{flip\_walk}.}
    \label{fig:dualgnn_4x6tri}
\end{figure}

dualGNN outperforms the baselines we tested, achieving a lower KL divergence (compared to a uniform/flat distribution) at $10^6$ samples. See \cref{fig:dualgnn_4x6tri} (for those unfamiliar with KL divergence, see a visualization in \cref{fig:rope_scarce,fig:rope_rankfreq} albeit for a different set of data). In fact, even with significantly reduced initial data (down to $<200$ triangulations total, $<0.5\%$ of all triangulations), dualGNN outperforms all methods other than \code{flip\_walk} which also showed strong performance. For dualGNN variants with reduced training data, we use the bootstrap procedure described in \cref{subsec:sampler}: every $500$ training steps, the current model generates $100$ candidate triangulations, and newly found valid triangulations are added to the training pool. This matters because real applications often involve polygons with astronomical counts of FTs (we study in \cref{subsubsec:multi} polygons with up to $10^{21}$ FTs), for which one necessarily begins with a very small subset of all triangulations. While this bootstrapping performance is impressive, it is not unique to dualGNN: \cite{yip2025transformingcalabiyauconstructionsgenerating} observed similar behavior with their transformer model and we recreate similar behavior with a transformer in \cref{sec:transformer}.

\begin{figure}
    \centering
    \includegraphics[width=0.8\textwidth]{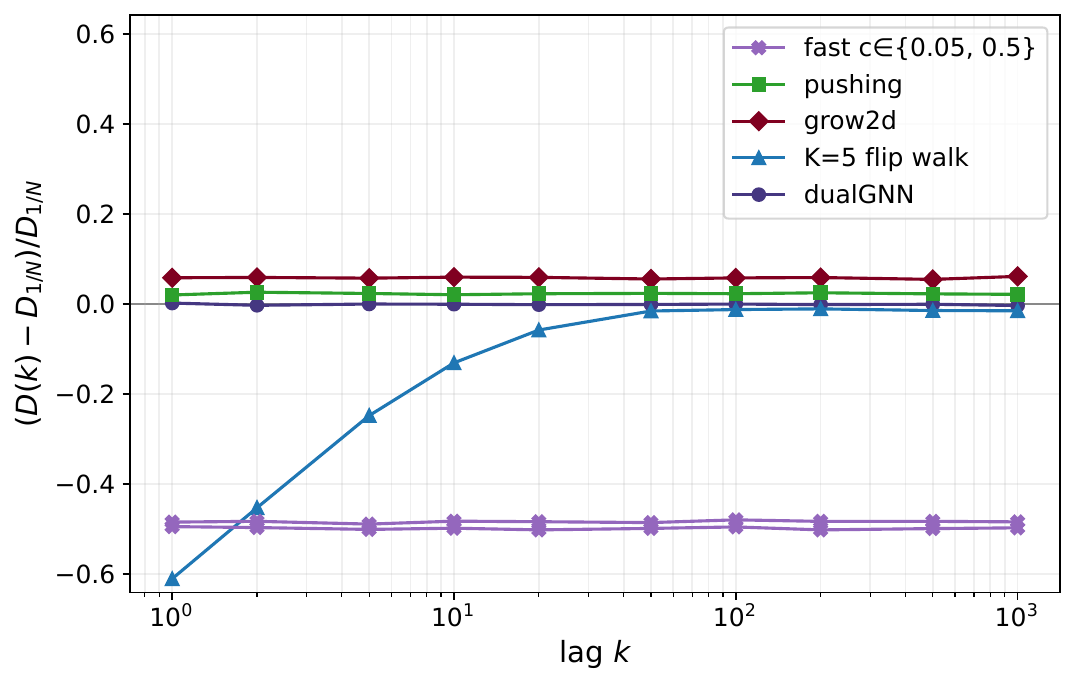}
    \caption{For the data in \cref{fig:dualgnn_4x6tri}, the autocorrelation between samples. Distance between samples is measured via $\mathrm{dist}(\mathcal{T}_i,\mathcal{T}_j)\geq\frac{1}{4}\left|(\mathcal{T}_i \cup \mathcal{T}_j) \setminus (\mathcal{T}_i \cap \mathcal{T}_j)\right|$; autocorrelation via $(\mathrm{dist}(\mathcal{T}_i,\mathcal{T}_{i+k}) - \mathrm{dist}_{1/N})/\mathrm{dist}_{1/N}$. Here, $\mathrm{dist}_{1/N}$ is the empirically-measured distance between samples of a uniform sampler. This is analogous to how correlation is diagnosed in random number generators \cite{Knuth1997-eo}. All methods other than \code{flip\_walk} show constant distances, with only dualGNN having a value agreeing with a uniform sampler ($0$). \code{flip\_walk}, on the other hand, shows strong correlation for samples up to $\sim20$ draws apart.}
    \label{fig:dualgnn_autocorrelation_4x6tri}
\end{figure}

KL divergence is not the entire story. When assessing pseudorandom number generators, one typically also studies the correlation between samples\cite{Knuth1997-eo}. We do the same here for our triangulation samplers. Explicitly, we use the symmetric difference between triangulations as a lower bound on flip distance\footnote{Every flip of a fine triangulation of a lattice polygon is a diagonal flip. That is, it replaces two simplices with two new ones.} $\mathrm{dist}(\mathcal{T}_i,\mathcal{T}_{i+k})\geq\frac{1}{4}\left|(\mathcal{T}_i \cup \mathcal{T}_{i+k}) \setminus (\mathcal{T}_i \cap \mathcal{T}_{i+k})\right|$ for samples $i$ and $i+k$. Ideally, on average, this flip distance should match that of a true $1/N$ sampler, so we plot the normalized difference from this baseline in \cref{fig:dualgnn_autocorrelation_4x6tri}.
The methods \code{pushing} and \code{grow2d} show constant distances, as expected from a uniform sampler, but they show unexpectedly \textit{high} flip distances, indicating these samplers under-sample nearby triangulations rather than over-sampling them. \code{fast}, on the other hand, shows constant distances consistently below the uniform expectation, indicating the sampler over-samples nearby triangulations. dualGNN is the only sampler whose flip distances are consistent with a true uniform distribution --- constant and at the expected value compared to a uniform sampler. Finally, \code{flip\_walk} shows strong correlation when $k$ is low (consistent with \cite{demirtas2020boundingkreuzerskarkelandscape}'s observations of long mixing times), leveling off for higher $k$. This \code{flip\_walk} correlation makes sense: the Markov-chain nature of \code{flip\_walk} implies that samples with lag-$k$ are at most $k$-flips apart from one-another. The explicit \code{random\_triangulations\_fair} algorithm discussed in~\cite{demirtas2020boundingkreuzerskarkelandscape} takes extra measures (modifications to heights) that may resolve this correlation, but this is anticipated to come at the expense of the already slow sampling rate, as we discuss below.

Sample rate splits the methods into four tiers (\cref{fig:dualgnn_4x6tri}): \code{pushing} is fastest at $\sim40,000$ samples/second (since it does not require a regularity check); \code{grow2d}/\code{fast} follow at $500-650$ samples/second; 
dualGNN is slower (regularity + simplex-selection) at $\sim$200 samples/second\footnote{\label{fn:timing}All timing numbers in this paper were measured with the experiment-era code (release \code{v0.0.1} in the repository). Subsequent releases include inference optimizations affecting only dualGNN: polygon sampling improved from $\sim$200 to $\sim$750 samples/second on the polygon of \cref{fig:4x6tri} (RTX 5060 Ti), CPU inference sped up by $\gtrsim4\times$, and end-to-end CY sampling (with the combination of FRTs into an FRST) now runs within $\sim5\times$ of \code{fast}. The comparisons reported in the text are therefore conservative.}; \code{flip\_walk} is slowest, at $\sim50$ samples/second. All reported times include regularity checking (except for \code{pushing} which has guaranteed-regular outputs). The speed of \code{flip\_walk} is set in part by the number of flips between samples; one could reduce this from $5$ to a lower number, but that would worsen the correlation between samples. Here, we note that the CYTools variant of \code{random\_triangulations\_fair} is significantly slower than even \code{flip\_walk}.

\begin{figure}
    \centering
    \includegraphics[width=1.0\textwidth]{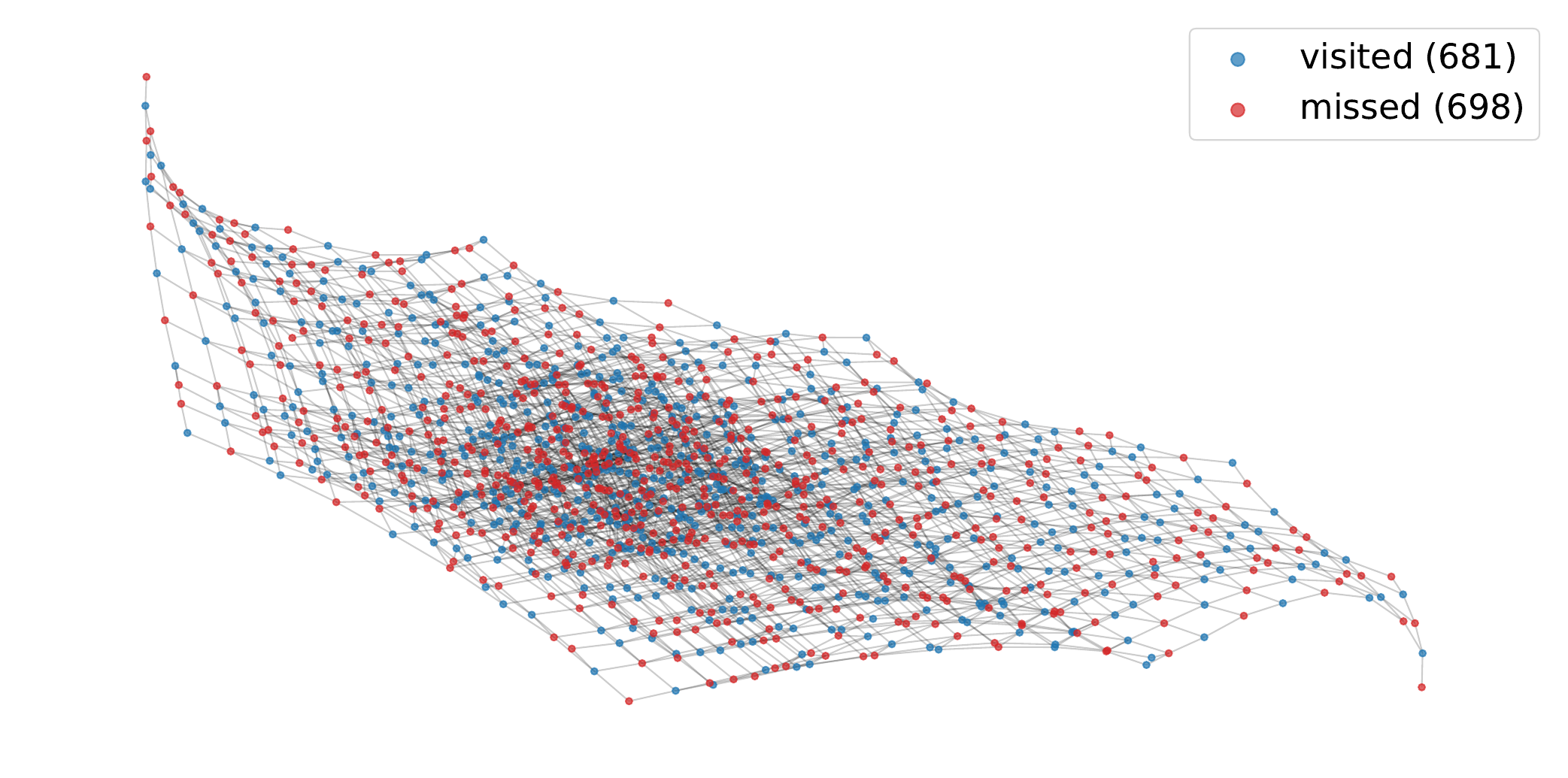}
    \caption{The flip graph of $\mathrm{conv}(\{(0,0),(2,3),(0,9)\})$, with nodes representing FRTs and edges representing flips. This graph is bipartite so a Markov-chain sampler such as \code{flip\_walk} with an even number of steps between samples would only sample from one of two colors (red or blue). This example was found by happenstance on the first polygon tested.}
    \label{fig:bipartite}
\end{figure}

A second concern with \code{flip\_walk} arises from the structure of the flip graph itself. Some flip graphs are bipartite (see \cref{fig:bipartite}), meaning a walk with an even number of steps between samples will be unable to generate $\sim50\%$ of all triangulations. This is resolvable by using an odd number of flips between samples. dualGNN does not face this problem: each inference call generates an independent sample.

Overall, \code{pushing} and \code{grow2d} are competitive samplers if bias is tolerated, especially \code{pushing}, which is very quick (but this method cannot, even in principle, generate some FRTs). \code{flip\_walk} is a very effective untrained model, but it has non-trivial sample correlation and a sensitivity to walk-length choice. \code{fast} is generally not recommended --- it is neither extremely fast nor very uniform. Finally, if uniformity is crucial, dualGNN outperforms all methods while having speed between \code{grow2d} and \code{flip\_walk} (dualGNN has since become significantly quicker --- see footnote~\ref{fn:timing}).

\subsubsection{Zero-shot Transfer}
\label{subsubsec:zeroshot}

As we show here and in \cref{subsubsec:multi}, dualGNN generalizes across polygons. To demonstrate this, we begin by applying the model trained in \cref{subsubsec:single} zero-shot to the polygon $[0,4]^2$. This polygon has $25$ lattice points and $4$ vertices; each triangulation of it consists of $32$ simplices. In all regards, then, sampling from $[0,4]^2$ is a harder task than that from the polygon in \cref{fig:4x6tri} which has $19$ lattice points and $3$ vertices; each triangulation of the polygon in \cref{fig:4x6tri} consists of $24$ simplices. As aforementioned, these polygons also differ significantly in triangulation count: $[0,4]^2$ has $735,430,548$ FRTs (out of $736,983,568$ FTs) while the polygon in \cref{fig:4x6tri} has only $405,706$ (out of $408,826$).

\begin{figure}
    \centering
    \includegraphics[width=0.8\textwidth]{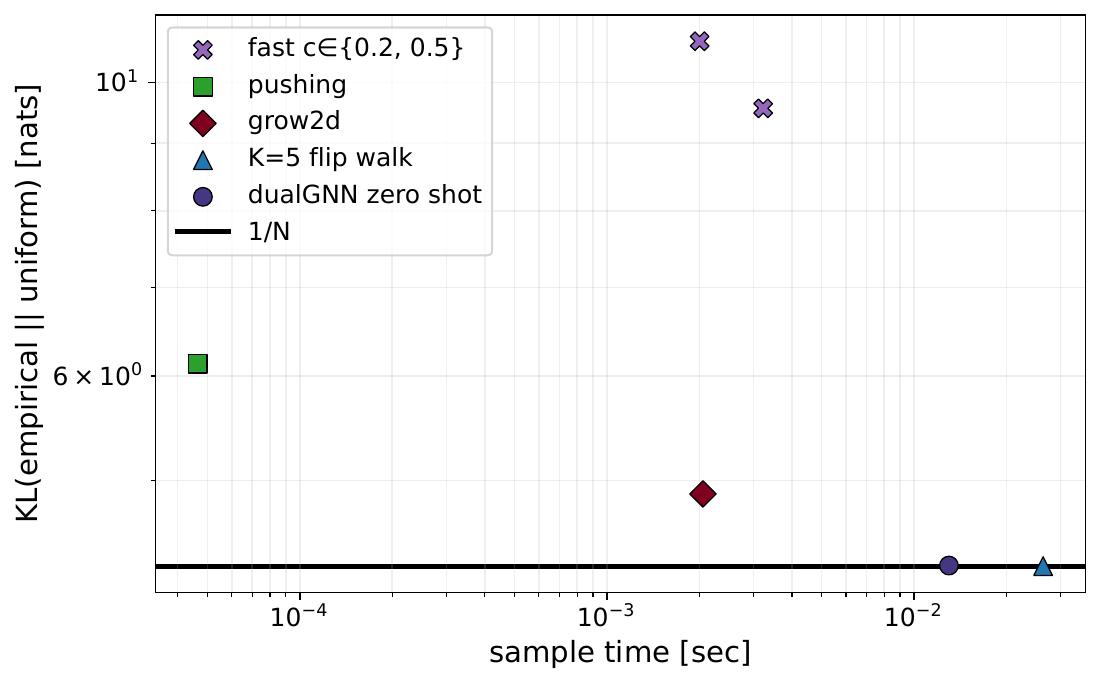}
    \caption{The dualGNN sampler, trained on the polygon in \cref{fig:4x6tri}, applied to $[0,4]^2$. This polygon has $735,430,548$ FRTs, many more than we could sample in our computational budget ($10^7$). Both \code{flip\_walk} and dualGNN achieve near uniform pools of samples, measured against the noise floor due to the finite number of samples. \code{grow2d} and \code{pushing} show moderate bias but significantly increased sampling rates. Finally, \code{fast} shows the largest bias despite being significantly slower than \code{pushing}.}
    \label{fig:dualgnn_4x4sq_zeroshot}
\end{figure}

Generating the billions of samples needed for a complete uniformity measurement is outside our compute budget, so we instead draw $10^7$ and assess uniformity within them. This is arguably a more representative test: for most applications in string theory, one wants a modest number of FRTs sampled from $2$-faces with astronomical counts of FRTs --- the large number of CYs come from different ways of grouping these samples together~\cite{berglund2024generatingtriangulationsfibrationsreinforcement,macfadden2025dnacalabiyauhypersurfaces}. The small sample pool compared to the total number of FRTs means that even a truly uniform (but finitely drawn) sampler shows significant KL divergence compared to the flat distribution, making KL divergence a less informative diagnostic in this case (see \cref{fig:dualgnn_4x4sq_zeroshot}). Still, \code{flip\_walk} and dualGNN show the lowest KL divergence, only $<0.005$ above the noise floor (see \cref{table:zero_shot}; set by our finite number of samples). \code{flip\_walk} has the lowest KL.

\begin{table}
    \centering
    \begin{tabular}{lrrr}
        \toprule
        Sampler & \# unique & \# collisions & excess KL \\
        \midrule
        \code{fast\_c0.2}  & $61,316$ & $8.12\times10^9$ & $6.44$ \\
        \code{fast\_c0.5}  & $499,829$ & $3.49\times10^9$ & $5.26$ \\
        \code{pushing}     & $3,779,093$ & $1.24\times10^8$ & $1.82$ \\
        \code{grow2d}      & $7,321,880$ & $2.19\times10^7$ & $0.58$ \\
        dualGNN              & $9,895,669$ & $1.05\times10^5$ & $0.005$ \\
        \code{flip\_walk}  & $9,924,852$ & $7.55\times10^{4}$ & $0.001$ \\
        \hline
        $1/N$                & $9,932,320$ & $6.80\times10^4$ & $0$ \\
        \bottomrule
    \end{tabular}
    \caption{Performance of FRT samplers on $[0,4]^2$, listed in order of increasing fairness. Notably, dualGNN here is the model trained on the polygon in \cref{fig:4x6tri} applied zero-shot to $[0,4]^2$. The bottom row contains theoretical predictions for a true uniform sampler, for which there are $\#\mathrm{unique}=N\big(1-(1-1/N)^M\big)$ and $\#\mathrm{collisions}=M(M-1)/(2N)$. \code{flip\_walk} is the most fair sampler, but dualGNN is not far behind despite being applied zero-shot.}
    \label{table:zero_shot}
\end{table}

To better discriminate bias, we also study the total number of collisions (different samples giving the same triangulation). For $M$ uniform samples out of a pool of $N$ items, any pair ${M\choose2}$ is expected to collide at probability $1/N$. Thus, we expect a total number of collisions
\begin{equation}
    \#\mathrm{collisions} = {M\choose2} \frac{1}{N} = \frac{M(M-1)}{2N}.
\end{equation}
For our case, with $M=10^7$ and $N=735,430,548$, we thus predict $67,987$ collisions (out of $\sim 5\times10^{13}$ possible ones). This diagnostic shows bias in all samplers (see \cref{table:zero_shot}), with \code{flip\_walk} and the zero-shot dualGNN having the closest number of collisions to the uniform prediction. Again, \code{flip\_walk} shows the lowest bias here, but the zero shot dualGNN is not far behind.

This is very strong performance from dualGNN given that this model had only been trained for $O(\text{hours})$ on a \textit{different} polygon with significantly different shape, point count, and triangulation count (i.e., that from \cref{fig:4x6tri}). This suggests that the architecture generalizes strongly across polygons.

\subsubsection{Multiple Polygons}
\label{subsubsec:multi}

The ultimate goal of dualGNN is to be a general-purpose FRT sampler. We describe such a model in this section. Trained across $271$ polygons and fine-tuned with REINFORCE\cite{Williams1992}, it achieves the most uniform sampling of any method tested across $N_\mathrm{pts} \leq 18$ and is consistent with uniformity at $N_\mathrm{pts}=40$ within the resolution granted by our sample budgets.

We maintain the same hyperparameters as before ($32$-dimensional feature vectors, $16$ message-passing rounds). What we change is the training: instead of training on (sometimes bootstrapped) pools of fine triangulations of a single polygon, we train on bootstrapped pools from \textit{randomly chosen} polygons. Explicitly, we generate $116$ randomly chosen polygons to start with $12\leq N_\mathrm{pts}\leq 40$ and build initial pools by either fully enumerating the triangulations (for $N_\mathrm{pts}\leq 17$) or sampling an initial pool of $10,000$ FTs using \code{grow2d}. To better enable multi-polygon learning, we also modify the exploration rounds: in each such round, with probability $80\%$, we pick an existing polygon from the training pool; otherwise ($20\%$ chance) we generate a new random polygon with $5\leq N_\mathrm{pts}\leq 40$. In our $\sim5.5$ hour training (on a single 5060 Ti) over $500,000$ steps, this led to $271$ polygons in total. Each new polygon is seeded with $2,000$ \code{grow2d} samples.

\input{tikz/ood_small}

\begin{figure}
    \centering
    \includegraphics[width=0.8\textwidth]{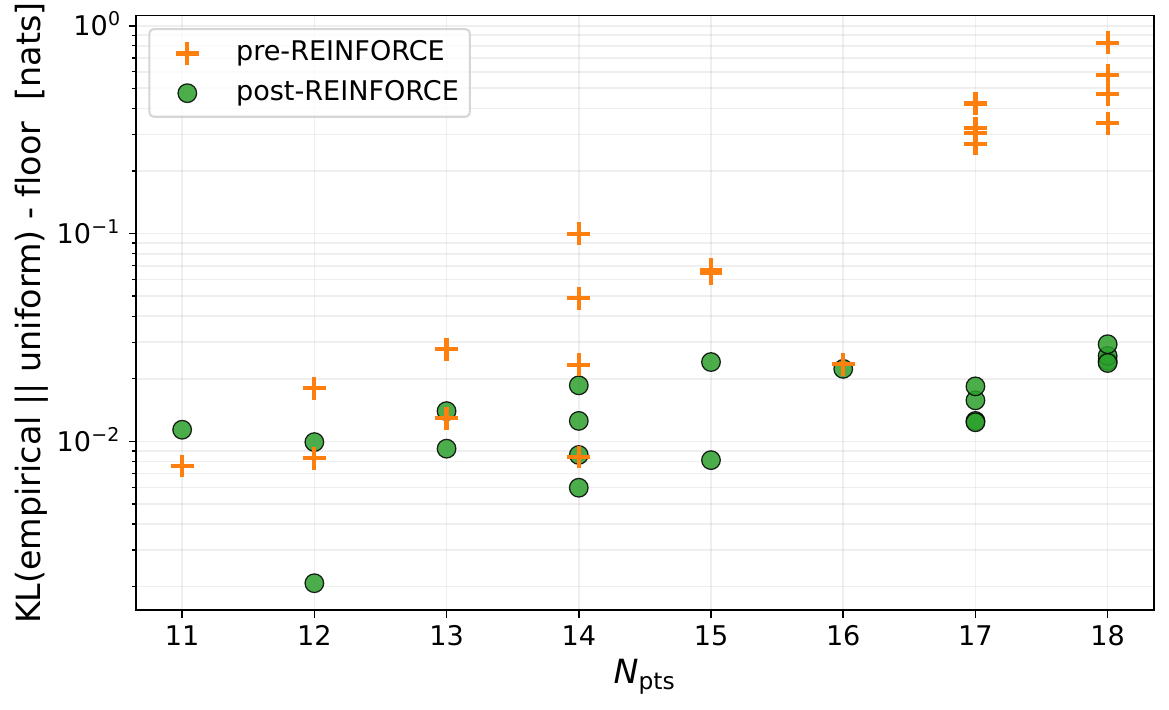}
    \caption{Pre- and post-REINFORCE finetuning on the dualGNN autoregressive sampler that was trained on multiple polygons. All of these markers correspond to polygons not seen in the training (see \cref{fig:ood-polygons-grid}), $200,000$ samples each. In all cases except the smallest polygon, REINFORCE caused more uniform samples, oftentimes significantly so.}
    \label{fig:need_for_rl}
\end{figure}

\begin{figure}
    \centering
    \includegraphics[width=1.0\textwidth]{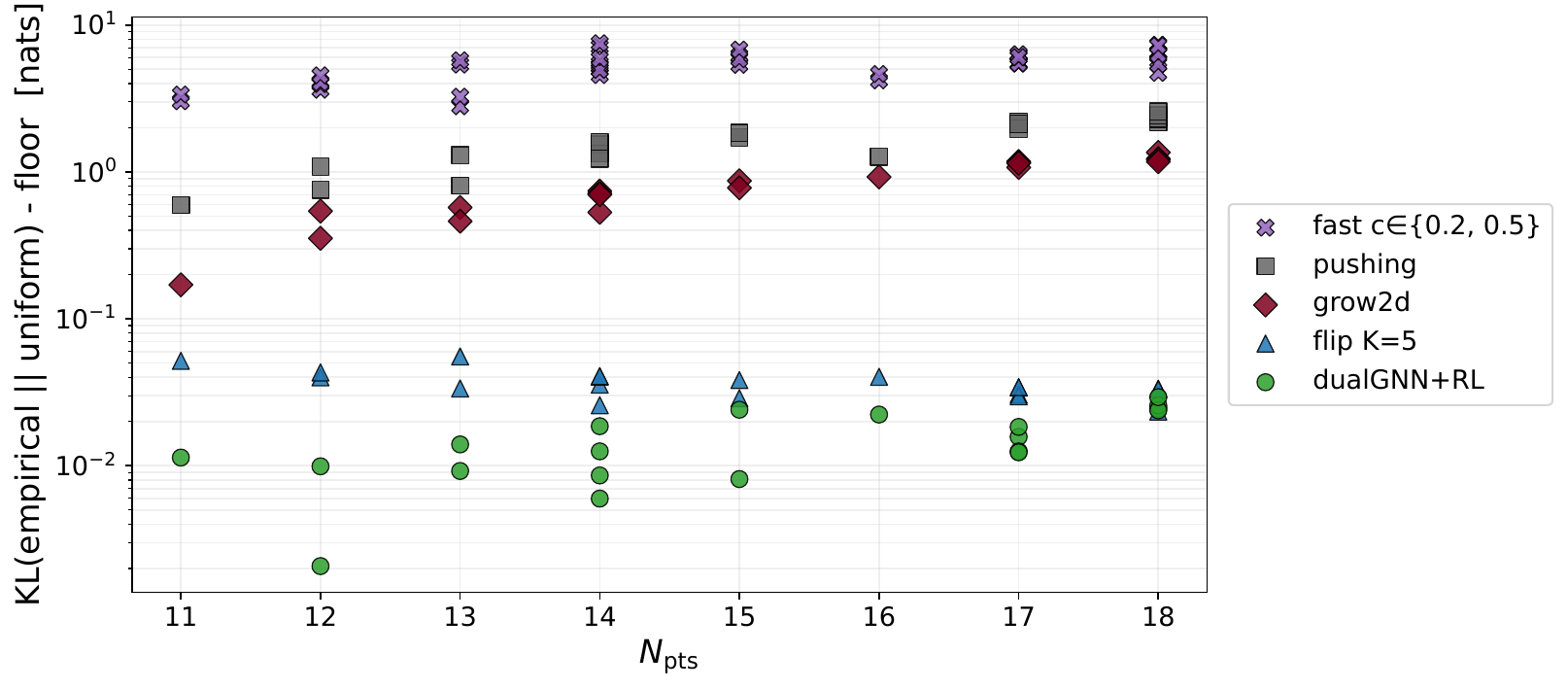}
    \caption{Comparison of the uniformity of samples for $200,000$ samples on the polygons in \cref{fig:ood-polygons-grid}. In all cases, dualGNN is the most uniform, except the largest $N_\mathrm{pts}=18$ at which it is equal to \code{flip\_walk} despite being $\sim4\times$ faster than \code{flip\_walk}.}
    \label{fig:multipoly}
\end{figure}

\begin{figure}
    \centering
    \includegraphics[width=0.8\textwidth]{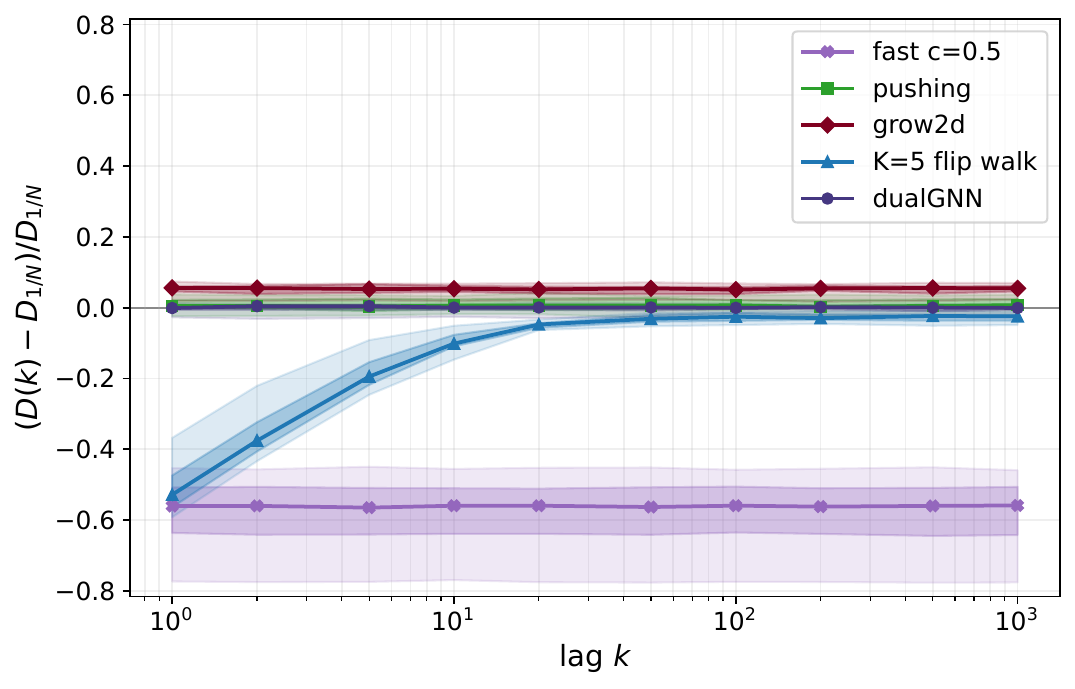}
    \caption{Autocorrelation of the various samplers over $200,000$ samples on the polygons in \cref{fig:ood-polygons-grid}. As with \cref{fig:dualgnn_autocorrelation_4x6tri}, we use the symmetric difference between the triangulations as a lower bound on the flip distance, with a curve for the mean value and bands indicating the spread. dualGNN and \code{pushing} are both consistent with a uniform distribution ($y=0$) for all polys while \code{flip\_walk} shows significant correlation for low lags. \code{grow2d} and \code{fast} both show constant flip distances, which would be consistent with uniform, but they average to a different distance than what a uniform sampler would predict. 
    }
    \label{fig:multipoly_autocorrelation}
\end{figure}

To evaluate this trained model, we apply it to $20$ polygons from outside the training pool (see \cref{fig:ood-polygons-grid}) with $11\leq N_\mathrm{pts}\leq18$, generating $200,000$ samples from each polygon. The supervised learning leads to moderate uniformity (orange pluses in \cref{fig:need_for_rl}), but it is not up to our standards. This is likely because the cross-entropy objective is only an indirect proxy for what we actually want (a uniform sampler over complete triangulations). Cross-entropy matches per-step conditional probabilities to estimated targets, but biases in the training pool \textit{do} affect the uniformity of the trained model. To directly optimize the triangulation distribution, we fine-tune dualGNN with REINFORCE~\cite{Williams1992} using an entropy-maximizing reward. Explicitly, we perform $10,000$ steps on the training polygons, each step a batch of four full rollouts. After the rollouts, the probability of each triangulation being sampled is computed as the product of each simplex's conditional probability. If the triangulation is regular, we then assign reward $-\log \mathrm{prob}(T)$, otherwise the triangulation receives reward $-2$ (penalty magnitude chosen arbitrarily since no regular triangulation gets a negative reward). This maximizes entropy, rewarding the policy for generating regular triangulations it deems unlikely, driving the distribution toward uniformity. We clip gradients by magnitude $1$ and use a learning rate of $3\times 10^{-5}$. This RL post-training takes $\sim2$ additional hours ($\sim7.5$ hours total training) and leads to significantly more uniform samples (green circles in \cref{fig:need_for_rl}). Comparing to other reference samplers post this fine-tuning, we find that dualGNN is the most uniform studied (see \cref{fig:multipoly}; tied with \code{flip\_walk} at $N_\mathrm{pts}=18$) and, in contrast to \code{flip\_walk}, shows no autocorrelation (\cref{fig:multipoly_autocorrelation}).

One subtlety deserves comment: our reward is defined over rollouts (orderings of simplex placements) while uniformity is desired over triangulations, and a single triangulation is reachable by many rollouts. GFlowNets~\cite{bengio2021flownetworkbasedgenerative} were designed for precisely this many-paths setting. In our 2D setting, however, this machinery is not strictly necessary: every fine triangulation of a fixed polygon contains the same number $n$ of simplices and every ordering of a triangulation's simplices is a feasible rollout, so each triangulation corresponds to exactly $n!$ rollouts. Uniformity over rollouts therefore implies uniformity over triangulations, and our entropy-reward REINFORCE targets the desired distribution directly. Note, however, that our architecture treats a triangulation's simplices as a sequence (in the selection order) rather than a set, and so does not build in their order-invariance; a GFlowNet would respect this symmetry structurally. This is a cost in inductive bias rather than in the uniformity itself. Additionally, GFlowNet objectives become attractive for non-uniform target distributions, higher dimensions, or learning non-uniform distributions (the purpose of \cite{berglund2024generatingtriangulationsfibrationsreinforcement,macfadden2025dnacalabiyauhypersurfaces}; a goal listed in \cite{yip2025transformingcalabiyauconstructionsgenerating}); we leave a direct comparison to future work.

\begin{figure}[t]
\centering
\tikzset{
  latticepoint/.style={draw, circle, inner sep=0.9pt, fill},
  trilinestyle/.style={line width=0.25mm},
}
\setlength{\tabcolsep}{8pt}
\renewcommand{\arraystretch}{1.0}
\begin{tabular}{cccc}
\begin{tabular}{c}\begin{tikzpicture}[baseline={(current bounding box.center)}]
\tikzmath{\scale = 0.3;}
\draw[trilinestyle] (0*\scale,1*\scale) -- (6*\scale,0*\scale) -- (6*\scale,5*\scale) -- (5*\scale,8*\scale) -- (2*\scale,7*\scale) -- cycle;
  \node[latticepoint] at (0*\scale,1*\scale) {};
  \node[latticepoint] at (1*\scale,1*\scale) {};
  \node[latticepoint] at (1*\scale,2*\scale) {};
  \node[latticepoint] at (1*\scale,3*\scale) {};
  \node[latticepoint] at (1*\scale,4*\scale) {};
  \node[latticepoint] at (2*\scale,1*\scale) {};
  \node[latticepoint] at (2*\scale,2*\scale) {};
  \node[latticepoint] at (2*\scale,3*\scale) {};
  \node[latticepoint] at (2*\scale,4*\scale) {};
  \node[latticepoint] at (2*\scale,5*\scale) {};
  \node[latticepoint] at (2*\scale,6*\scale) {};
  \node[latticepoint] at (2*\scale,7*\scale) {};
  \node[latticepoint] at (3*\scale,1*\scale) {};
  \node[latticepoint] at (3*\scale,2*\scale) {};
  \node[latticepoint] at (3*\scale,3*\scale) {};
  \node[latticepoint] at (3*\scale,4*\scale) {};
  \node[latticepoint] at (3*\scale,5*\scale) {};
  \node[latticepoint] at (3*\scale,6*\scale) {};
  \node[latticepoint] at (3*\scale,7*\scale) {};
  \node[latticepoint] at (4*\scale,1*\scale) {};
  \node[latticepoint] at (4*\scale,2*\scale) {};
  \node[latticepoint] at (4*\scale,3*\scale) {};
  \node[latticepoint] at (4*\scale,4*\scale) {};
  \node[latticepoint] at (4*\scale,5*\scale) {};
  \node[latticepoint] at (4*\scale,6*\scale) {};
  \node[latticepoint] at (4*\scale,7*\scale) {};
  \node[latticepoint] at (5*\scale,1*\scale) {};
  \node[latticepoint] at (5*\scale,2*\scale) {};
  \node[latticepoint] at (5*\scale,3*\scale) {};
  \node[latticepoint] at (5*\scale,4*\scale) {};
  \node[latticepoint] at (5*\scale,5*\scale) {};
  \node[latticepoint] at (5*\scale,6*\scale) {};
  \node[latticepoint] at (5*\scale,7*\scale) {};
  \node[latticepoint] at (5*\scale,8*\scale) {};
  \node[latticepoint] at (6*\scale,0*\scale) {};
  \node[latticepoint] at (6*\scale,1*\scale) {};
  \node[latticepoint] at (6*\scale,2*\scale) {};
  \node[latticepoint] at (6*\scale,3*\scale) {};
  \node[latticepoint] at (6*\scale,4*\scale) {};
  \node[latticepoint] at (6*\scale,5*\scale) {};
\end{tikzpicture}\\\footnotesize \end{tabular} &
\begin{tabular}{c}\begin{tikzpicture}[baseline={(current bounding box.center)}]
\tikzmath{\scale = 0.3;}
\draw[trilinestyle] (2*\scale,8*\scale) -- (0*\scale,0*\scale) -- (7*\scale,5*\scale) -- (9*\scale,8*\scale) -- cycle;
  \node[latticepoint] at (0*\scale,0*\scale) {};
  \node[latticepoint] at (1*\scale,1*\scale) {};
  \node[latticepoint] at (1*\scale,2*\scale) {};
  \node[latticepoint] at (1*\scale,3*\scale) {};
  \node[latticepoint] at (1*\scale,4*\scale) {};
  \node[latticepoint] at (2*\scale,2*\scale) {};
  \node[latticepoint] at (2*\scale,3*\scale) {};
  \node[latticepoint] at (2*\scale,4*\scale) {};
  \node[latticepoint] at (2*\scale,5*\scale) {};
  \node[latticepoint] at (2*\scale,6*\scale) {};
  \node[latticepoint] at (2*\scale,7*\scale) {};
  \node[latticepoint] at (2*\scale,8*\scale) {};
  \node[latticepoint] at (3*\scale,3*\scale) {};
  \node[latticepoint] at (3*\scale,4*\scale) {};
  \node[latticepoint] at (3*\scale,5*\scale) {};
  \node[latticepoint] at (3*\scale,6*\scale) {};
  \node[latticepoint] at (3*\scale,7*\scale) {};
  \node[latticepoint] at (3*\scale,8*\scale) {};
  \node[latticepoint] at (4*\scale,3*\scale) {};
  \node[latticepoint] at (4*\scale,4*\scale) {};
  \node[latticepoint] at (4*\scale,5*\scale) {};
  \node[latticepoint] at (4*\scale,6*\scale) {};
  \node[latticepoint] at (4*\scale,7*\scale) {};
  \node[latticepoint] at (4*\scale,8*\scale) {};
  \node[latticepoint] at (5*\scale,4*\scale) {};
  \node[latticepoint] at (5*\scale,5*\scale) {};
  \node[latticepoint] at (5*\scale,6*\scale) {};
  \node[latticepoint] at (5*\scale,7*\scale) {};
  \node[latticepoint] at (5*\scale,8*\scale) {};
  \node[latticepoint] at (6*\scale,5*\scale) {};
  \node[latticepoint] at (6*\scale,6*\scale) {};
  \node[latticepoint] at (6*\scale,7*\scale) {};
  \node[latticepoint] at (6*\scale,8*\scale) {};
  \node[latticepoint] at (7*\scale,5*\scale) {};
  \node[latticepoint] at (7*\scale,6*\scale) {};
  \node[latticepoint] at (7*\scale,7*\scale) {};
  \node[latticepoint] at (7*\scale,8*\scale) {};
  \node[latticepoint] at (8*\scale,7*\scale) {};
  \node[latticepoint] at (8*\scale,8*\scale) {};
  \node[latticepoint] at (9*\scale,8*\scale) {};
\end{tikzpicture}\\\footnotesize \end{tabular} &
\begin{tabular}{c}\begin{tikzpicture}[baseline={(current bounding box.center)}]
\tikzmath{\scale = 0.3;}
\draw[trilinestyle] (0*\scale,0*\scale) -- (7*\scale,1*\scale) -- (6*\scale,7*\scale) -- (2*\scale,7*\scale) -- cycle;
  \node[latticepoint] at (1*\scale,1*\scale) {};
  \node[latticepoint] at (1*\scale,2*\scale) {};
  \node[latticepoint] at (1*\scale,3*\scale) {};
  \node[latticepoint] at (2*\scale,1*\scale) {};
  \node[latticepoint] at (2*\scale,2*\scale) {};
  \node[latticepoint] at (2*\scale,3*\scale) {};
  \node[latticepoint] at (2*\scale,4*\scale) {};
  \node[latticepoint] at (2*\scale,5*\scale) {};
  \node[latticepoint] at (2*\scale,6*\scale) {};
  \node[latticepoint] at (3*\scale,1*\scale) {};
  \node[latticepoint] at (3*\scale,2*\scale) {};
  \node[latticepoint] at (3*\scale,3*\scale) {};
  \node[latticepoint] at (3*\scale,4*\scale) {};
  \node[latticepoint] at (3*\scale,5*\scale) {};
  \node[latticepoint] at (3*\scale,6*\scale) {};
  \node[latticepoint] at (4*\scale,1*\scale) {};
  \node[latticepoint] at (4*\scale,2*\scale) {};
  \node[latticepoint] at (4*\scale,3*\scale) {};
  \node[latticepoint] at (4*\scale,4*\scale) {};
  \node[latticepoint] at (4*\scale,5*\scale) {};
  \node[latticepoint] at (4*\scale,6*\scale) {};
  \node[latticepoint] at (5*\scale,1*\scale) {};
  \node[latticepoint] at (5*\scale,2*\scale) {};
  \node[latticepoint] at (5*\scale,3*\scale) {};
  \node[latticepoint] at (5*\scale,4*\scale) {};
  \node[latticepoint] at (5*\scale,5*\scale) {};
  \node[latticepoint] at (5*\scale,6*\scale) {};
  \node[latticepoint] at (6*\scale,1*\scale) {};
  \node[latticepoint] at (6*\scale,2*\scale) {};
  \node[latticepoint] at (6*\scale,3*\scale) {};
  \node[latticepoint] at (6*\scale,4*\scale) {};
  \node[latticepoint] at (6*\scale,5*\scale) {};
  \node[latticepoint] at (6*\scale,6*\scale) {};
  \node[latticepoint] at (0*\scale,0*\scale) {};
  \node[latticepoint] at (2*\scale,7*\scale) {};
  \node[latticepoint] at (6*\scale,7*\scale) {};
  \node[latticepoint] at (7*\scale,1*\scale) {};
  \node[latticepoint] at (3*\scale,7*\scale) {};
  \node[latticepoint] at (4*\scale,7*\scale) {};
  \node[latticepoint] at (5*\scale,7*\scale) {};
\end{tikzpicture}\\\footnotesize \end{tabular} &
\begin{tabular}{c}\begin{tikzpicture}[baseline={(current bounding box.center)}]
\tikzmath{\scale = 0.3;}
\draw[trilinestyle] (5*\scale,9*\scale) -- (0*\scale,3*\scale) -- (6*\scale,0*\scale) -- (8*\scale,2*\scale) -- cycle;
  \node[latticepoint] at (1*\scale,3*\scale) {};
  \node[latticepoint] at (1*\scale,4*\scale) {};
  \node[latticepoint] at (2*\scale,3*\scale) {};
  \node[latticepoint] at (2*\scale,4*\scale) {};
  \node[latticepoint] at (2*\scale,5*\scale) {};
  \node[latticepoint] at (3*\scale,2*\scale) {};
  \node[latticepoint] at (3*\scale,3*\scale) {};
  \node[latticepoint] at (3*\scale,4*\scale) {};
  \node[latticepoint] at (3*\scale,5*\scale) {};
  \node[latticepoint] at (3*\scale,6*\scale) {};
  \node[latticepoint] at (4*\scale,2*\scale) {};
  \node[latticepoint] at (4*\scale,3*\scale) {};
  \node[latticepoint] at (4*\scale,4*\scale) {};
  \node[latticepoint] at (4*\scale,5*\scale) {};
  \node[latticepoint] at (4*\scale,6*\scale) {};
  \node[latticepoint] at (4*\scale,7*\scale) {};
  \node[latticepoint] at (5*\scale,1*\scale) {};
  \node[latticepoint] at (5*\scale,2*\scale) {};
  \node[latticepoint] at (5*\scale,3*\scale) {};
  \node[latticepoint] at (5*\scale,4*\scale) {};
  \node[latticepoint] at (5*\scale,5*\scale) {};
  \node[latticepoint] at (5*\scale,6*\scale) {};
  \node[latticepoint] at (5*\scale,7*\scale) {};
  \node[latticepoint] at (5*\scale,8*\scale) {};
  \node[latticepoint] at (6*\scale,1*\scale) {};
  \node[latticepoint] at (6*\scale,2*\scale) {};
  \node[latticepoint] at (6*\scale,3*\scale) {};
  \node[latticepoint] at (6*\scale,4*\scale) {};
  \node[latticepoint] at (6*\scale,5*\scale) {};
  \node[latticepoint] at (6*\scale,6*\scale) {};
  \node[latticepoint] at (7*\scale,2*\scale) {};
  \node[latticepoint] at (7*\scale,3*\scale) {};
  \node[latticepoint] at (7*\scale,4*\scale) {};
  \node[latticepoint] at (0*\scale,3*\scale) {};
  \node[latticepoint] at (5*\scale,9*\scale) {};
  \node[latticepoint] at (6*\scale,0*\scale) {};
  \node[latticepoint] at (8*\scale,2*\scale) {};
  \node[latticepoint] at (2*\scale,2*\scale) {};
  \node[latticepoint] at (4*\scale,1*\scale) {};
  \node[latticepoint] at (7*\scale,1*\scale) {};
\end{tikzpicture}\\\footnotesize \end{tabular} \\
\end{tabular}
\caption{Four out-of-distributions polygons at $N_{\mathrm{pts}}=40$ used for post-REINFORCE inference.}
\label{fig:npts40-ood-polygons-grid}
\end{figure}
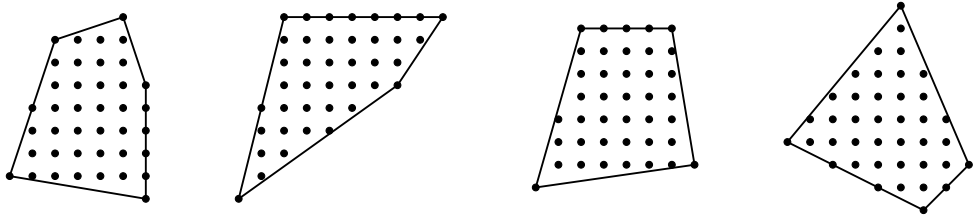

We also apply this same model to larger polygons, each with $N_\mathrm{pts}=40$ (see \cref{fig:npts40-ood-polygons-grid}). The large triangulation counts for such large polygons (with upper bounds ranging from $3.7\times10^{19}$ to $5.9\times 10^{20}$ FTs\cite{anclin2003}) make it significantly more difficult to measure bias. Even significantly biased samplers can show low collision counts, meaning that they would have KL divergences nearing the noise floor (due to our finite number of samples). We address this, in part, by again using the collision count itself as a diagnostic since it was better suited to sparse samples in \cref{subsubsec:zeroshot}.

Across the four $N_\mathrm{pts}=40$ polygons, dualGNN is the only sampler to have no collisions across the $100,000$ draws (see \cref{table:multi_polygon}). This is consistent with the above rough estimates on the number of FRTs for these polygons, $\gg10^9$. This lack of collisions is a necessary, not sufficient, test for dualGNN's uniformity. No other method passes this test: \code{grow2d} is the only other sampler that achieves $0$ collisions for one polygon, but it has nonzero collisions for all other polygons. All other methods collide on all polygons. By inverting the predicted number of uniques for a uniform sampler, $\#\mathrm{unique} = N\big(1 - (1 - 1/N)^M\big)$, we can get a rough estimate of an `effective' population that each sampler is sampling out of (if it were uniform). For the biased samplers (\code{fast}), this is as low as $10^5$; for other samplers, this is consistently $\lesssim10^9$. We stress: the lack of collisions is not a proof that dualGNN is uniform, just the passing of a necessary test that no other samplers passed, not even often-competitive \code{flip\_walk}.

\begin{table}[h]
    \centering
    \hspace*{-0.1\textwidth}\resizebox{1.2\textwidth}{!}{%
    \begin{tabular}{l rrr rrr rrr rrr}
        \toprule
         & \multicolumn{3}{c}{Polygon $P_1$} & \multicolumn{3}{c}{Polygon $P_2$} & \multicolumn{3}{c}{Polygon $P_3$} & \multicolumn{3}{c}{Polygon $P_4$} \\
        \cmidrule(lr){2-4} \cmidrule(lr){5-7} \cmidrule(lr){8-10} \cmidrule(lr){11-13}
        Sampler & \# unique & \# coll. & $N_\mathrm{eff}$ & \# unique & \# coll. & $N_\mathrm{eff}$ & \# unique & \# coll. & $N_\mathrm{eff}$ & \# unique & \# coll. & $N_\mathrm{eff}$ \\
        \midrule
        \code{fast\_c0.2}  & 94{,}702 & 5{,}298 & $9.1\times10^{5}$ & 81{,}164 & 18{,}836 & $2.3\times10^{5}$ & 87{,}123 & 12{,}877 & $3.5\times10^{5}$ & 75{,}995 & 24{,}005 & $1.7\times10^{5}$ \\
        \code{fast\_c0.5}  & 98{,}373 & 1{,}627 & $3.0\times10^{6}$ & 96{,}584 & 3{,}416 & $1.4\times10^{6}$ & 99{,}019 & 981 & $5.1\times10^{6}$ & 94{,}968 & 5{,}032 & $9.6\times10^{5}$ \\
        \code{pushing}     & 99{,}993 & 7 & $7.1\times10^{8}$ & 99{,}989 & 11 & $4.5\times10^{8}$ & 99{,}991 & 9 & $5.6\times10^{8}$ & 99{,}988 & 12 & $4.2\times10^{8}$ \\
        \code{grow2d}      & 100{,}000 & 0 & $\infty$ & 99{,}997 & 3 & $1.7\times10^{9}$ & 99{,}996 & 4 & $1.3\times10^{9}$ & 99{,}998 & 2 & $2.5\times10^{9}$ \\
        \code{flip\_walk}  & 99{,}998 & 2 & $2.5\times10^{9}$ & 99{,}995 & 5 & $1.0\times10^{9}$ & 99{,}997 & 3 & $1.7\times10^{9}$ & 99{,}997 & 3 & $1.7\times10^{9}$ \\
        dualGNN              & 100{,}000 & 0 & $\infty$ & 100{,}000 & 0 & $\infty$ & 100{,}000 & 0 & $\infty$ & 100{,}000 & 0 & $\infty$ \\
        \bottomrule
    \end{tabular}%
    }
    \caption{Performance of FRT samplers on the $N_\mathrm{pts}=40$ polygons. Each polygon block reports the number of unique triangulations recovered (\# unique), number of collisions (\# coll.), and the effective population $N_\mathrm{eff}$ obtained by inverting $\#\mathrm{unique} = N\big(1 - (1 - 1/N)^M\big)$ for $N$ at $M = 100{,}000$. Note: for $0$ collisions, the inversion gives $N_\mathrm{eff}=\infty$, which is only an artifact of our small sample count. Recall that there are strict upper bounds ranging from $3.7\times10^{19}$ to $5.9\times 10^{20}$ FTs (hence FRTs) for these polygons.}
    \label{table:multi_polygon}
\end{table}

\begin{figure}[ht]
    \centering
    \includegraphics[width=0.8\textwidth]{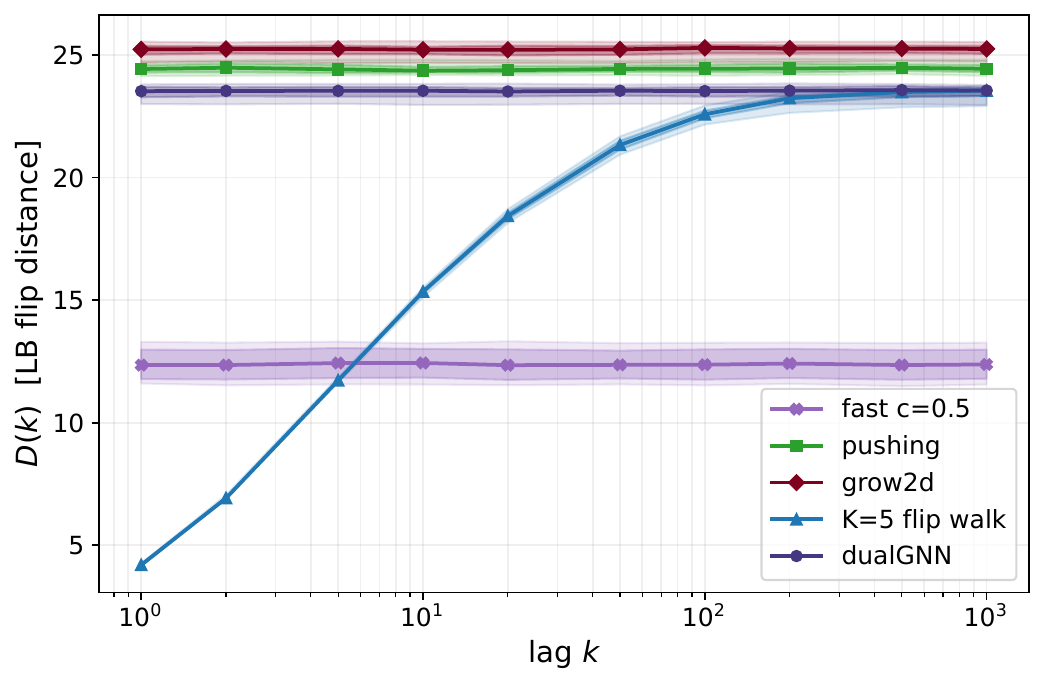}
    \caption{Autocorrelation of the various samplers over $100,000$ samples on the polygons in \cref{fig:npts40-ood-polygons-grid}. As with previous plots, we use the symmetric difference between the triangulations as a lower bound on the flip distance. Since we do not have counts of FRTs for these polygons, we cannot compute $D_{1/N}$ so we just plot the raw lower bounds $D(k)$ on the y-axis. This means that we cannot definitively say that dualGNN is consistent with uniform sampling (since we do not have $D_{1/N}$) but we can (a) identify that \code{flip\_walk} still shows the correlation for small lags, (b) see that the flip distances across methods show the same relative pattern as in \cref{fig:dualgnn_autocorrelation_4x6tri,fig:multipoly_autocorrelation}, and (c) see that the large-$k$ limit of \code{flip\_walk} (presumed uniform) matches dualGNN. This is consistent with expectations if dualGNN were uniform.
    }
    \label{fig:multipoly_autocorrelation_npts40}
\end{figure}

We also return to the autocorrelation of the samplers. This is arguably more important for these large polygons: as polygons get more FTs, their flip graphs become correspondingly larger and there is increased risk for a sampler that makes local explorations (i.e., \code{flip\_walk}) to only explore a small region of triangulations. As can be seen in \cref{fig:multipoly_autocorrelation_npts40}, only dualGNN is consistent with a uniform sampler. \code{flip\_walk} again shows significant correlation for low lags while the other methods show constant distances, but inconsistent with dualGNN and the large-$k$ limit of \code{flip\_walk} (which are consistently the most uniform samplers tested so far).

Altogether, the multi-polygon dualGNN model is the only sampler consistent with uniform sampling across all our diagnostics. In all tests above, dualGNN has consistently lower KL divergences than the other methods (tying with \code{flip\_walk} in some cases), consistent collision rates, and consistent flip distances with a uniform sampler. We stress: these results are all for polygons unseen in the training. dualGNN's uniformity on unseen polygons suggests that it learns the geometry of the problem rather than memorizing a training distribution.

\section{Application: CY Sampling}
\label{sec:cy}

First we recall from \cref{sec:intro} the connection to Calabi-Yau threefolds (CYs). This paper was motivated by the application of generating CYs using Batyrev's construction~\cite{Batyrev:1993oya}. This construction maps a fine, regular, and `star' triangulation (FRST) of a $4$D reflexive polytope (all such polytopes enumerated in the Kreuzer-Skarke database~\cite{kreuzer2000completeclassificationreflexivepolyhedra}, directly accessible in CYTools) to a CY. As argued in the introduction, this map is many-to-one due to any two FRSTs with the same $2$-face restrictions mapping to homotopy-equivalent CYs. This motivated our prior development of the `NTFE algorithm' in \cite{macfadden2026efficientalgorithmgeneratinghomotopy} which constructs FRSTs via their $2$-face triangulations, thus sidestepping the redundancy. This construction and the NTFE algorithm are both implemented in CYTools\cite{Demirtas:2022hqf}.

Since uniformity in the CY samples is a key goal, we review the uniformity of rejection sampling \`a la \cite{macfadden2026efficientalgorithmgeneratinghomotopy} here. For a given polytope $\Delta$ with $2$-faces $f_1,f_2,\dots,f_{N_{2D}}$, the NTFE algorithm~\cite{macfadden2026efficientalgorithmgeneratinghomotopy} accepts triangulations $\mathcal{T}_1,\mathcal{T}_2,\dots,\mathcal{T}_{N_{2D}}$ and outputs an FRST with said $2$-face restrictions if one exists. If an FRST exists, one says $\mathcal{T}_1,\mathcal{T}_2,\dots,\mathcal{T}_{N_{2D}}$ `extends'~\cite{macfadden2026efficientalgorithmgeneratinghomotopy}. This FRST can be interpreted as a representative of a $2$-face equivalence class (FRSTs mod identical $2$-face restrictions) which \cite{macfadden2026efficientalgorithmgeneratinghomotopy} originally called an `NTFE'; it will be convenient to also follow \cite{macfadden2025dnacalabiyauhypersurfaces} and call $\mathcal{T}_1,\mathcal{T}_2,\dots,\mathcal{T}_{N_{2D}}$ the `DNA' of this NTFE. We want to show that
\begin{equation}
    P(\mathcal{T}_i) = \frac{1}{\mathcal{N}_i} \implies P(\mathrm{NTFE}_j | \text{DNA extends}) = \frac{1}{\#\mathrm{NTFEs}},
\end{equation}
using notation where $\mathcal{N}_i$ is the number of FRTs for the $2$-face $f_i$. Observe that $P(\mathrm{NTFE}_j \cap \mathrm{extends}) = P(\mathrm{DNA}_j)$ for $\mathrm{DNA}_j$ the DNA associated to $\mathrm{NTFE}_j$ since the algorithm is a bijection between extendable DNA and NTFEs. Thus $P(\mathrm{NTFE}_j \cap \mathrm{extends}) =\prod_{1\leq i\leq N_{2D}} P(\mathcal{T}_i)$ for $\mathcal{T}_i$ the triangulation of $f_i$ corresponding to $\mathrm{NTFE}_j$. This, under our assumption on $P(\mathcal{T}_i)$, immediately gives us our result
\begin{equation}
     P(\mathrm{NTFE}_j | \mathrm{extends}) = \frac{1}{P(\mathrm{extends})} \prod_{1\leq i\leq N_{2D}} \frac{1}{\mathcal{N}_i} = \frac{1}{P(\mathrm{extends})\cdot \#\mathrm{DNAs}} = \frac{1}{\#\mathrm{NTFEs}}.
\end{equation}
This is why \cite{macfadden2026efficientalgorithmgeneratinghomotopy} is a reduction: by uniformly sampling $2$-face triangulations, it directly gives you uniform samples over NTFEs/CYs.

\subsection{Comparison to CYTransformer}
\label{subsec:comparison_to_cytransformer}

We begin by briefly comparing to CYTransformer~\cite{yip2025transformingcalabiyauconstructionsgenerating}\footnote{We use slightly different diagnostics: \cite{yip2025transformingcalabiyauconstructionsgenerating} assess sampling quality via height-space representativeness while we hold dualGNN to the stronger criterion of per-triangulation uniformity, which implies representativeness but not conversely.}. Since CYTransformer generates $4$-simplices directly, its output space includes the exponentially redundant\cite{demirtas2020boundingkreuzerskarkelandscape} collection of FRSTs sharing $2$-face restrictions. dualGNN, in contrast, sidesteps this redundancy by generating DNA directly. This leads to dualGNN showing strong performance in \cite{yip2025transformingcalabiyauconstructionsgenerating}'s diagnostic of the average number of NTFEs generated at different $h^{1,1}$. We emphasize that this comparison primarily measures the value of the $2$-face reduction rather than relative model quality: any $4$D sampler must contend with the exponential FRST redundancy that dualGNN sidesteps. We demonstrate dualGNN's performance on the $200$ polytopes that \cite{yip2025transformingcalabiyauconstructionsgenerating} studied ($5\leq h^{1,1}\leq 10$) and on $200$ more for each of $h^{1,1}=12,14,16$ (first $200$ at each $h^{1,1}$ in Kreuzer-Skarke order). These Hodge numbers are relatively small, so we can enumerate all NTFEs using \cite{macfadden2026efficientalgorithmgeneratinghomotopy} to give a true $1/\#\mathrm{NTFE}$ uniform sampler comparison.

\begin{figure}
    \makebox[\textwidth][c]{%
        \includegraphics[width=1.1\textwidth]{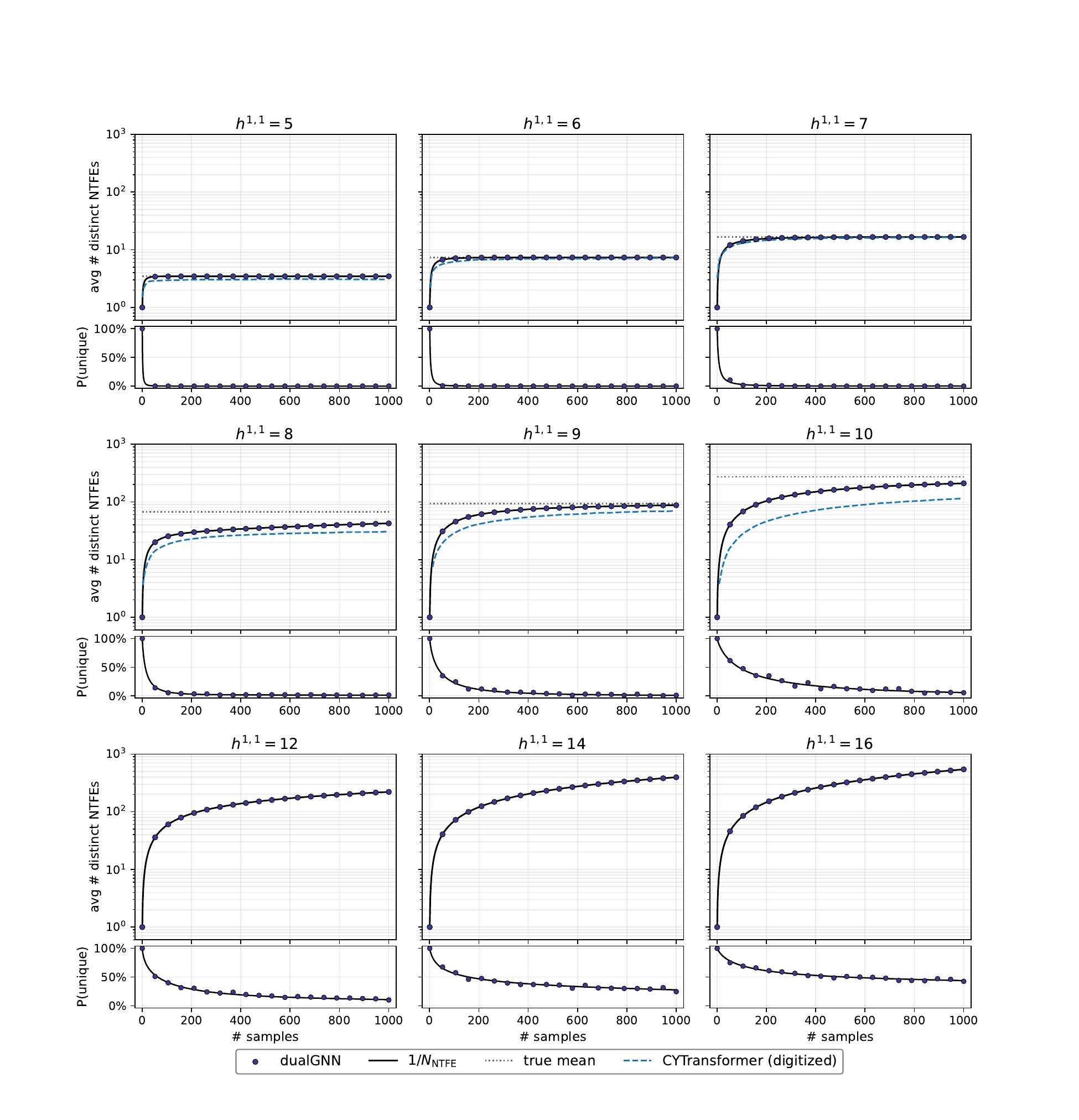}%
    }
    \caption{Recreation of \cite{yip2025transformingcalabiyauconstructionsgenerating}'s figure 4, extended to larger Hodge numbers $h^{1,1} =12, 14, 16$. At all Hodge numbers, dualGNN generates NTFEs at a rate (purple dots) consistent with a true $1/N_\mathrm{NTFE}$ sampler (black line). For Hodge numbers $h^{1,1}\leq 10$, we compare directly to the polytopes studied in \cite{yip2025transformingcalabiyauconstructionsgenerating} (obtained via private communication), consistently obtaining more NTFEs at every sample count. This advantage is primarily attributed to the $2$-face encoding rather than the relative quality of the underlying models. The CYTransformer curves come from their digitized figure $4$.}
    \label{fig:cytransformer_fig4}
\end{figure}

In all cases for which \cite{yip2025transformingcalabiyauconstructionsgenerating} presented data ($h^{1,1}\leq 10$), dualGNN generates more NTFEs with fewer samples (NTFE curves from \cite{yip2025transformingcalabiyauconstructionsgenerating} come from their digitized figure). See \cref{fig:cytransformer_fig4}. In fact, across all Hodge numbers, dualGNN is consistent with a uniform sampler (within noise) while CYTransformer undersamples NTFEs relative to the uniform reference (particularly at $h^{1,1}=8,9,$ and $10$). This makes sense: because CYTransformer samples in the redundant FRST space rather than via a 2-face encoding, the fraction of triangulations that are NTFE-distinct falls as $h^{1,1}$ grows, since FRSTs are exponentially-with-$h^{1,1}$ redundant compared to NTFEs. This is precisely the regime ($h^{1,1}\gg10$) where such samplers are most needed, and it is exactly what the 2-face reduction of \cite{macfadden2026efficientalgorithmgeneratinghomotopy} is designed to address.

\subsection{Sampling at High-$h^{1,1}$}

\begin{figure}[t]
\centering
\tikzset{
  latticepoint/.style={draw, circle, inner sep=0.9pt, fill},
  trilinestyle/.style={line width=0.25mm},
}
\setlength{\tabcolsep}{8pt}
\renewcommand{\arraystretch}{1.0}
\begin{tabular}{cc}
\begin{tabular}{c}\begin{tikzpicture}[baseline={(current bounding box.center)}]
\tikzmath{\scale = 0.6;}
\draw[trilinestyle] (0*\scale,0*\scale) -- (4*\scale,0*\scale) -- (0*\scale,4*\scale) -- cycle;
  \node[latticepoint] at (0*\scale,0*\scale) {};
  \node[latticepoint] at (0*\scale,4*\scale) {};
  \node[latticepoint] at (4*\scale,0*\scale) {};
  \node[latticepoint] at (0*\scale,1*\scale) {};
  \node[latticepoint] at (0*\scale,2*\scale) {};
  \node[latticepoint] at (0*\scale,3*\scale) {};
  \node[latticepoint] at (1*\scale,0*\scale) {};
  \node[latticepoint] at (1*\scale,3*\scale) {};
  \node[latticepoint] at (2*\scale,0*\scale) {};
  \node[latticepoint] at (2*\scale,2*\scale) {};
  \node[latticepoint] at (3*\scale,0*\scale) {};
  \node[latticepoint] at (3*\scale,1*\scale) {};
  \node[latticepoint] at (1*\scale,1*\scale) {};
  \node[latticepoint] at (1*\scale,2*\scale) {};
  \node[latticepoint] at (2*\scale,1*\scale) {};
\end{tikzpicture}\\\footnotesize $\times 8$\end{tabular} &
\begin{tabular}{c}\begin{tikzpicture}[baseline={(current bounding box.center)}]
\tikzmath{\scale = 0.6;}
\draw[trilinestyle] (0*\scale,0*\scale) -- (4*\scale,0*\scale) -- (4*\scale,2*\scale) -- (0*\scale,2*\scale) -- cycle;
  \node[latticepoint] at (0*\scale,0*\scale) {};
  \node[latticepoint] at (4*\scale,0*\scale) {};
  \node[latticepoint] at (0*\scale,2*\scale) {};
  \node[latticepoint] at (4*\scale,2*\scale) {};
  \node[latticepoint] at (1*\scale,0*\scale) {};
  \node[latticepoint] at (2*\scale,0*\scale) {};
  \node[latticepoint] at (3*\scale,0*\scale) {};
  \node[latticepoint] at (0*\scale,1*\scale) {};
  \node[latticepoint] at (4*\scale,1*\scale) {};
  \node[latticepoint] at (1*\scale,2*\scale) {};
  \node[latticepoint] at (2*\scale,2*\scale) {};
  \node[latticepoint] at (3*\scale,2*\scale) {};
  \node[latticepoint] at (1*\scale,1*\scale) {};
  \node[latticepoint] at (2*\scale,1*\scale) {};
  \node[latticepoint] at (3*\scale,1*\scale) {};
\end{tikzpicture}\\\footnotesize $\times 6$\end{tabular} \\
\end{tabular}
\caption{The two distinct $2$-faces of the $h^{1,1}=86$ polytope, labeled with their multiplicity in the polytope. The triangle (left) has $7,422$ FRTs; the rectangle (right) has $12,170$.}
\label{fig:two-faces-h86}
\end{figure}
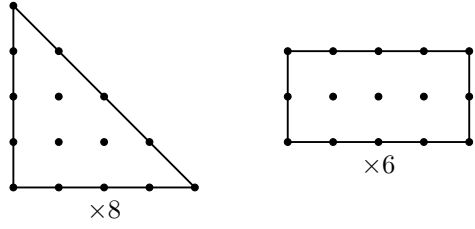

We also push the CY sampling to higher Hodge numbers for which \cite{macfadden2026efficientalgorithmgeneratinghomotopy} cannot exhaustively enumerate all NTFEs. First, as a proof of concept, we study the following $h^{1,1}=86$ polytope 
\begin{equation}
    \Delta_{86} = \mathrm{conv}\begin{bmatrix}
        -1 & -1 & -1 & -1 &  1 &  1 &  1 &  1 \\
        -1 & -1 & -1 &  3 & -1 & -1 & -1 &  3 \\
        -1 & -1 &  3 & -1 & -1 & -1 &  3 & -1 \\
        -1 &  3 & -1 & -1 & -1 &  3 & -1 & -1
    \end{bmatrix}.
\end{equation}
We choose $\Delta_{86}$ because each of its $2$-faces contains exactly $N_\mathrm{pts}=15$, so we can validate dualGNN's uniformity against full enumeration. While this is a non-negligible $h^{1,1}$, significantly higher than what previous works could sample (especially because the number of triangulations depends exponentially on $h^{1,1}$), this is not a very demanding application of dualGNN. One can fully enumerate the FRTs of these $2$-faces in short order: there are two distinct geometries, $\mathrm{conv}(\{(0,0), (0,4), (4,0)\})$ and $\mathrm{conv}(\{(0,0), (0,4), (2,0), (2,4)\})$ with $7,422$ and $12,170$ FRTs respectively (see \cref{fig:two-faces-h86}). As a more demanding test, we also study the following $h^{1,1}=128$ polytope 
\begin{equation}
    \Delta_{128} = \mathrm{conv}\begin{bmatrix}
          1 & -11 & -11 & 1 & 1 & 1 & 13 \\
          0 &  -4 &  -4 & 0 & 0 & 2 &  6 \\
          0 &  -6 &  -6 & 2 & 2 & 0 &  8 \\
          0 & -12 &  -6 & 0 & 6 & 0 & 12
    \end{bmatrix}
\end{equation}
for which each $2$-face has $N_\mathrm{pts}\leq 35$ (see \cref{fig:h11-128-2face-frts}). In principle there is nothing stopping us from studying arbitrarily large $h^{1,1}$; we only stop at $h^{1,1}=128$ since our uniformity diagnostics already begin to struggle at $N_\mathrm{pts}\approx40$, so it would be significantly harder to validate the uniformity of larger polytopes.

\input{tikz/h11_128}

We begin with $\Delta_{86}$, which only has two distinct $2$-face geometries (\cref{fig:two-faces-h86}). We generate $10^6$ FRTs of each $2$-face with the dualGNN model from \cref{subsubsec:multi}. For $2$-faces with $N_\mathrm{pts}\leq 17$, we find a maximum KL divergence (compared to a flat distribution) of $0.016$ across such $2$-faces; a uniform, but still $10^6$-draw sampler achieves a maximum KL divergence of $0.006$. This suggests that dualGNN is indeed nearly uniformly sampling FRTs of these $2$-faces, as should be expected from \cref{subsubsec:multi}.

From the dualGNN samples, we apply \cite{macfadden2026efficientalgorithmgeneratinghomotopy} to generate $10,000$ FRSTs of $\Delta_{86}$, all of which happen to have distinct $2$-face restrictions. By our arguments in \cref{sec:cy} and by the demonstrated uniformity of the $2$-face triangulations, this is a uniform sample of CYs mod $2$-face equivalences (out of the pool of $< 2.99\times10^{55}$). No prior work has uniformly sampled such CYs at Hodge numbers close to $h^{1,1}=86$ before. These CYs are potentially inequivalent, not provably inequivalent; certifying full homotopy-inequivalence is a strictly harder problem\cite{gendler2023countingcalabiyauthreefolds,chandra2023enumeratingcalabiyaumanifoldsplacing}, only done up to $h^{1,1}=5$ with partial results at $h^{1,1}=6$.

\begin{figure}
    \centering
    \includegraphics[width=1.0\textwidth]{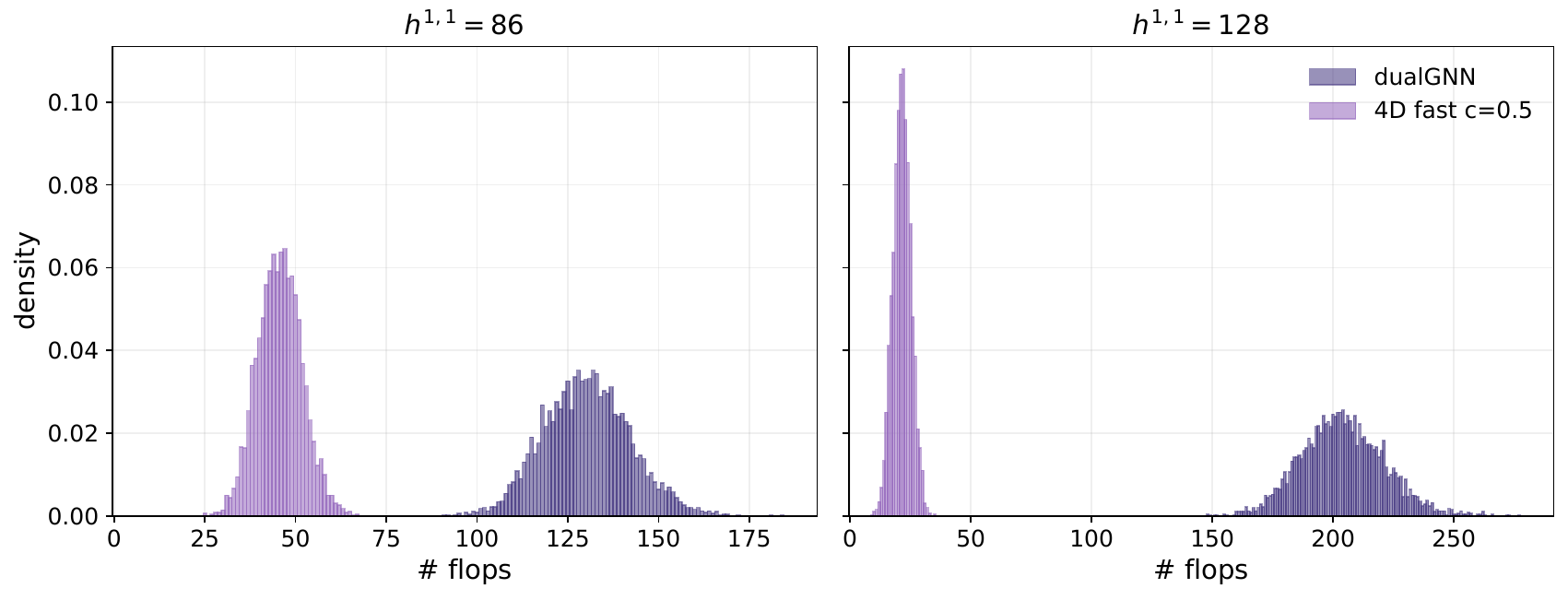}
    \caption{Diversity of sampled CYs, measured by pairwise flop distance (upper bound via a linear path through height-space). In dark purple, we plot the samples from dualGNN, combined via \cite{macfadden2026efficientalgorithmgeneratinghomotopy}. In light purple, we plot the samples from \code{fast} applied to the $4$D polytope. Left: for $\Delta_{86}$, \code{fast} generates CYs that are between $25$ and $67$ flops from one-another (mean $45.7$). In contrast, the dualGNN alternative generates samples that are between $91$ and $184$ flops from one-another (mean $130.2$). Right: $\Delta_{128}$, \code{fast} generates CYs that are between $9$ and $35$ flops from one-another (mean $21.5$) while dualGNN generates samples between $148$ and $277$ flops from one-another (mean $204.0$). Larger flop distances indicate less concentrated samples; this comparison probes diversity, not uniformity.
    }
    \label{fig:flops_hist}
\end{figure}

With uniformity addressed above, we turn to a physically motivated measure of sample diversity: flop distance. A `flop' is a local geometric transition between two Calabi-Yau threefolds; the flop distance between two CYs is the minimum number of such transitions needed to transform one into the other, providing a discrete metric on the space of CYs. Explicitly, we measure an upper bound on the number of flops between these CYs, obtained via taking a linear trajectory in height-space using \code{regfans}~\cite{regfans,macfadden2025calabiyauthreefoldsvextriangulations} and counting the flop transitions. We compare to a sample of $10,000$ FRSTs from \code{fast} in the left of \cref{fig:flops_hist}: from the dualGNN samples, we observe a mean flop count of $130.2$, significantly higher than the mean $45.7$ of the \code{fast} samples. That is, the dualGNN-sampled CYs are much more diverse than the de facto standard of \code{fast}-sampled CYs.

The more interesting example is that of $\Delta_{128}$, for which one cannot fully enumerate the $2$-face FRTs ($2$-faces have up to $N_\mathrm{pts}=35$; see \cref{fig:h11-128-2face-frts}). Here, we perform analogous tests: we generate $10^4$ FRTs of each $2$-face and validate, for each $2$-face with $N_\mathrm{pts}\leq 17$, that the KL divergence is low (max is $0.012$; uniform sampler with the same sample count has max KL divergence of $0.008$). For the larger $2$-faces, we confirmed that the dualGNN samples had no collisions, consistent with uniformity (but not sufficient to show it). With the uniformity of each $2$-face argued, we likewise construct $10,000$ FRSTs of $\Delta_{128}$ by \cite{macfadden2026efficientalgorithmgeneratinghomotopy}. Again comparing to \code{fast}, we see (right side of \cref{fig:flops_hist}) that the dualGNN samples have a mean flop count of $204.0$ while the biased \code{fast} reference has only $21.5$.

\clearpage

\code{fast} is significantly faster than the dualGNN construction ($>7$ CYs/sec vs $\sim0.1$ CYs/sec, but we remind that dualGNN has since become significantly quicker --- see footnote~\ref{fn:timing}) but (a) typically the CY generation is not rate limiting due to the high cost of most downstream applications and (b) good data (uniform samples) is typically much more valuable than biased data. We expose this sampler as \code{random\_triangulations\_gnn} in CYTools, so one can easily choose between these samplers. To our knowledge, this is the first demonstration of uniform CY sampling at Hodge numbers at $h^{1,1}=86$, significantly above the previous $h^{1,1}=10$. We hesitate to also claim this for $h^{1,1}=128$, despite showing near uniform diagnostics, because our sampling resolution is worse here.

\begin{figure}
    \centering
    \tikzset{
        latticepoint/.style={draw, circle, inner sep=1.0pt, fill}
    }
    \begin{tikzpicture}[xscale=0.13, yscale=0.5]
        \foreach \y in {0,...,7}{
            \pgfmathtruncatemacro{\xmax}{84 - 12*\y}
            \foreach \x in {0,...,\xmax}{
                \node[latticepoint] at (\x,\y) {};
            }
        }
        \draw[thick] (0,0) -- (84,0) -- (0,7) -- cycle;
    \end{tikzpicture}
    \caption{The polygon $\mathrm{conv}(\{(0,0),(84,0),(0,7)\})$, the largest polygon occurring in our string theory applications (as a $2$-face of the only $h^{1,1}=491$ $4$D reflexive polytope). This polygon has between $3.90\times10^{167}$ and $1.96\times10^{180}$ FRTs.}
    \label{fig:extreme_polygon}
\end{figure}
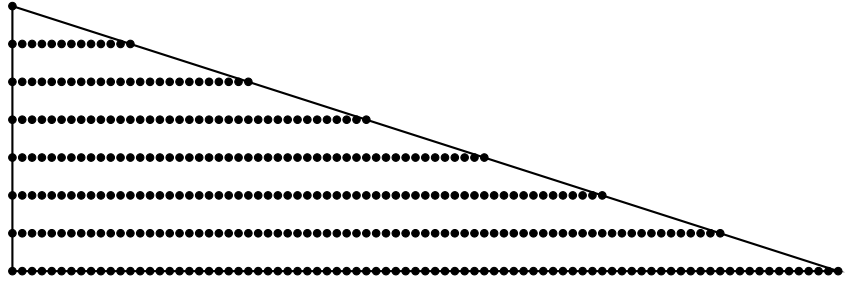

\section{Limitations}
\label{sec:limits}

dualGNN's architecture scales to polygons of any practical size --- even the largest polygon of interest (the largest $2$-face of $h^{1,1}=491$, \cref{fig:extreme_polygon}) generates a graph of only $69{,}416$ nodes. One can even run the code in the associated repo on this polygon, though it is far out of distribution and the $K=16$ message-passing rounds are likely too few. The primary limit is in \textit{validating} uniformity at scale: this polygon has $N_\mathrm{pts}=344$ while our diagnostics start to lose resolution around $N_\mathrm{pts}\approx 40$.

Additionally, dualGNN's strongest architectural guarantee, that every forward pass produces a valid and fine triangulation, is specific to $2$D. The construction relies on the fact that any uncovered region of a partial $2$D triangulation always admits a fine completion using existing lattice points; this is not true in higher dimensions. dualGNN can still operate in higher dimensions: circuits generalize naturally as the underlying combinatorial object, and the symmetry invariance follows. The specific $\mathbf{C}_{ab}$ encoding would need to be rebuilt (the $4$D vector with `left'/`right' ordering is $2$D-specific), but that only requires engineering. The every-rollout-produces-an-FT guarantee is simply lost --- it relies on a geometric fact specific to $2$D. This matters for applications like vex triangulations~\cite{macfadden2025calabiyauthreefoldsvextriangulations} which currently require operating on $4$D geometries directly (no algorithm like \cite{macfadden2026efficientalgorithmgeneratinghomotopy} exists yet for these geometries).

Relatedly, the circuit encoding is sufficient to expose regularity, as demonstrated by the classifier of \cref{subsec:classifier}, but the autoregressive sampler does not perfectly enforce it. The strategy to best ensure regular triangulations is unclear: improve training or architecture? \Cref{subsec:classifier} had a nearly identical (strictly simpler) architecture and could easily extract regularity. That said, the architecture could likely also be improved by adding global attention \`a la Graphormer~\cite{graphormer}. Evidence for this is that our transformer experiments in \cref{sec:transformer} used a similar training regime to our dualGNN autoregressive sampler but were better at targeting regularity (see \cref{fig:rope_irreg} for a good irregular sampler).

Finally, dualGNN is empirically uniform, with no theoretical guarantees. There is a provably uniform algorithm for sampling FTs (and hence FRTs via rejection sampling): \cite{Kaibel_Ziegler_2003}. This algorithm is not generally applicable due to its exponential scaling on the polygon width. For example, in \cite{macfadden2026boundingkreuzerskarkelandscape}, we applied the counting variant of this algorithm to the polygon in \cref{fig:extreme_polygon}, but that took $\sim57$ core days (serially). While dualGNN currently is not applicable for this polygon ($K=16$ message passing rounds are too few), its time scales linearly with $K$ and one must set $K$ at least as large as the maximum diameter of a dual graph of the polygon. This is significantly milder scaling than exponentially with the width. In practice, using the existing dualGNN code in the associated repo (albeit with the optimistic $K=16$), one can sample a triangulation in $\sim60s$ for graph creation (amortized) and $\sim75s$ for rollout (although the rollout time is expected to be an underestimate) on this $h^{1,1}=491$ polytope.

\section{Conclusion}
\label{sec:conclusion}

We introduced dualGNN, an autoregressive GNN which encodes all fine triangulations of a lattice polytope $\Delta$ via a generalization of the dual graph of a triangulation. By encoding certain `dependency vectors' $\lambda$ corresponding to `circuits', one both better represents the combinatorics of the triangulation (the oriented matroid) and the regularity of the associated triangulations. We showed that such a GNN can extract the regularity signal in its edges and can operate as an autoregressive sampler of fine, regular triangulations consistent with uniformity. This autoregressive sampler generalizes zero-shot to other polytopes due to its symmetry invariance and $N_\mathrm{pts}$-independent structure, making a single $\sim92$k parameter model generally applicable to unseen polygons. The small size of the model enables very quick training on consumer hardware, $O(\text{hours})$. Overall, dualGNN was the only sampler consistent with uniformity across all diagnostics, with an architecture that scales to any polygon size of practical interest.

We applied this model to the task of generating CYs at Hodge numbers $h^{1,1}=86$ and $h^{1,1}=128$. For $h^{1,1}=86$ our diagnostics are strong enough to validate uniformity (but this case does not require dualGNN since one can fully enumerate its FRTs for the $2$-faces); for $h^{1,1}=128$ dualGNN passes our uniformity diagnostics (within their resolution limits), but we stress that our diagnostics are weaker here. This reach is enabled both by the $2$-face reduction of \cite{macfadden2026efficientalgorithmgeneratinghomotopy} and by dualGNN's polygon-general uniform sampling.

More broadly, this model demonstrates a GNN-compatible encoding for realizable oriented matroids (those representable as point or vector configurations~\cite{Ziegler1999-bp,DeLoera2010}). Oriented matroids are common as the underlying structure of various disparate mathematical domains\cite{Ziegler1999-bp}, including directed graphs, linear programming, hyperplane arrangements, convex polytopes, and polyhedral fans. In this way, the studied architecture may be applicable more broadly; however most immediate applications are in string theory. As a concrete example, dualGNN naturally extends to vex triangulations~\cite{macfadden2025calabiyauthreefoldsvextriangulations}, which generate a strictly broader class of Calabi-Yau threefolds than the Batyrev construction. The vex setting is, combinatorially, simpler: its symmetry group is $\mathrm{GL}(d, \mathbb{Z})$ rather than $\mathrm{GL}(d, \mathbb{Z}) \ltimes \mathbb{Z}^d$, and its circuits satisfy a purely linear condition $\mathbf{A}\lambda = 0$ rather than the affine $[\mathbf{A}; \mathbf{1}]\lambda = 0$. dualGNN's circuit encoding handles this case directly, with the $\sum_i \lambda_i = 0$ constraint relaxed. The argument for such an application is identical to what we discuss here: by using a matroid to respect a problem's symmetries, one can obtain significantly stronger generalization and uniformity than direct sequence modeling.

To ease future comparison, the $20$ held-out benchmark polygons, their exact FRT counts, and the uniformity-scoring protocol (collision counts and KL against the finite-sample noise floor) ship with the code in the repository's \code{eval/} directory, and dualGNN itself is available as an opt-in sampler in CYTools.

\section{Acknowledgments}

We would like to acknowledge the authors of \cite{yip2025transformingcalabiyauconstructionsgenerating} for inspiring this project, and Jacky Yip in particular for providing the polytopes used in \cref{fig:cytransformer_fig4}. We would also like to acknowledge Jim Halverson, Liam McAllister, Andreas Schachner, and Elijah Sheridan for reading and providing feedback on this paper. Finally, I would like to acknowledge my wife Guin Gunter for her love and support.

This work was funded in part by NSF grant PHY-2309456.

Software development was assisted by Claude Opus 4.7 (Anthropic).

\bibliographystyle{utphys}
\bibliography{refs}

@article{demirtas2020boundingkreuzerskarkelandscape,
  title = {Bounding the Kreuzer‐Skarke Landscape},
  volume = {68},
  ISSN = {1521-3978},
  url = {http://dx.doi.org/10.1002/prop.202000086},
  DOI = {10.1002/prop.202000086},
  number = {11-12},
  journal = {Fortschritte der Physik},
  publisher = {Wiley},
  author = {Demirtas,  Mehmet and McAllister,  Liam and Rios‐Tascon,  Andres},
  year = {2020},
  month = Oct 
}

@misc{macfadden2026boundingkreuzerskarkelandscape,
      title={Further Bounding the Kreuzer-Skarke Landscape}, 
      author={Nate MacFadden and Stepan Yu. Orevkov and Michael Stepniczka},
      year={2026},
      eprint={2602.16909},
      archivePrefix={arXiv},
      primaryClass={hep-th},
      url={https://arxiv.org/abs/2602.16909}, 
}

@misc{macfadden2026efficientalgorithmgeneratinghomotopy,
      title={Efficient Algorithm for Generating Homotopy Inequivalent Calabi-Yaus}, 
      author={Nate MacFadden},
      year={2023},
      eprint={2309.10855},
      archivePrefix={arXiv},
      primaryClass={hep-th},
      url={https://arxiv.org/abs/2309.10855}, 
}

@misc{macfadden2025calabiyauthreefoldsvextriangulations,
      title={Calabi-Yau Threefolds from Vex Triangulations}, 
      author={Nate MacFadden and Elijah Sheridan},
      year={2025},
      eprint={2512.14817},
      archivePrefix={arXiv},
      primaryClass={hep-th},
      url={https://arxiv.org/abs/2512.14817}, 
}

@misc{yip2025transformingcalabiyauconstructionsgenerating,
      title={Transforming Calabi-Yau Constructions: Generating New Calabi-Yau Manifolds with Transformers}, 
      author={Jacky H. T. Yip and Charles Arnal and François Charton and Gary Shiu},
      year={2025},
      eprint={2507.03732},
      archivePrefix={arXiv},
      primaryClass={hep-th},
      url={https://arxiv.org/abs/2507.03732}, 
}

@book{DeLoera2010,
  title = {Triangulations},
  ISBN = {9783642129711},
  ISSN = {1431-1550},
  url = {http://dx.doi.org/10.1007/978-3-642-12971-1},
  DOI = {10.1007/978-3-642-12971-1},
  journal = {Algorithms and Computation in Mathematics},
  publisher = {Springer Berlin Heidelberg},
  author = {De Loera,  Jesús A. and Rambau,  J\"{o}rg and Santos,  Francisco},
  year = {2010}
}

@article{Batyrev:1993oya,
    author = "Batyrev, Victor V.",
    title = "{Dual Polyhedra and Mirror Symmetry for Calabi-Yau Hypersurfaces in Toric Varieties}",
    eprint = "alg-geom/9310003",
    archivePrefix = "arXiv",
    journal = "J. Alg. Geom.",
    volume = "3",
    pages = "493--545",
    year = "1994"
}

@article{Demirtas:2022hqf,
    author = "Demirtas, Mehmet and Rios-Tascon, Andres and McAllister, Liam",
    title = "{CYTools: A Software Package for Analyzing Calabi-Yau Manifolds}",
    eprint = "2211.03823",
    archivePrefix = "arXiv",
    primaryClass = "hep-th",
    month = "11",
    year = "2022"
}

@techreport{Rambau2002,
    author      = {J{\"o}rg Rambau},
    title       = "{TOPCOM: Triangulations of Point Configurations and Oriented Matroids}",
    institution = {ZIB},
    address     = {Takustr. 7, 14195 Berlin},
    number      = {02-17},
    language    = {eng},
    year        = {2002}
}

@article{berglund2024generatingtriangulationsfibrationsreinforcement,
    title = {Generating triangulations and fibrations with reinforcement learning},
    journal = {Physics Letters B},
    volume = {860},
    pages = {139158},
    year = {2025},
    issn = {0370-2693},
    doi = {https://doi.org/10.1016/j.physletb.2024.139158},
    url = {https://www.sciencedirect.com/science/article/pii/S0370269324007160},
    author = {Per Berglund and Giorgi Butbaia and Yang-Hui He and Elli Heyes and Edward Hirst and Vishnu Jejjala},
}

@article{macfadden2025dnacalabiyauhypersurfaces,
  title = {The DNA of Calabi–Yau Hypersurfaces: A Genetic Algorithm for Polytope Triangulations},
  volume = {74},
  ISSN = {1521-3978},
  url = {http://dx.doi.org/10.1002/prop.70060},
  DOI = {10.1002/prop.70060},
  number = {2},
  journal = {Fortschritte der Physik},
  publisher = {Wiley},
  author = {MacFadden,  Nate and Schachner,  Andreas and Sheridan,  Elijah},
  year = {2025},
  month = Nov 
}

@InProceedings{bodnar2021weisfeilerlehmantopologicalmessage,
  title = 	 {Weisfeiler and Lehman Go Topological: Message Passing Simplicial Networks},
  author =       {Bodnar, Cristian and Frasca, Fabrizio and Wang, Yuguang and Otter, Nina and Montufar, Guido F and Li{\'o}, Pietro and Bronstein, Michael},
  booktitle = 	 {Proceedings of the 38th International Conference on Machine Learning},
  pages = 	 {1026--1037},
  year = 	 {2021},
  editor = 	 {Meila, Marina and Zhang, Tong},
  volume = 	 {139},
  series = 	 {Proceedings of Machine Learning Research},
  month = 	 {18--24 Jul},
  publisher =    {PMLR},
  url = 	 {https://proceedings.mlr.press/v139/bodnar21a.html},
}

@article{su2023roformerenhancedtransformerrotary,
    author = {Su, Jianlin and Ahmed, Murtadha and Lu, Yu and Pan, Shengfeng and Bo, Wen and Liu, Yunfeng},
    title = {RoFormer: Enhanced transformer with Rotary Position Embedding},
    year = {2024},
    issue_date = {Feb 2024},
    publisher = {Elsevier Science Publishers B. V.},
    address = {NLD},
    volume = {568},
    number = {C},
    issn = {0925-2312},
    url = {https://doi.org/10.1016/j.neucom.2023.127063},
    doi = {10.1016/j.neucom.2023.127063},
    journal = {Neurocomput.},
    month = feb,
    numpages = {12}
}

@inproceedings{vaswani2023attentionneed,
 author = {Vaswani, Ashish and Shazeer, Noam and Parmar, Niki and Uszkoreit, Jakob and Jones, Llion and Gomez, Aidan N and Kaiser, \L ukasz and Polosukhin, Illia},
 booktitle = {Advances in Neural Information Processing Systems},
 editor = {I. Guyon and U. Von Luxburg and S. Bengio and H. Wallach and R. Fergus and S. Vishwanathan and R. Garnett},
 pages = {},
 publisher = {Curran Associates, Inc.},
 title = {Attention is All you Need},
 url = {https://proceedings.neurips.cc/paper_files/paper/2017/file/3f5ee243547dee91fbd053c1c4a845aa-Paper.pdf},
 volume = {30},
 year = {2017}
}

@inbook{Kaibel_Ziegler_2003, 
    place={Cambridge}, 
    series={London Mathematical Society Lecture Note Series}, 
    title="{Counting Lattice Triangulations}", 
    booktitle="{Surveys in Combinatorics 2003}", 
    publisher={Cambridge University Press}, 
    author="{Kaibel, V. and Ziegler, G.M.}", 
    year={2003}, pages={277–308}, 
    collection={London Mathematical Society Lecture Note Series},
    eprint={math/0211268},
    archivePrefix={arXiv},
    primaryClass={math.CO}
}

@book{Ziegler1999-bp,
    place={Cambridge},
    edition={2},
    series={Encyclopedia of Mathematics and its Applications},
    title={Oriented Matroids},
    publisher={Cambridge University Press},
    author={Björner, Anders and Las Vergnas, Michel and Sturmfels, Bernd and White, Neil and Ziegler, Gunter M.},
    year={1999},
    collection={Encyclopedia of Mathematics and its Applications}
}

@inproceedings{gilmer2017neuralmessagepassingquantum,
    author = {Gilmer, Justin and Schoenholz, Samuel S. and Riley, Patrick F. and Vinyals, Oriol and Dahl, George E.},
    title = {Neural message passing for Quantum chemistry},
    year = {2017},
    publisher = {JMLR.org},
    booktitle = {Proceedings of the 34th International Conference on Machine Learning - Volume 70},
    pages = {1263–1272},
    numpages = {10},
    location = {Sydney, NSW, Australia},
    series = {ICML'17}
}

@misc{bronstein2021geometricdeeplearninggrids,
      title={Geometric Deep Learning: Grids, Groups, Graphs, Geodesics, and Gauges}, 
      author={Michael M. Bronstein and Joan Bruna and Taco Cohen and Petar Veli\v{c}kovi\'{c}},
      year={2021},
      eprint={2104.13478},
      archivePrefix={arXiv},
      primaryClass={cs.LG},
      url={https://arxiv.org/abs/2104.13478}, 
}

@inproceedings{vinyals2017pointernetworks,
     author = {Vinyals, Oriol and Fortunato, Meire and Jaitly, Navdeep},
     booktitle = {Advances in Neural Information Processing Systems},
     editor = {C. Cortes and N. Lawrence and D. Lee and M. Sugiyama and R. Garnett},
     pages = {},
     publisher = {Curran Associates, Inc.},
     title = {Pointer Networks},
     url = {https://proceedings.neurips.cc/paper_files/paper/2015/file/29921001f2f04bd3baee84a12e98098f-Paper.pdf},
     volume = {28},
     year = {2015}
}

@BOOK{Knuth1997-eo,
  title     = "Art of computer programming, volume 2",
  author    = "Knuth, Donald E",
  publisher = "Addison Wesley",
  edition   =  3,
  month     =  nov,
  year      =  1997,
  address   = "Boston, MA",
  language  = "en"
}

@misc{mcallister2024candidatesittervacua,
      title={Candidate de Sitter Vacua}, 
      author={Liam McAllister and Jakob Moritz and Richard Nally and Andreas Schachner},
      year={2024},
      eprint={2406.13751},
      archivePrefix={arXiv},
      primaryClass={hep-th},
      url={https://arxiv.org/abs/2406.13751}, 
}

@article{Schnhardt1928,
  title = {\"{U}ber die Zerlegung von Dreieckspolyedern in Tetraeder},
  volume = {98},
  ISSN = {1432-1807},
  url = {http://dx.doi.org/10.1007/BF01451597},
  DOI = {10.1007/bf01451597},
  number = {1},
  journal = {Mathematische Annalen},
  publisher = {Springer Science and Business Media LLC},
  author = {Sch\"{o}nhardt,  E.},
  year = {1928},
  month = Mar,
  pages = {309–312}
}

@article{Williams1992,
  title = {Simple statistical gradient-following algorithms for connectionist reinforcement learning},
  volume = {8},
  ISSN = {1573-0565},
  url = {http://dx.doi.org/10.1007/BF00992696},
  DOI = {10.1007/bf00992696},
  number = {3-4},
  journal = {Machine Learning},
  publisher = {Springer Science and Business Media LLC},
  author = {Williams,  Ronald J.},
  year = {1992},
  month = May,
  pages = {229–256}
}

@inproceedings{tran2019discreteflowsinvertiblegenerative,
     author = {Tran, Dustin and Vafa, Keyon and Agrawal, Kumar and Dinh, Laurent and Poole, Ben},
     booktitle = {Advances in Neural Information Processing Systems},
     editor = {H. Wallach and H. Larochelle and A. Beygelzimer and F. d\textquotesingle Alch\'{e}-Buc and E. Fox and R. Garnett},
     pages = {},
     publisher = {Curran Associates, Inc.},
     title = {Discrete Flows: Invertible Generative Models of Discrete Data},
     url = {https://proceedings.neurips.cc/paper_files/paper/2019/file/e046ede63264b10130007afca077877f-Paper.pdf},
     volume = {32},
     year = {2019}
}

@inproceedings{bengio2021flownetworkbasedgenerative,
    author = {Bengio, Emmanuel and Jain, Moksh and Korablyov, Maksym and Precup, Doina and Bengio, Yoshua},
    title = {Flow network based generative models for non-iterative diverse candidate generation},
    year = {2021},
    isbn = {9781713845393},
    publisher = {Curran Associates Inc.},
    address = {Red Hook, NY, USA},
    booktitle = {Proceedings of the 35th International Conference on Neural Information Processing Systems},
    articleno = {2097},
    numpages = {14},
    series = {NIPS '21}
}

@inproceedings{hoogeboom2019integerdiscreteflowslossless,
    author = {Hoogeboom, Emiel and Peters, Jorn W.T. and van den Berg, Rianne and Welling, Max},
    title = {Integer discrete flows and lossless compression},
    year = {2019},
    publisher = {Curran Associates Inc.},
    address = {Red Hook, NY, USA},
    booktitle = {Proceedings of the 33rd International Conference on Neural Information Processing Systems},
    articleno = {1088},
    numpages = {11}
}

@article{kreuzer2000completeclassificationreflexivepolyhedra,
  title = {Complete classification of reflexive polyhedra in four dimensions},
  volume = {4},
  ISSN = {1095-0753},
  url = {http://dx.doi.org/10.4310/ATMP.2000.v4.n6.a2},
  DOI = {10.4310/atmp.2000.v4.n6.a2},
  number = {6},
  journal = {Advances in Theoretical and Mathematical Physics},
  publisher = {International Press of Boston},
  author = {Kreuzer,  Maximilian and Skarke,  Harald},
  year = {2000},
  pages = {1209–1230}
}

@misc{gendler2023countingcalabiyauthreefolds,
      title={Counting Calabi-Yau Threefolds}, 
      author={Naomi Gendler and Nate MacFadden and Liam McAllister and Jakob Moritz and Richard Nally and Andreas Schachner and Mike Stillman},
      year={2023},
      eprint={2310.06820},
      archivePrefix={arXiv},
      primaryClass={hep-th},
      url={https://arxiv.org/abs/2310.06820}, 
}

@article{chandra2023enumeratingcalabiyaumanifoldsplacing,
  title = {Enumerating Calabi‐Yau Manifolds: Placing Bounds on the Number of Diffeomorphism Classes in the Kreuzer‐Skarke List},
  volume = {72},
  ISSN = {1521-3978},
  url = {http://dx.doi.org/10.1002/prop.202300264},
  DOI = {10.1002/prop.202300264},
  number = {5},
  journal = {Fortschritte der Physik},
  publisher = {Wiley},
  author = {Chandra,  Aditi and Constantin,  Andrei and Fraser‐Taliente,  Cristofero S. and Harvey,  Thomas R. and Lukas,  Andre},
  year = {2024},
  month = Mar 
}

@misc{regfans,
  doi = {10.5281/ZENODO.19406101},
  url = {https://zenodo.org/doi/10.5281/zenodo.19406101},
  author = {MacFadden,  Nate},
  title = {regfans},
  publisher = {Zenodo},
  year = {2026},
  copyright = {GNU General Public License v3.0 or later}
}

@article{anclin2003,
    title = "{An upper bound for the number of planar lattice triangulations}",
    journal = {Journal of Combinatorial Theory, Series A},
    volume = {103},
    number = {2},
    pages = {383-386},
    year = {2003},
    issn = {0097-3165},
    doi = {https://doi.org/10.1016/S0097-3165(03)00097-9},
    author = {Emile E. Anclin},
}

@inproceedings{loshchilov2019decoupledweightdecayregularization,
    title={Decoupled Weight Decay Regularization},
    author={Ilya Loshchilov and Frank Hutter},
    booktitle={International Conference on Learning Representations},
    year={2019},
    url={https://openreview.net/forum?id=Bkg6RiCqY7},
}

@article{Cveti__2001,
   title={Three-Family Supersymmetric Standardlike Models from Intersecting Brane Worlds},
   volume={87},
   ISSN={1079-7114},
   url={http://dx.doi.org/10.1103/PhysRevLett.87.201801},
   DOI={10.1103/physrevlett.87.201801},
   number={20},
   journal={Physical Review Letters},
   publisher={American Physical Society (APS)},
   author={Cvetič, Mirjam and Shiu, Gary and Uranga, Angel M.},
   year={2001},
   month=Oct }

@article{Aldazabal_2000,
   title={D-Branes at singularities: a bottom-up approach to the string embedding of the standard model},
   volume={2000},
   ISSN={1029-8479},
   url={http://dx.doi.org/10.1088/1126-6708/2000/08/002},
   DOI={10.1088/1126-6708/2000/08/002},
   number={08},
   journal={Journal of High Energy Physics},
   publisher={Springer Science and Business Media LLC},
   author={Aldazabal, Gerardo and Ibáñez, Luis E and Quevedo, Fernando and Uranga, Angel M},
   year={2000},
   month=Aug, pages={002–002} }

@article{Cveti__2019,
   title={Quadrillion {$F$}-Theory Compactifications with the Exact Chiral Spectrum of the Standard Model},
   volume={123},
   ISSN={1079-7114},
   url={http://dx.doi.org/10.1103/PhysRevLett.123.101601},
   DOI={10.1103/physrevlett.123.101601},
   number={10},
   journal={Physical Review Letters},
   publisher={American Physical Society (APS)},
   author={Cvetič, Mirjam and Halverson, James and Lin, Ling and Liu, Muyang and Tian, Jiahua},
   year={2019},
   month=Sept }

@article{Blumenhagen_2009,
   title={{GUTs} in type {IIB} orientifold compactifications},
   volume={815},
   ISSN={0550-3213},
   url={http://dx.doi.org/10.1016/j.nuclphysb.2009.02.011},
   DOI={10.1016/j.nuclphysb.2009.02.011},
   number={1-2},
   journal={Nuclear Physics B},
   publisher={Elsevier BV},
   author={Blumenhagen, Ralph and Braun, Volker and Grimm, Thomas W. and Weigand, Timo},
   year={2009},
   month=July, pages={1–94} }

@inproceedings{graphormer,
author = {Ying, Chengxuan and Cai, Tianle and Luo, Shengjie and Zheng, Shuxin and Ke, Guolin and He, Di and Shen, Yanming and Liu, Tie-Yan},
title = {Do transformers really perform bad for graph representation?},
year = {2021},
isbn = {9781713845393},
publisher = {Curran Associates Inc.},
address = {Red Hook, NY, USA},
booktitle = {Proceedings of the 35th International Conference on Neural Information Processing Systems},
articleno = {2212},
numpages = {12},
series = {NIPS '21}
}

@misc{wang2026trisearchlearningoptimizetriangulations,
      title={TriSearch: Learning to Optimize Triangulations via Bistellar Flips}, 
      author={Yiran Wang and Guido Montúfar},
      year={2026},
      eprint={2605.30220},
      archivePrefix={arXiv},
      primaryClass={cs.LG},
      url={https://arxiv.org/abs/2605.30220}, 
}

@mastersthesis{meyer2002enumeration,
  author = {Meyer, S.},
  title  = {Enumeration von Triangulierungen},
  school = {Technische Universit\"at Berlin},
  type   = {Diplomarbeit},
  year   = {2002},
  note   = {87 pages}
}

@inproceedings{314161.314262,
author = {Felsner, Stefan and Wernisch, Lorenz},
title = {Markov chains for linear extensions, the two-dimensional case},
year = {1997},
isbn = {0898713900},
publisher = {Society for Industrial and Applied Mathematics},
address = {USA},
booktitle = {Proceedings of the Eighth Annual ACM-SIAM Symposium on Discrete Algorithms},
pages = {239–247},
numpages = {9},
location = {New Orleans, Louisiana, USA},
series = {SODA '97}
}

@article{orevkov2022counting,
    author  = {S. Yu. Orevkov},
    title   = "{Counting lattice triangulations: Fredholm equations in combinatorics}",
    journal = {Sb. Math.},
    year    = {2022},
    volume  = {213},
    number  = {11},
    pages   = {1530--1558},
    doi     = {10.4213/sm9727e},
    eprint={2201.12827v2},
    archivePrefix={arXiv},
    primaryClass={math.CO}
}

\newpage
\appendix
\crefalias{section}{appendix}

\section{dualGNN Inference Visualization}
\label{sec:inference}

\begin{figure}[H]
    \centering
    \includegraphics[width=0.9\textwidth]{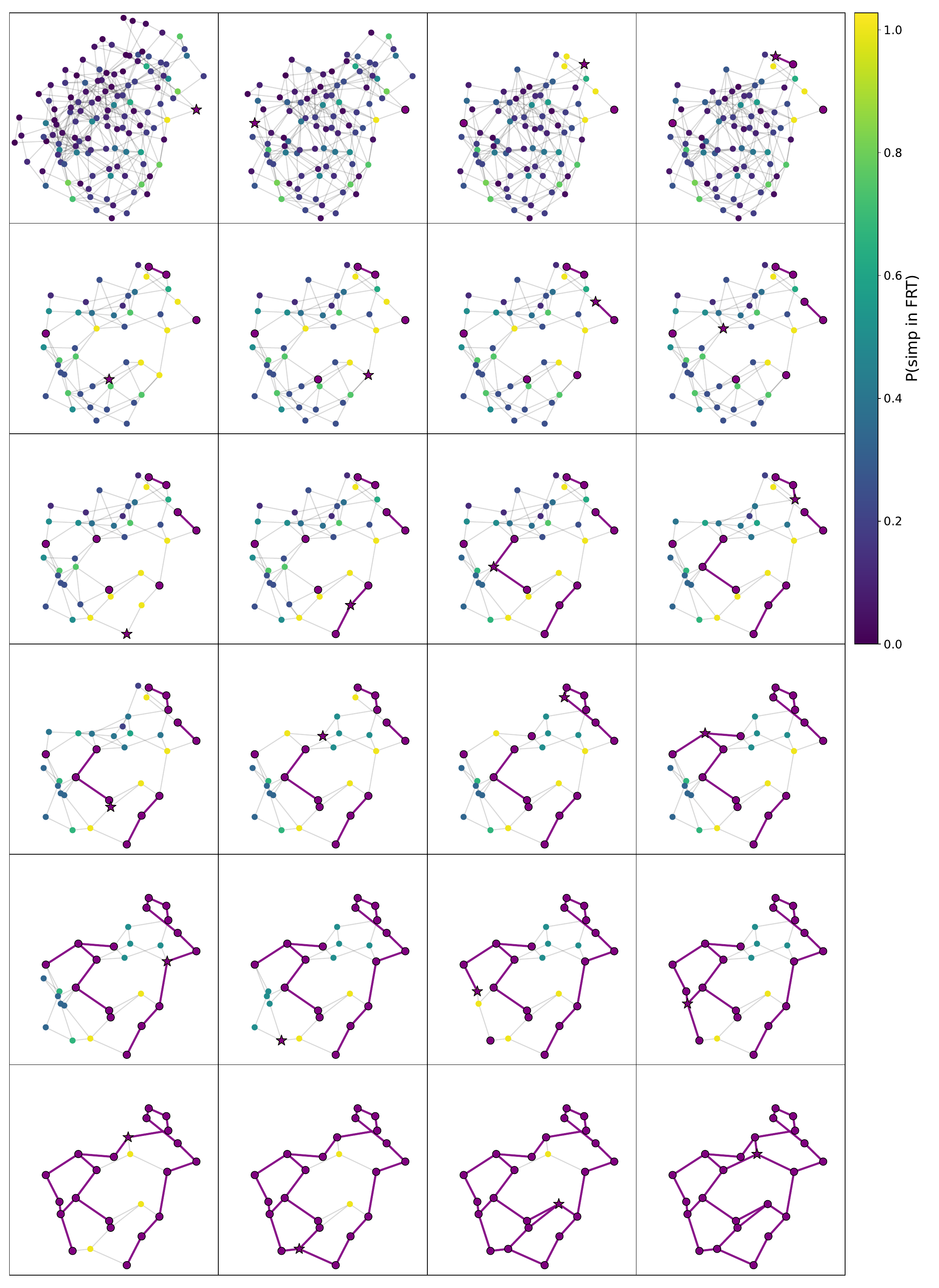}
    \caption{The inference rollout of dualGNN applied to the polygon in \cref{fig:4x6tri}. Nodes are colored with the probabilities dualGNN assigns them, with a purple star indicating the chosen simplex. Selected simplices are drawn in purple with edges between them also drawn in purple.}
    \label{fig:inference_prob}
\end{figure}

\begin{figure}[H]
    \centering
    \includegraphics[width=0.9\textwidth]{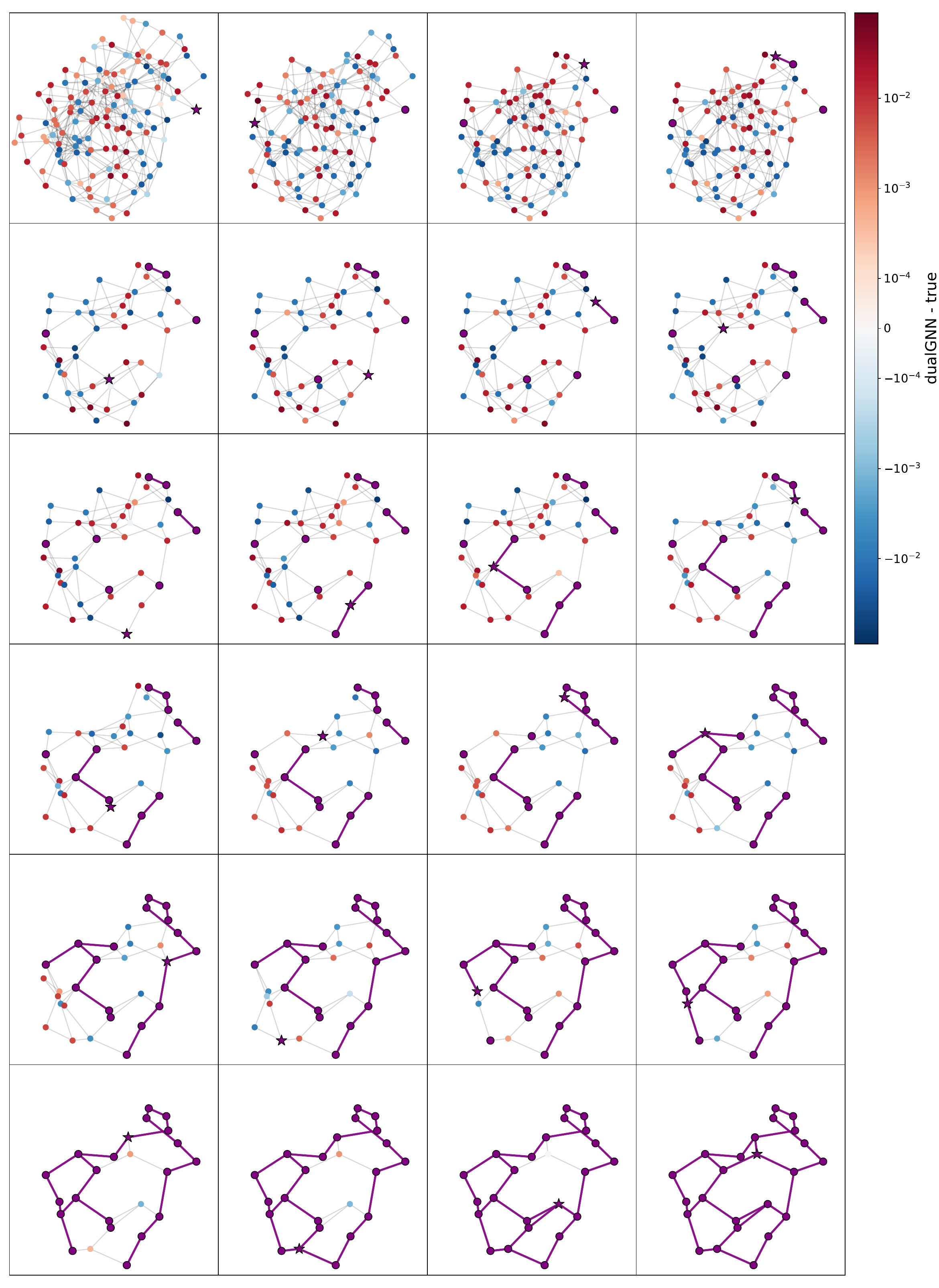}
    \caption{The inference rollout of dualGNN applied to the polygon in \cref{fig:4x6tri}. Nodes are colored with the difference in probability that dualGNN assigns them compared to the true probability computed from an exhaustive enumeration of FRTs. Selected simplices are drawn in purple with edges between them also drawn in purple.}
    \label{fig:inference_diff}
\end{figure}

\section{Classical Algorithms to Sample Triangulations}
\label{sec:classical}

There are a large number of classical algorithms for randomly sampling fine triangulations. We provide a partial enumeration of algorithms in \cref{table:classical}, but do not claim this is complete.

\begin{table}[!htp]
    \centering
    \begin{threeparttable}
        \begin{tabularx}{\textwidth}{l X c X}
            \toprule
            Strategy & Procedure & Only regular? & Notes \\
            \midrule
            \code{uniform} & Enumerate all triang. and then sample uniformly & No & Efficient for many samples or few triangulations \\
            \code{flip\_walk} & MCMC random walk on flip graph from seed triang. & No & See \cite{demirtas2020boundingkreuzerskarkelandscape}. Also see \cite{Kaibel_Ziegler_2003} despite minor differences. Reported long mixing times \\
            \code{grow2d} & Greedily add simplices to partial triangulation & No & See CYTools. Fast, semi-fair, optimized for $2$D \\
            \code{pushing} & Greedily `place' points in the partial triangulation & Yes & See def 4.3.3 of \cite{DeLoera2010}. Misses some regular triangulations \\
            \code{fast}\tnote{a} & Lift by heights sampled near a seed height vector & Yes & See \cite{demirtas2020boundingkreuzerskarkelandscape,Demirtas:2022hqf}. Fast, local \\
            \code{fair}\tnote{b} & Hybrid MCMC-like approach & Yes & See \cite{demirtas2020boundingkreuzerskarkelandscape,Demirtas:2022hqf}. Not used in this work. \\
            \bottomrule
        \end{tabularx}
        \begin{tablenotes}
            \footnotesize
            \item[a] Standing for \code{random\_triangulations\_fast} as in the CYTools implementation.
            \item[b] Standing for \code{random\_triangulations\_fair} as in the CYTools implementation.
        \end{tablenotes}
    \end{threeparttable}
    \caption{Classical methods for sampling fine triangulations of lattice polygons.}
    \label{table:classical}
\end{table}

Here we provide some of these algorithms, particularly those that we use as comparisons or those that have not appeared previously in the literature. First, we list the two from \cite{demirtas2020boundingkreuzerskarkelandscape} that we use in this work.

\begin{algorithm}[H]
\DontPrintSemicolon
\SetKwInOut{Input}{Input}
\SetKwInOut{Output}{Output}
\Input{Seed height vector $h_0 \in \mathbb{R}^{N_{\mathrm{pts}}}$}
\Input{Standard deviation $c > 0$}
\Output{Triangulation $\mathcal{T}$ of $\Delta$}
\BlankLine
Sample $\boldsymbol{\epsilon} \sim \mathcal{N}(\mathbf{0}, c^2 \mathbf{I})$\;
$h \gets h_0 + \boldsymbol{\epsilon}$\;
$\mathcal{T} \gets \code{lift}(\Delta, h)$\;
\Return $\mathcal{T}$\;
\caption{\code{random\_triangulations\_fast}}
\label{alg:fast}
\end{algorithm}

\begin{algorithm}[H]
\DontPrintSemicolon
\SetKwInOut{Input}{Input}
\SetKwInOut{Output}{Output}
\SetKwFunction{Flip}{RandomFlip}
\SetKwFunction{Neighbors}{Neighbors}
\Input{Seed triangulation $\mathcal{T}_0$ of $\Delta$}
\Input{Number of steps $k$}
\Output{Triangulation $\mathcal{T}$ of $\Delta$}
\BlankLine
$\mathcal{T} \gets \mathcal{T}_0$\;
\For{$i \gets 1$ \KwTo $k$}{
    $\mathcal{N} \gets \Neighbors(\mathcal{T})$ \tcp*{triangulations reachable by one flip}
    $\mathcal{T} \gets$ uniformly random element of $\mathcal{N}$\;
}
\Return $\mathcal{T}$\;
\caption{\code{MCMC} \code{flip\_walk}}
\label{alg:flipwalk}
\end{algorithm}

We also list the other two \code{grow2d} and \code{pushing} that we use heavily. The former was released with CYTools with \cite{macfadden2026efficientalgorithmgeneratinghomotopy}; the latter is heavily inspired by TOPCOM\cite{Rambau2002}. Both operate by adding simplices incrementally in a way that enables a fine triangulation (i.e., do not add simplices that cover $>d+1$ lattice points for a $d$-dimensional polytope). These latter algorithms always converge in $2$D but not more generally. \code{pushing} has one more guarantee: it always generates regular triangulations. This fact enables \code{pushing}, for the task of generating FRTs, to be, by far, the quickest algorithm. Not every regular triangulation is a pushing triangulation, though, so there are some FRTs that \code{pushing} cannot generate.

\begin{algorithm}[H]
\DontPrintSemicolon
\SetKwInOut{Input}{Input}
\SetKwInOut{Output}{Output}
\SetKwFunction{RandomSimplex}{RandomSeedSimplex}
\SetKwFunction{OpenFacets}{OpenFacets}
\SetKwFunction{RandomOrder}{RandomOrder}
\Input{$d$-dimensional lattice polytope $\Delta$ with lattice points $\mathbf{A}$}
\Output{Fine triangulation $\mathcal{T}$ of $\Delta$}
\BlankLine
$\sigma_0 \gets \RandomSimplex(\Delta)$\;
$\mathcal{T} \gets \{\sigma_0\}$\;
\While{$\mathcal{T}$ does not triangulate $\Delta$}{
    pick a facet $f$ from $\OpenFacets(\mathcal{T})$ \tcp*{$f \not\subset \partial\Delta$, unmatched}
    $[x_{P(0)}, x_{P(1)}, \dots] \gets \RandomOrder(\mathbf{A})$\;
    \ForEach{$x_{P(i)}$ in order}{
        $\sigma \gets f \cup \{x_{P(i)}\}$\;
        \If{$|\sigma \cap \mathbf{A}| = d+1$}{
            $\mathcal{T} \gets \mathcal{T} \cup \{\sigma\}$\;
            \textbf{break}\;
        }
    }
}
\Return $\mathcal{T}$\;
\caption{\code{grow2d}}
\label{alg:grow2d}
\end{algorithm}

\begin{algorithm}[H]
\DontPrintSemicolon
\SetKwInOut{Input}{Input}
\SetKwInOut{Output}{Output}
\SetKwFunction{RandomSimplex}{RandomSeedSimplex}
\SetKwFunction{VisibleFacets}{VisibleFacets}
\SetKwFunction{RandomOrder}{RandomOrder}
\Input{$d$-dimensional lattice polytope $\Delta$ with lattice points $\mathbf{A}$}
\Output{Fine triangulation $\mathcal{T}$ of $\Delta$}
\BlankLine
$\sigma_0 \gets \RandomSimplex(\Delta)$\;
$\mathcal{T} \gets \{\sigma_0\}$\;
\While{$\mathcal{T}$ does not triangulate $\Delta$}{
    $\mathbf{A}_\mathrm{rem} \gets \{x \in \mathbf{A} : x \text{ not in any } \sigma \in \mathcal{T}\}$\;
    $[x_{P(0)}, x_{P(1)}, \dots] \gets \RandomOrder(\mathbf{A}_\mathrm{rem})$\;
    \ForEach{$x_{P(i)}$ in order}{
        $\mathcal{S} \gets \{f \cup \{x_{P(i)}\} : f \in \VisibleFacets(\mathcal{T}, x_{P(i)})\}$\;
        \If{$|\sigma \cap \mathbf{A}| = d+1$ for every $\sigma \in \mathcal{S}$}{
            $\mathcal{T} \gets \mathcal{T} \cup \mathcal{S}$\;
            \textbf{break}\;
        }
    }
}
\Return $\mathcal{T}$\;
\caption{\code{pushing}}
\label{alg:pushing}
\end{algorithm}

\section{Transformer Baselines}
\label{sec:transformer}

As a baseline for dualGNN, inspired by~\cite{yip2025transformingcalabiyauconstructionsgenerating}, we apply two transformer models to the task of generating FRTs of lattice polygons. These models differ slightly from CYTransformer: most notably, CYTransformer is encoder-decoder while, for simplicity, we study decoder-only models. Our two variants are similar (e.g., both use the standard attention mechanism~\cite{vaswani2023attentionneed}), differing primarily in how lattice geometry is exposed to the model.

The first variant is the simpler of the two, using the serialization (semi-analogous to \cite{yip2025transformingcalabiyauconstructionsgenerating})
\begin{equation}
    \code{x0,y0 x1,y1 \ldots\ | a0,b0,c0 a1,b1,c1 \ldots}.
\end{equation}
The vocabulary here consists of
\begin{enumerate}
    \item the special characters \code{PAD}, \code{SPACE}, \code{EOS}, and delimiters (\code{,}, \code{|}) as well as
    \item a token for each integer $0,\dots,C$ for $C$ the maximum integer needed, as described below.
\end{enumerate}
Points are input coordinate-wise $(x_i,y_i)$, translated so $\min_i(x_i)=\min_i(y_i)=0$, while simplices are lists of point-indices $(i,j,k)$. This means that $C$ is the larger of the maximum coordinate value (across all $x_i$ and $y_i$) and $N_\mathrm{pts}-1$. Positional information is supplied by a standard learned absolute positional embedding tied to sequence index. This serialization, in principle, can describe any polygon with $N_\mathrm{pts}\leq C+1$ and within the bounding box $[0,C]^2$, but we find it does not generalize strongly across polygons.

The second variant uses the same input/output stream, differing primarily in how point coordinates are exposed. Namely, points are directly input via their labels $0,\dots,N_\mathrm{pts}-1$ with their coordinates $(x_i,y_i)$ exposed via a variant of Rotary Position Embeddings~\cite{su2023roformerenhancedtransformerrotary} (RoPE). With \code{SPACE} no longer used to separate coordinate pairs, the sequences become more compact:
\begin{equation}
    \code{012\ldots|{$a_0$}{$b_0$}{$c_0$}\ldots}.
\end{equation}
To encode the coordinates, we use a variant of the RoPE encoding for which we encode a $3$D position $(x_i, y_i, s)$ for $s$ a `simplex index', taking value $s=0$ for the input points and $s=j+1$ for the three tokens forming the $j$th simplex in the output. This is achieved by splitting each attention head's $d_\mathrm{head}$ dimensions into three chunks $(d_x, d_y, d_s)$, with each chunk receiving a geometric-frequency rotation along its own coordinate. The \code{|} delimiter and special tokens (\code{PAD}, \code{EOS}) take sentinel position $(-1,-1,-1)$. The shared vocabulary makes this variant polygon-agnostic in principle.

\begin{figure}
    \centering
    \includegraphics[width=0.8\textwidth]{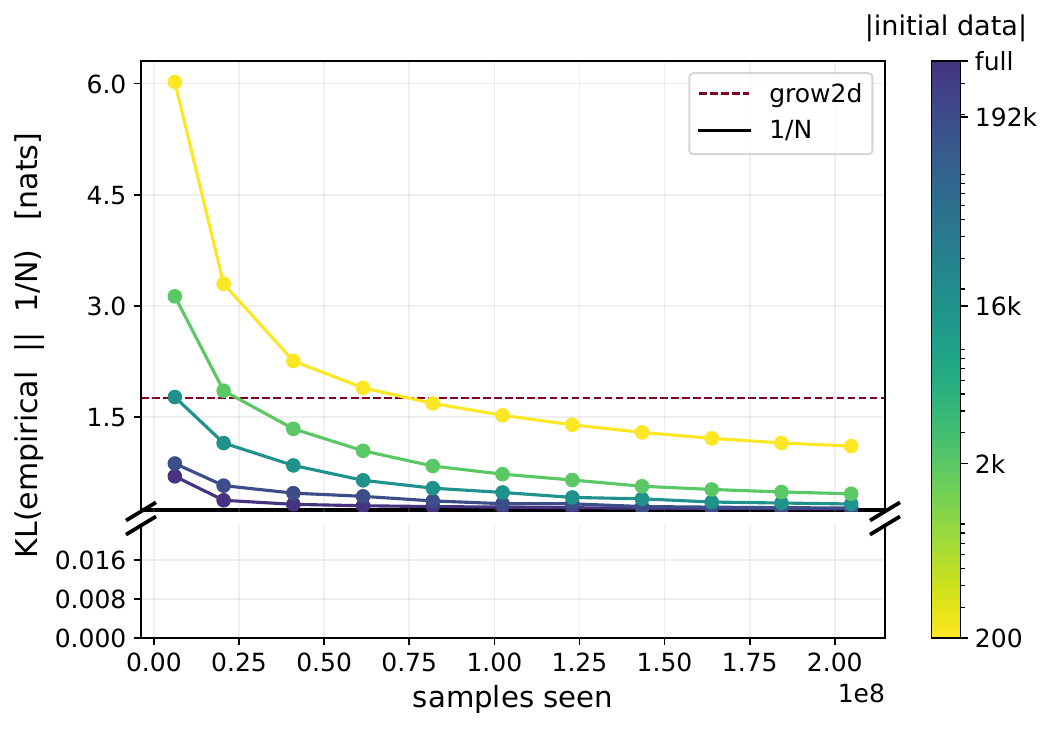}
    \caption{KL divergence to the uniform distribution over the course of training for the simple-encoding transformer on the polygon in \cref{fig:4x6tri}, at varying levels of initial training data scarcity. We only compare against \code{grow2d} since it generates samples quickly and somewhat uniformly.}
    \label{fig:simple_scarce}
\end{figure}

\begin{figure}
    \centering
    \includegraphics[width=0.8\textwidth]{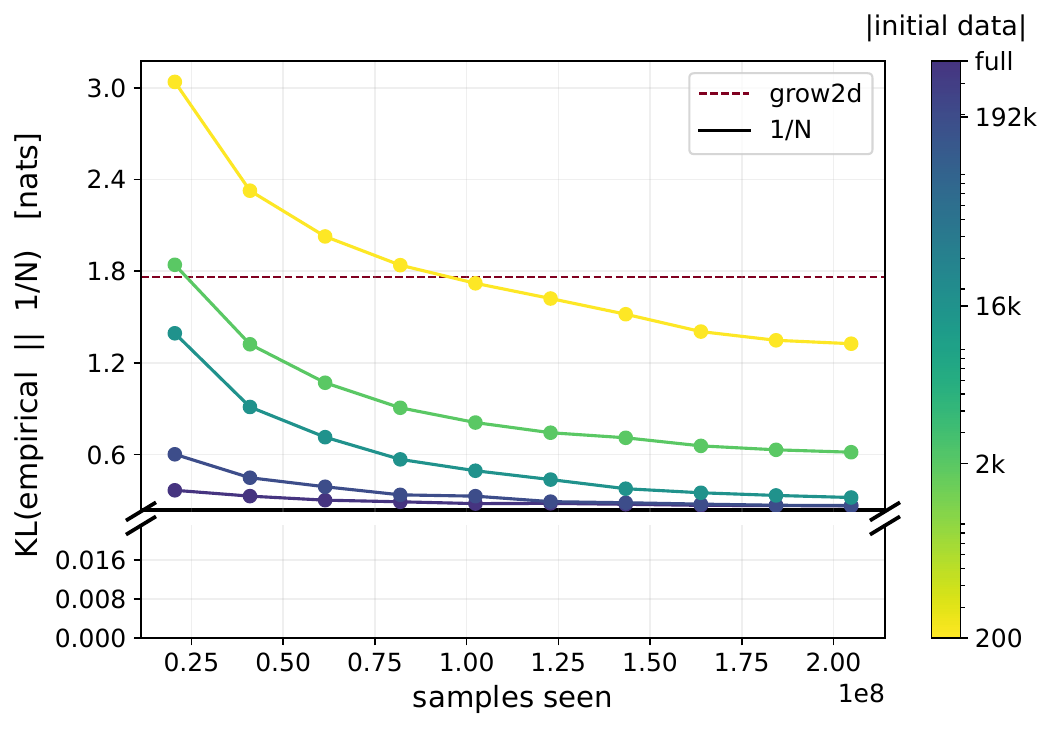}
    \caption{Same as \cref{fig:simple_scarce}, but for the RoPE-encoded transformer.}
    \label{fig:rope_scarce}
\end{figure}

\begin{figure}
    \centering
    \includegraphics[width=0.8\textwidth]{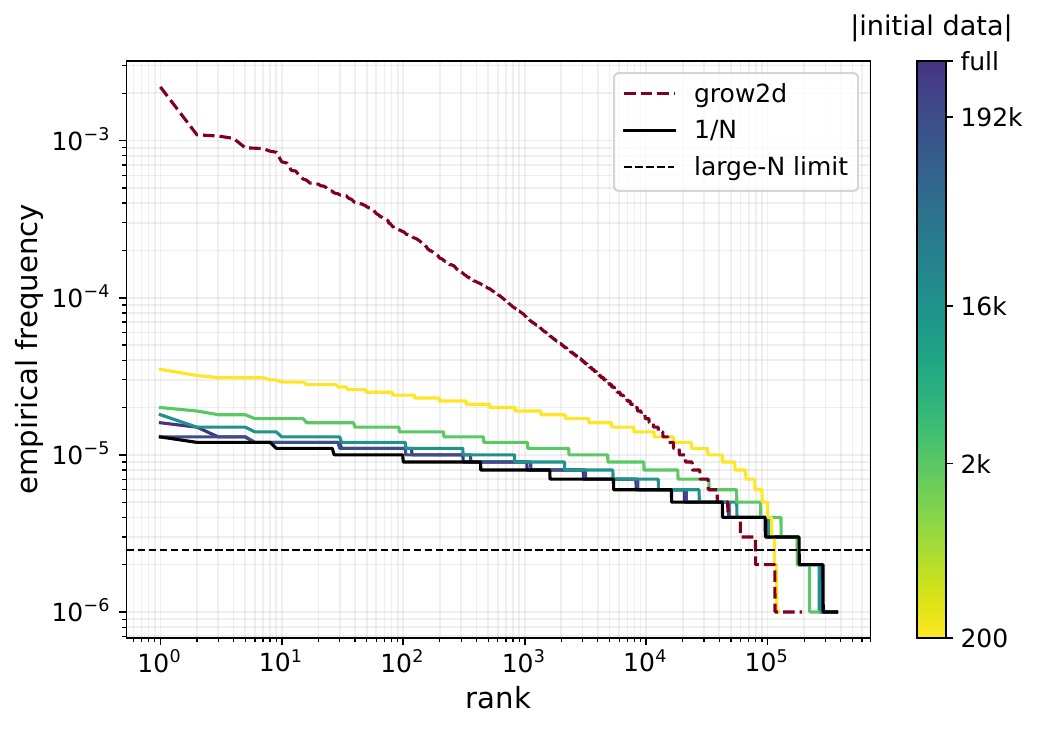}
    \caption{Rank-frequency plot of $10^6$ samples drawn from the trained RoPE transformer 
    on the polygon in \cref{fig:4x6tri}.}
    \label{fig:rope_rankfreq}
\end{figure}

We train both models on the polygon in \cref{fig:4x6tri} at varying levels of initial training data and sample at temperature $1$, allowing us to interpret the model's sampling behavior as the distribution it learned during training. \Cref{fig:simple_scarce,fig:rope_scarce} show KL divergence to the uniform distribution as a function of training, for the simple and RoPE encodings respectively. To visualize the final trained RoPE model, we draw $10^6$ samples and plot the frequency of each sampled triangulation versus its rank (in terms of highest frequency) in \cref{fig:rope_rankfreq}.

\begin{figure}[H]
    \centering
    \includegraphics[width=1.0\textwidth]{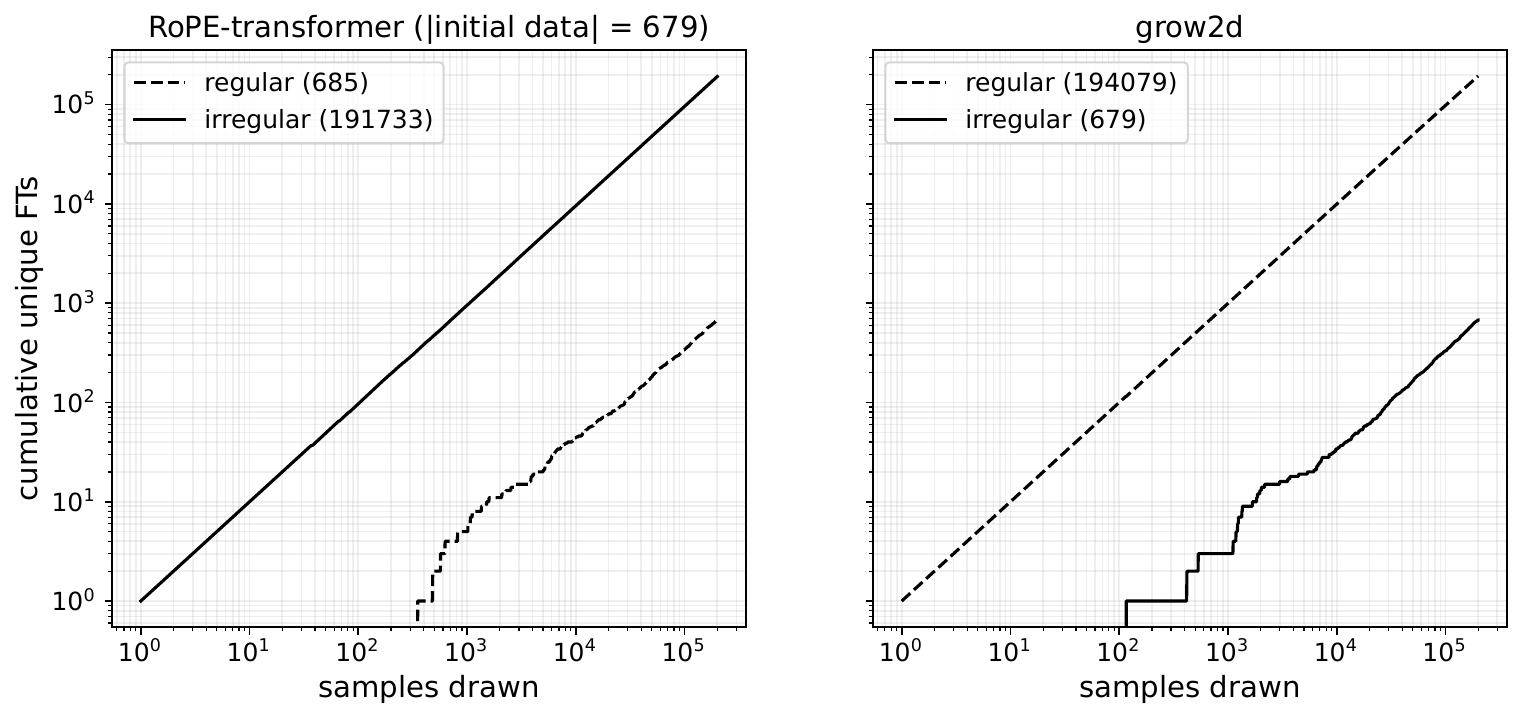}
    \caption{Number of regular and irregular triangulations generated for the polygon $[0,4]^2$ by the RoPE-based transformer trained on $679$ irregular triangulations (left) and \code{grow2d} (right). The RoPE transformer learns to sample irregular triangulations effectively, with $>95\%$ of the samples being unique irregular triangulations. Contrast this to \code{grow2d}, which generates only $\sim0.34\%$ irregular triangulations, roughly matching the $\sim0.2\%$ true fraction of irregular triangulations in $[0,4]^2$.}
    \label{fig:rope_irreg}
\end{figure}

These were the first models we studied (before even dualGNN). Both achieve strong single-polygon performance, especially the RoPE-variant which trains very quickly due to its compact serialization. These models are able to learn regularity, too --- see \cref{fig:rope_irreg} where we apply a model trained on only $679$ irregular triangulations of $[0,4]^2$ and compare its success in generating irregular triangulations to \code{grow2d}. 
We ultimately moved away from these models because they did not generalize across polygons, even when trained on many different ones simultaneously. We took this as motivation to better integrate the problem's symmetries into the architecture, leading to dualGNN.

\end{document}